\documentclass{aa}
\usepackage{txfonts}
\usepackage{graphicx,rotate,psfig,subfigure,supertabular,times}
\begin{document}
   \title{The kinematics in the pc-scale jets of AGN}
  \subtitle{The case of S5 1803+784}

   \author{S. Britzen\inst{1,2} \and N.A. Kudryavtseva\inst{1,3,4}\fnmsep\thanks{Member of the International Max Planck Research School (IMPRS) for Radio and Infrared Astronomy at the Universities of Bonn and Cologne} \and A. Witzel\inst{1} \and R.M. Campbell\inst{5}
       \and E. Ros\inst{1} \and M. Karouzos\inst{1,\star} \and A. Mehta\inst{6} \and M.F. Aller\inst{7} \and H.D. Aller\inst{7} \and T. Beckert\inst{1} \and J.A. Zensus\inst{1}}

   \offprints{Silke Britzen}

   \institute{Max-Planck-Institut f\"ur Radioastronomie, Auf dem H\"ugel 69,
      D-53121 Bonn, Germany\\
         \email{sbritzen@mpifr-bonn.mpg.de} \and Landessternwarte, K\"onigstuhl, D-69117 Heidelberg, Germany \and St. Petersburg State University, Petrodvoretz, St.-Petersburg, Russia \and Physics Department, University College Cork, Cork City, Cork, Ireland \and Joint Institute for VLBI in Europe, Oude Hoogeveensedijk 4, 7991 PD Dwingeloo, The Netherlands \and International University Bremen, Postfach 750561, D-28725, Bremen, Germany \and Astronomy Department, University of Michigan, Ann Arbor, MI 48109-1042, USA}

   \date{\today}

  \abstract
   {BL Lac objects show core-jet structures with features moving outwards along
the jet. We present a kinematic analysis of jet component motion in the
pc-scale jet of the BL Lac object S5 1803+784, which does not reveal long-term
outward motion for most of the components.}
   {S5 1803+784 shows complex kinematic phenomena; understanding these provides
new insights into the emission processes in BL Lac objects and possibly into
the differences between quasars and BL Lac objects.}
   {The blazar S5 1803+784 has been studied with VLBI at $\nu$=1.6, 2.3, 5,
8.4, and 15 GHz between 1993.88 and 2005.68 in 26 observing runs. We
(re)analyzed the data and present Gaussian model-fits. We collected the already
published kinematic information for this source from the literature and
re-identified the components according to the new scenario presented in this
paper. Altogether, 94 epochs of observations have been investigated.}
   {A careful study of the long-term kinematics reveals a new picture for jet
component motion in S5 1803+784. In contrast to previously discussed motion
scenarios, we find that the jet structure within 12 mas of the core can most
easily be described by the coexistence of several bright jet features that
remain on the long-term at roughly constant core separations (in addition to the already known
``stationary'' jet component $\sim$ 1.4 mas) and one faint component moving
with an apparent superluminal speed ($\sim$19$c$, based on 3 epochs). While
most of the components maintain long-term roughly constant distances from the core, we
observe significant, smooth changes in their position angles. We report on an
evolution of the whole jet ridge line with time over the almost 12 years of
observations. The width of the jet changes periodically with a period of
$\sim$8--9 years. We find a correlation between changes in the position angle
and maxima in the total flux-density light-curves. We present evidence for a
geometric origin of the observed phenomena and discuss possible models.}
{We find evidence for a significantly different scenario of jet component
motion in S5 1803+784 compared to the generally accepted one of outwardly
moving jet components, and conclude that the observed phenomena (evolution of
the jet ridge line, roughly constant component core separations but with
significant position angle changes) can most easily be explained within a
geometric model.}

   \maketitle

\section{Introduction}


Jet component motion in most Active Galactic Nuclei (AGN) has been successfully
described by relativistic shock-in-jet motion directed {\sl away} from the
presumed central engine (e.g., Marscher \& Gear 1985). Bulk relativistic motion
explains many observed jet-related phenomena and is also important with regard
to the observed short-term radio variability (IDV, Wagner \& Witzel 1995 and
references therein) providing the necessary Doppler factors to reduce the
observed brightness temperatures to the Compton limit (Kellermann \&
Pauliny-Toth 1969). However, as more detailed VLBI-monitoring reveals, this
model can not easily explain all the observed motion phenomena in a growing
number of AGN. In many sources {\it stationary} features or even motion towards
the core have been observed (e.g., Alberdi et al. 2000; Kellermann et al. 2004; Britzen et al. 2007).\\

Taking the core as the easternmost component, the brightest jet feature (in
8.4~GHz observations) at $r=1.4$ mas used to be one of the most prominent
candidates for a stationary component (e.g., Eckart et al. 1986; Schalinski et
al. 1988). However, as we showed in an earlier work, this so-called stationary
component has a varying core-separation (Britzen et al. 2005a). Based on a
more detailed analysis of new data, we show in this paper that most of the jet
components in the pc-scale jet of S5 1803+784 (out to $r\sim 14$ mas) seem to
remain at roughly constant core separations on the long term, but their position
angles undergo significant change. We present the results of a detailed
kinematic analysis of the pc-scale jet in this source and discuss possible
explanations of these phenomena. \\

\subsection{The blazar S5 1803+784}

As a member of the complete S5 sample of 13 flat-spectrum radio sources at high
declinations (Witzel 1987), S5 1803+784 has been observed repeatedly at many
frequencies and angular resolutions since the late 1970s (e.g., Eckart et
al. 1986, 1987; Schalinski et al. 1988; Witzel et al. 1988; Krichbaum et
al. 1993; Britzen \& Krichbaum 1995). These observations revealed the complex
morphology of the milli-arcsecond-jet of this intra-day variable source (Wagner
\& Witzel 1995 and references therein). The source shows a pronounced jet with
prominent jet components located at core separations of $r = 1.4$, 5, and
12~mas (e.g., Eckart et al. 1986). Geodetic and astronomical VLBI data
gathered at 8.4 and 5~GHz between 1979--1987 showed that the component located
at $r=1.4$~mas appears stationary (e.g., Schalinski et al. 1988, Witzel et
al. 1988). Several authors confirm this constant core separation, leading to a
subluminal velocity of $\beta_{{\rm app\/}}<0.74h^{-1}$ (e.g., Cawthorne et
al. 1993). Britzen et al. (2005a) presented an overview over the pc-scale
structure of S5 1803+784, including discussion of the kinematics of this {\it
stationary} $r=1.4$~mas component. Based on a jet-component identification
procedure in which
the brightest jet component in every epoch is associated with this one, this
$r=1.4$~mas component approaches the core whenever a new jet component was 
or was about to be ejected
from the core. Thus this {\it stationary} component was found to have
non-constant core separation (``oscillatory'' type behaviour). Britzen et al. (2005a) explain this motion
scenario within a model of reconfinement shocks. In addition, three jet
components were reported to move with apparent superluminal speeds of
7--10$c$.\\

S5 1803+784 shows curved jet structure on all scales probed with VLBI. The
curvature found in the cm-regime is described in Britzen et al. (2005b). In
the mm-regime, the inner jet components move superluminally with expansion
rates of $0.14\pm0.04$ mas yr$^{-1}$ and $0.07\pm0.05$ mas yr$^{-1}$, with
possibly variable proper motions (0.02-0.28 mas yr$^{-1}$) along a curved path,
suggesting helical motion (e.g., Krichbaum 1990). This source also belongs to
the group of misaligned AGN (Antonucci et al. 1986; Appl et al. 1996), with
the pc- and kpc-scale jet aligned almost perpendicularly. The probable
transition region between the pc- and kpc-scale jet has been investigated in
world array observations by Britzen et al. (2005b). A summary of the
observations of the kpc-scale structure can be found therein.\\

Velocities computed in the previously cited papers have been determined with
differing sets of cosmological parameters. Here, we use the following: $H_0=71$
km s$^{-1}$Mpc$^{-1}$, $\Omega_{\rm M}$= 0.27, $\Omega_{\rm \Lambda}$= 0.73,
and take $z=0.68$ (Stickel et al. 1993; Lawrence et al. 1987) 

\subsection{New scenario for jet motion in S5 1803+784}

This is the third paper in a series of papers on the radio structure of S5
1803+784 and presents an alternative  explanation for the observed kinematics
in the pc-scale jet compared to the scenario presented in Britzen et
al. (2005a). Preliminary results have already been published in Kudryavtseva
et al. (2006).\\ 

An obvious difference between the motion scenario presented in this paper
compared to that in Britzen et al. (2005a) is the significantly larger
time span covered by the observations upon which the present one is based. By
combining observations covering more than 20 years (in the case of the 1.6 GHz
observations), it is possible to investigate the long-term motion. This
long-term behaviour shows no significant outward motion of most of the
components in this source.


The most prominent jet component in the pc-scale jet of S5 1803+784 at
frequencies between 5 and 22 GHz is the 1.4mas component to the west of the
easternmost, brightest component.  This $r=1.4$~mas component has repeatedly
been characterized as {\it
stationary}. As densely sampled geodetic observations have shown, this {\it
stationary} component approaches the core from time to time, showing a
sort of {\it oscillatory} behavior (Britzen et al. 2005a). In this paper we
take note of the fact that since Eckart et al.\ (1986) other jet components
have repeatedly been observed at similar core separations of e.g., 6 and 12 mas
(Britzen et al. 2005b and references therein). \\

Despite the fact that some jet components have been observed at roughly
constant core separations over long periods, they have been identified in the
literature with components
showing long-term outward motion.  We here propose a new
component identification assuming that the components appearing repeatedly at
the same distance from the core indeed represent the same component. We compare
our results with results published in the literature and apply our
identification scheme to these data and find that the literature data are fully
consistent with our new identification scenario.\\

Several {\it stationary} components have been reported for a number of other 
AGN in the literature. These components remain at a similar position in core
separation and position angle. However, here we encounter jet components
that remain at roughly constant core separations but show significant changes
in their position angle.\\ 

We use multi-frequency data to study the kinematics of the components, present
a correlation analysis among a number of properties including the flux-density
evolution, propose an alternative identification scenario, and conclude with
the resulting implications. The structure of this paper is organized as
follows:  section~\ref{observations} overviews the data sets that have been
observed and analyzed or re-analyzed.  Section~\ref{results} presents the
results of a detailed model-fitting analysis of the 26 epochs of observation
and shows evidence for an alternative identification scenario and unusual jet
kinematics in S5 1803+784 based on 94 epochs altogether (data re-analyzed plus literature data). 
In addition, we compare our results with kinematic information derived from the literature.
Section~\ref{summary} summarizes the kinematic results. Section~5 presents
correlations and periodicities from our analysis. Finally,
section~\ref{discussion} briefly discusses the implications for our
understanding of kinematics and emission processes in AGN.

\section{Observations and data reduction}
\label{observations}

We have (re-)analyzed 26 multi-frequency ($\nu$=1.6, 2.3, 5, 8.4, and 15 GHz) VLBI observations of 
S5~1803+784, performed between 1993.88 and 2001.87 with heterogeneous arrays:
VLBI observations at 1.6 and 2.3 GHz conducted 
by Marcaide et al. (1995a, 1995b, 1997); at 5 GHz by Marcaide et al.
(1995a, 1995b, 1997), Guirado et al. (2001), and Gurvits et al. (priv. comm.);
at 8.4 GHz by P$\acute{\rm e}$rez-Torres et al. (2000), Ros et al. (2000, 2001,
in prep.); and at 15 GHz by P$\acute{\rm e}$rez-Torres et al. (2000, in prep),
Kellermann et al. (1998), Zensus et al. (2002), Lister \& Homan 2005. The data
have been fringe-fitted and calibrated before by the individual observers. For
calibration details we refer the reader to the original publications. \\

In addition, we performed space-VLBI observations at 1.6 GHz in 2001.87. These
observations were scheduled for the VLBA plus HALCA. Unfortunately we did not
obtain any data from ground-space baselines and use this observation as a
purely ground-based experiment. The data have been fringe-fitted within AIPS.\\

We fit circular Gaussian components to each of the data sets at each frequency
using the {\it difmap} package (v.2.4b, Shepherd 1997). In
order to find the optimum set of components and parameters, we fit every
data set starting from a point-like model. We used circular Gaussian components 
in order to avoid the effects of correlations among weakly-constrained
axial-ratio estimates and those of other parameters of the Gaussian components.

\subsection{Determination of the uncertainties}

The determination of model-fit parameter uncertainties is a complex topic and
no satisfying solution to this problem has been presented so far. A detailed
discussion of model-fitting uncertainties and their influence on the parameters
is presented in Britzen et al. (2007, 2008). The uncertainties of each data set in
general are influenced by different {\it u-v} coverages, calibrations, sensitivities,
etc. The relative positional errors are different at different core separations
and the brightness of the individual component affects the fitted position as
well. Uncertainties should reflect all these different error sources. In order
to obtain reliable error estimates we used two different algorithms to determine
the model-fitting uncertainties. 
In the first, the uncertainties for the core separation and the position angle are given by
the formulas $\Delta r = (d \sigma_{rms} \sqrt{1 + S_{peak} / \sigma_{rms}}) /
2 S_{peak}$ and $\Delta \theta = \arctan (\Delta r / r)$, where $\sigma_{rms}$
is the residual noise of the map after the subtraction of the model, $d$ is the
full width at half maximum (FWHM) of the component and $S_{peak}$ is the peak
flux density (Fomalont 1999). However, this formula tends to underestimate the
uncertainty if the peak flux-density is very high or the width of the component is
small. In the second, we calculated all uncertainties by comparing
different model fits ($\pm$1 component) obtained for the same set of data.
These uncertainties reflect the possible parameter ranges for the individual
components within model-fitting. To be conservative, we selected the maximum 
of the uncertainties from the two methods as the value of the uncertainty to
use in subsequent analysis.

\subsection{Data table}

\begin{figure*}[h]
\subfigure[]{\psfig{figure=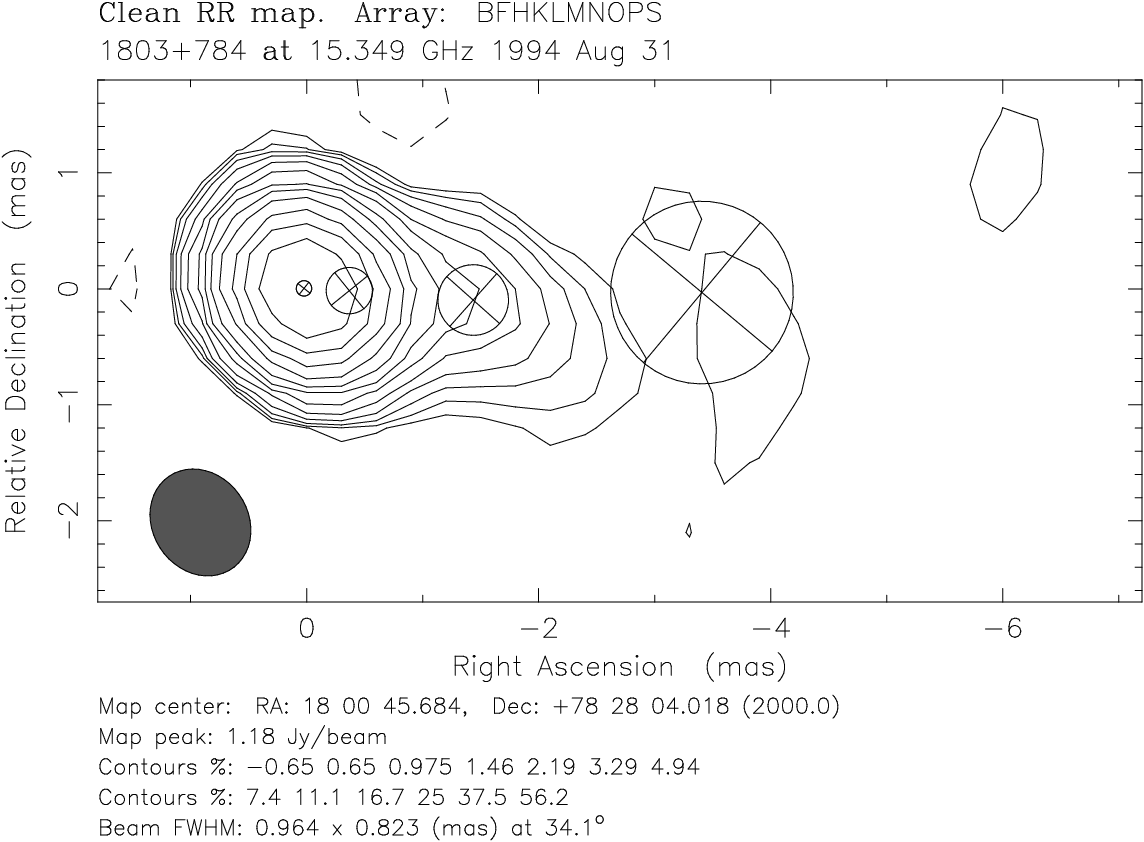,width=7.5cm}}
\hspace*{2cm}\subfigure[]{\psfig{figure=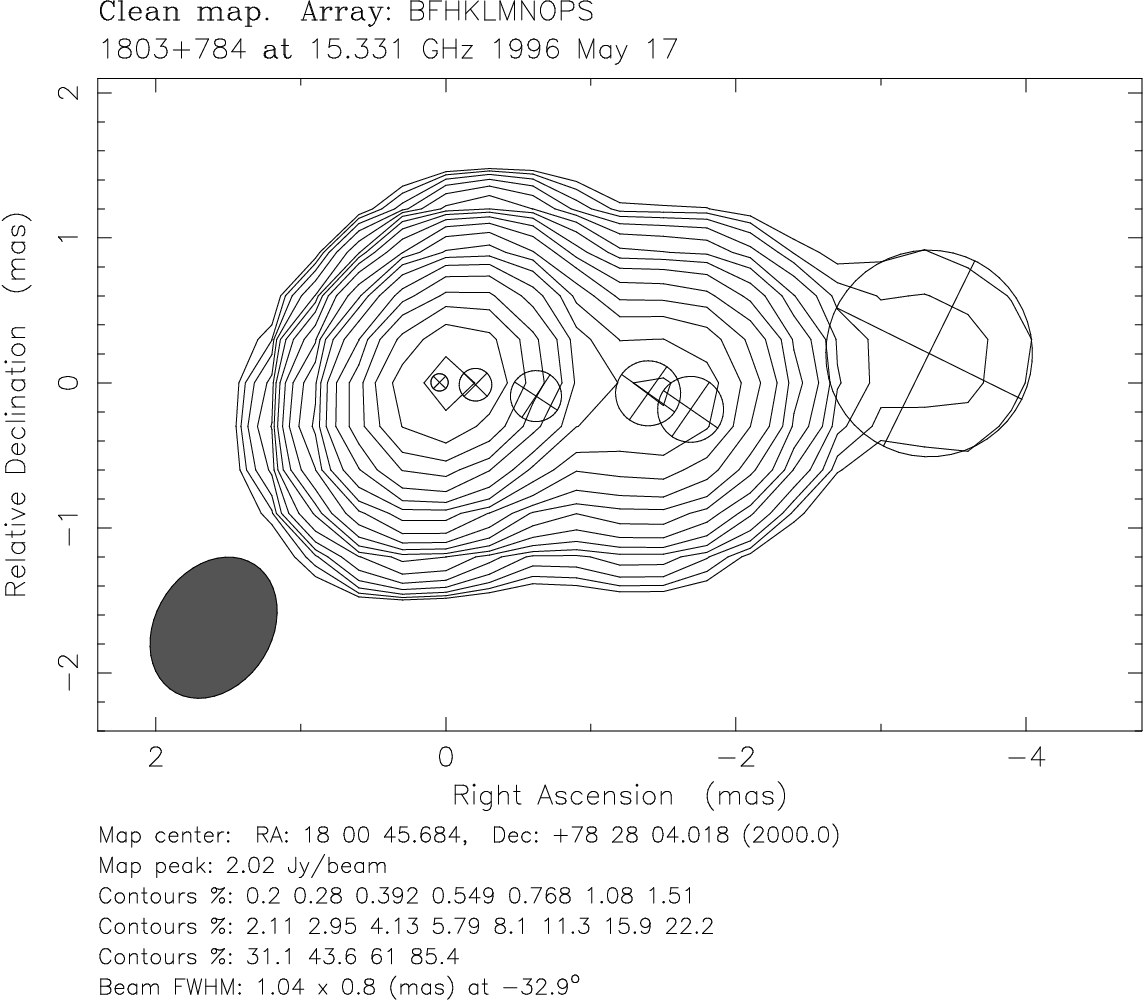,width=7.5cm}}\\
\subfigure[]{\psfig{figure=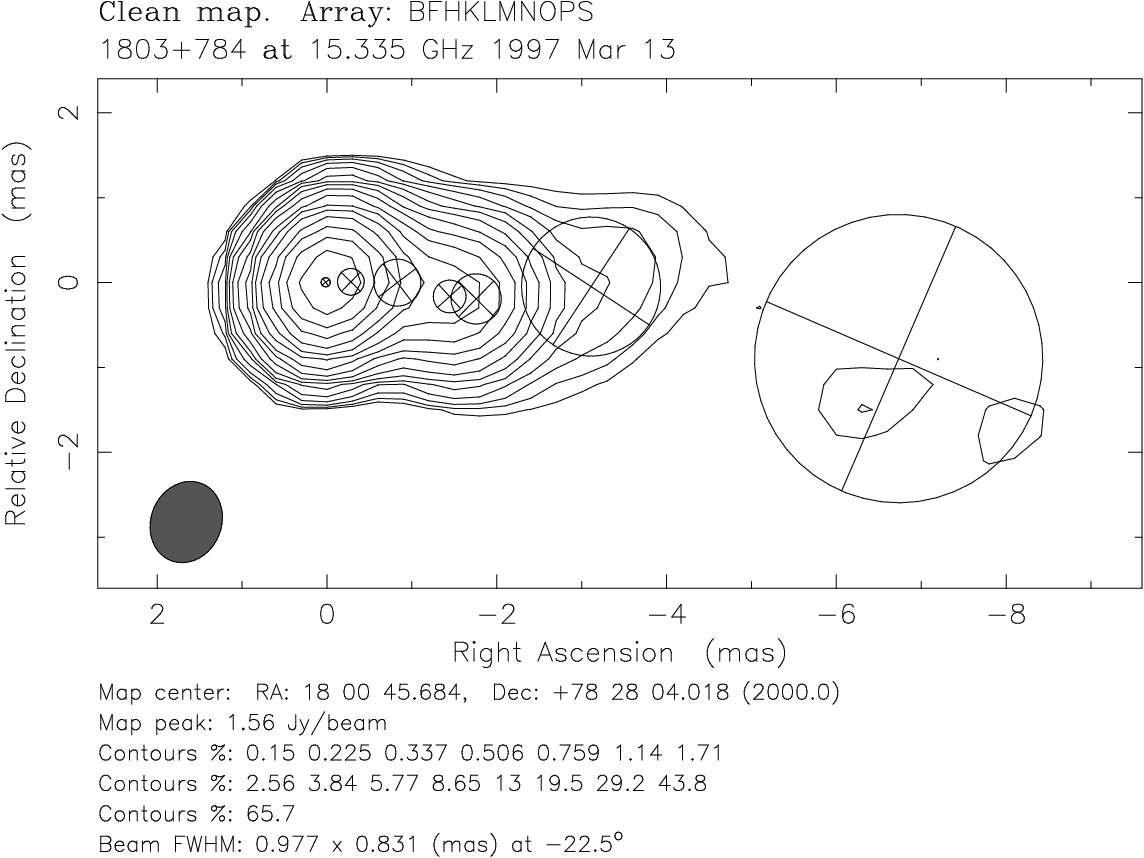,width=7.5cm}}
\hspace*{2cm}\subfigure[]{\psfig{figure=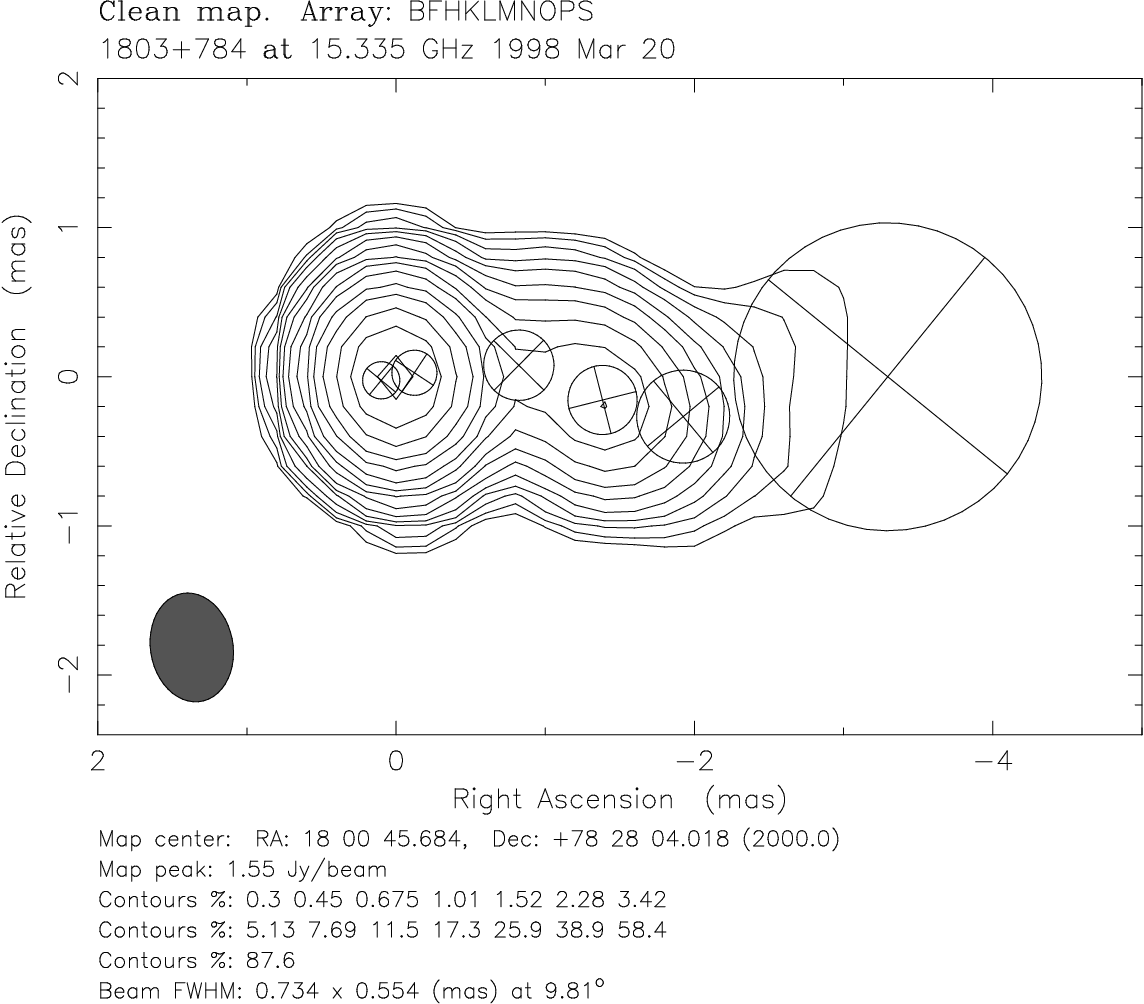,width=7.5cm}}
\subfigure[]{\psfig{figure=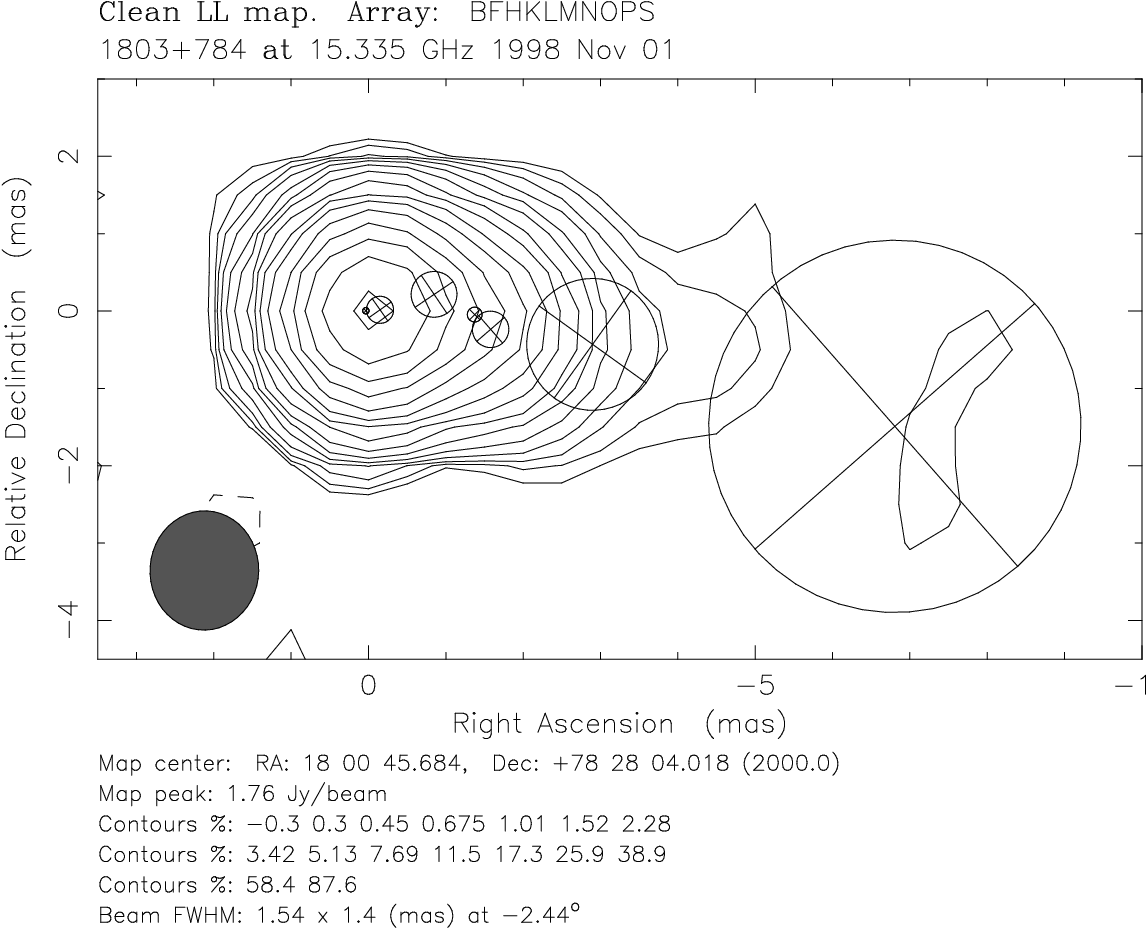,width=7.5cm}}
\hspace*{2cm}\subfigure[]{\psfig{figure=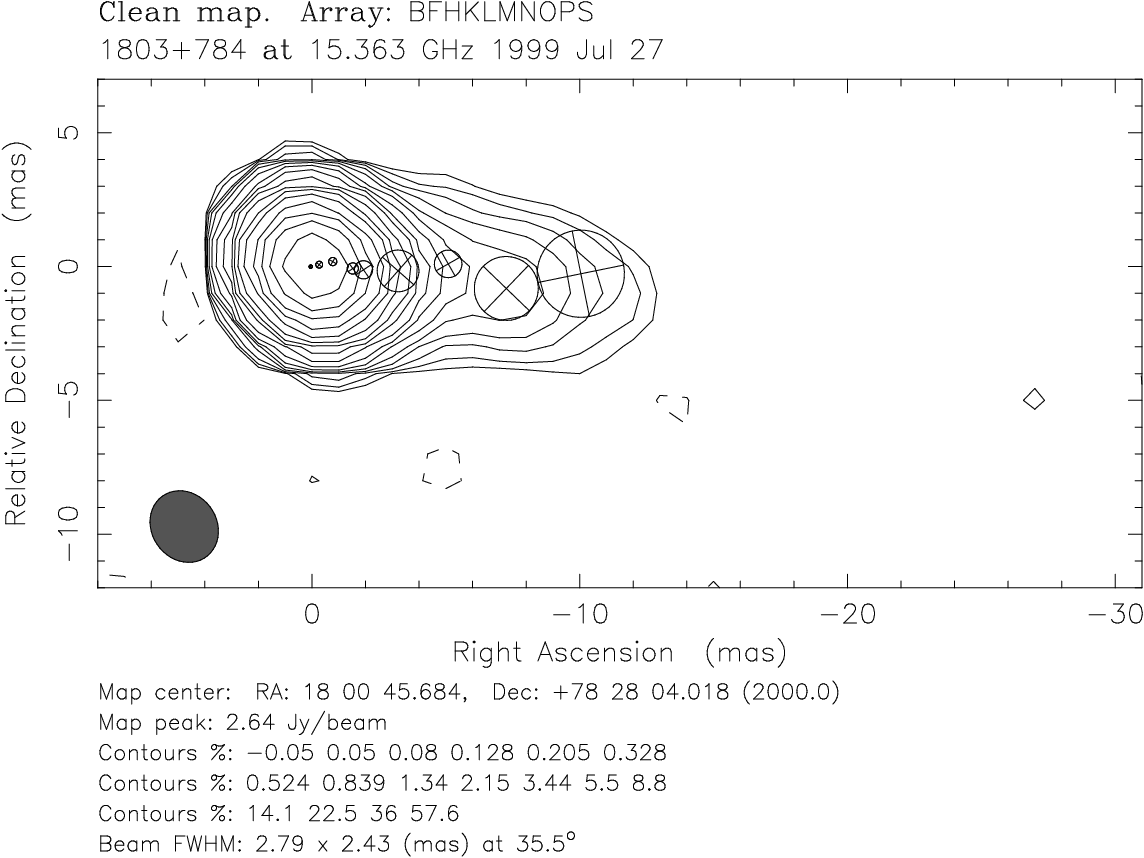,width=7.5cm}}
\caption{Results of hybrid imaging (contours) and circular Gaussian model fitting (circles with crosses) for S5 1803+784 at 15 GHz.}
\label{bilder1}
\end{figure*}

\begin{figure*}[h]                                                                                               
\subfigure[]{\psfig{figure=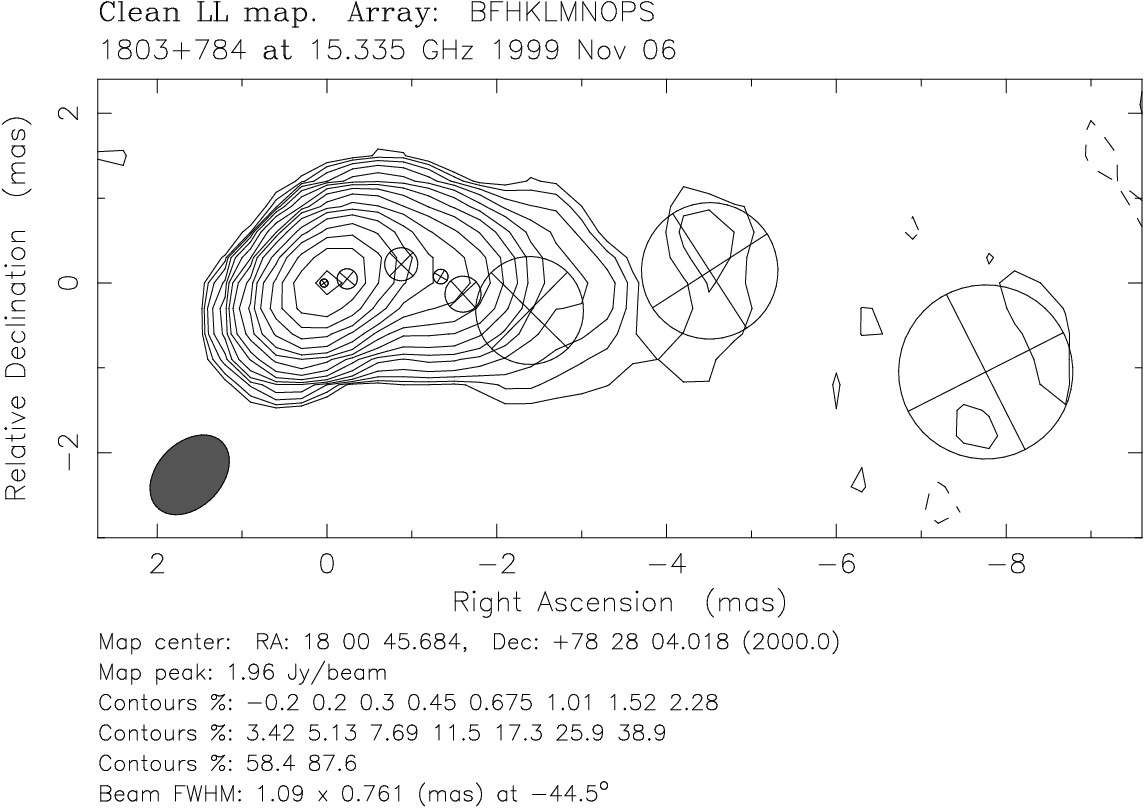,width=7.5cm}}
\hspace*{2cm}\subfigure[]{\psfig{figure=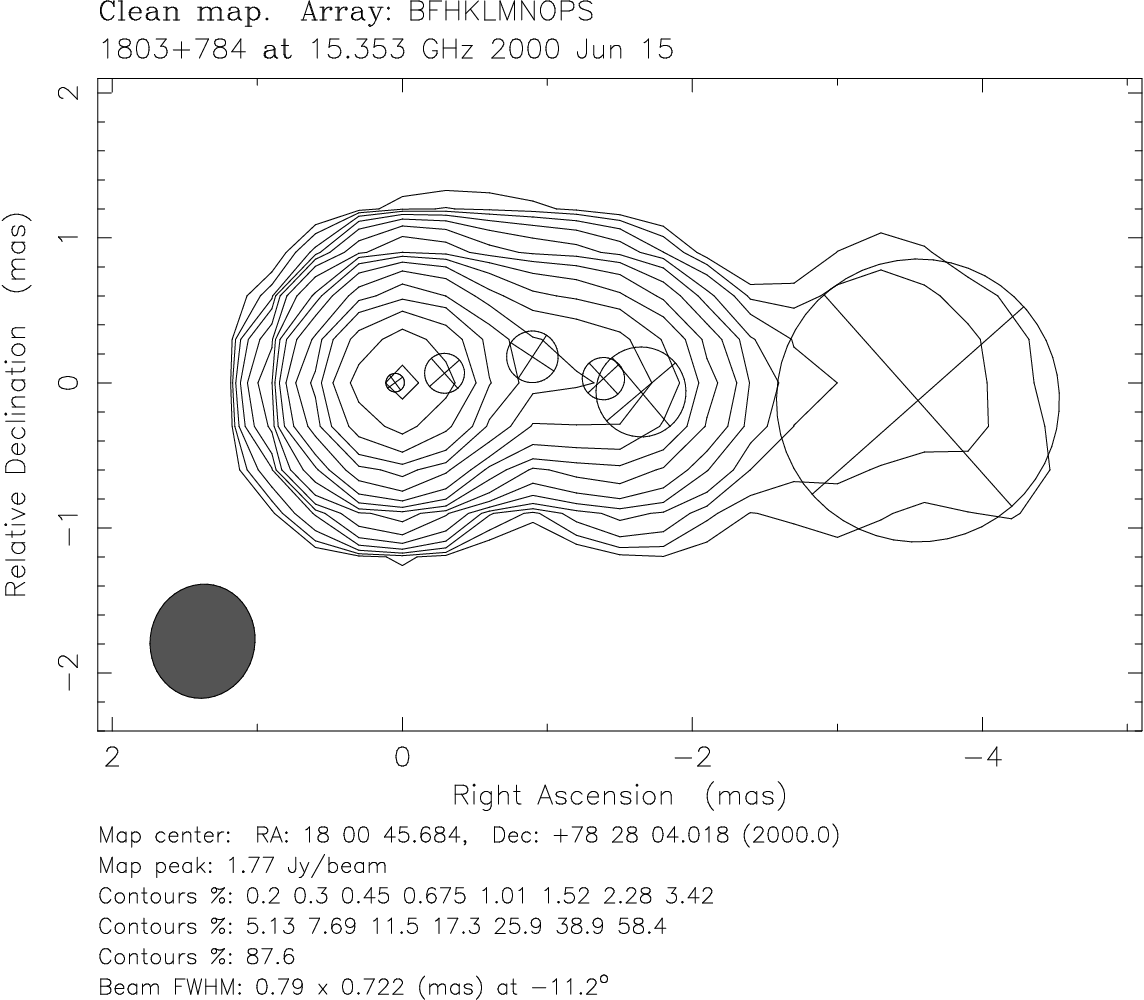,width=7.5cm}}
\subfigure[]{\psfig{figure=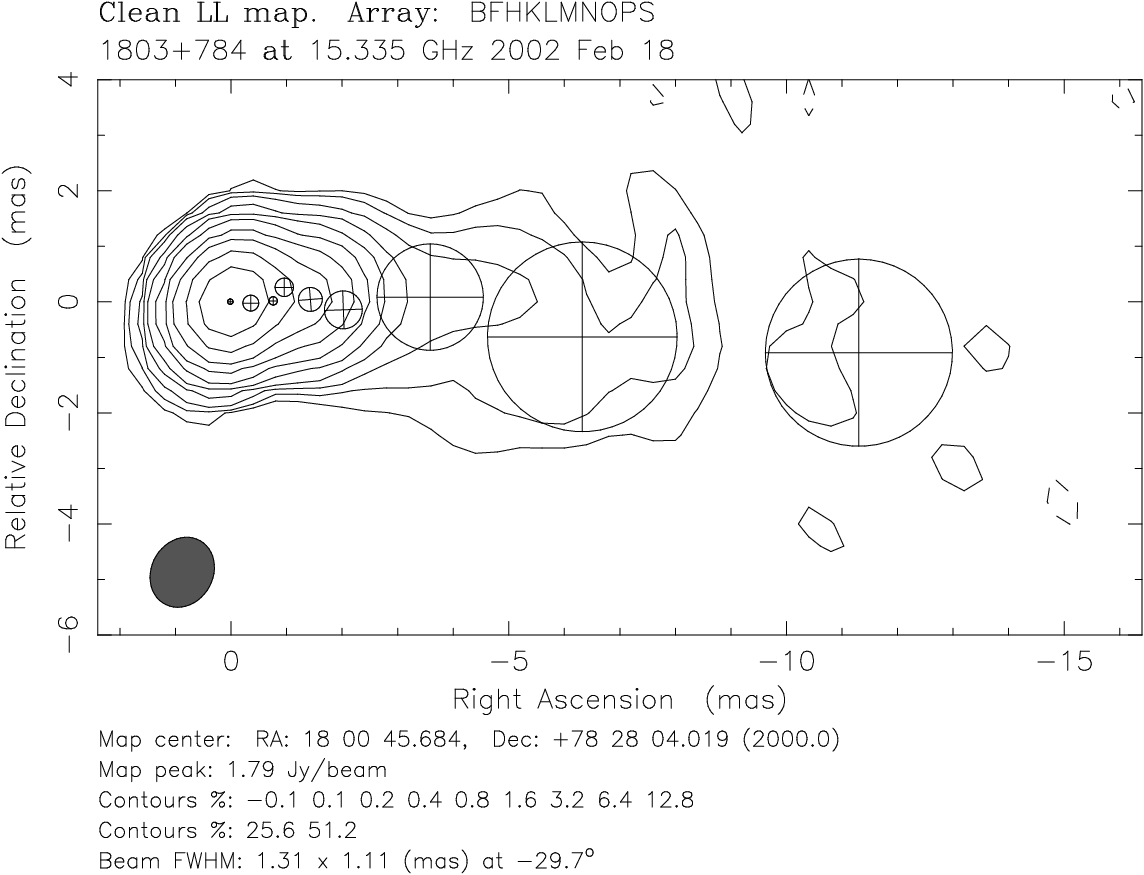,width=7.5cm}}
\hspace*{2cm}\subfigure[]{\psfig{figure=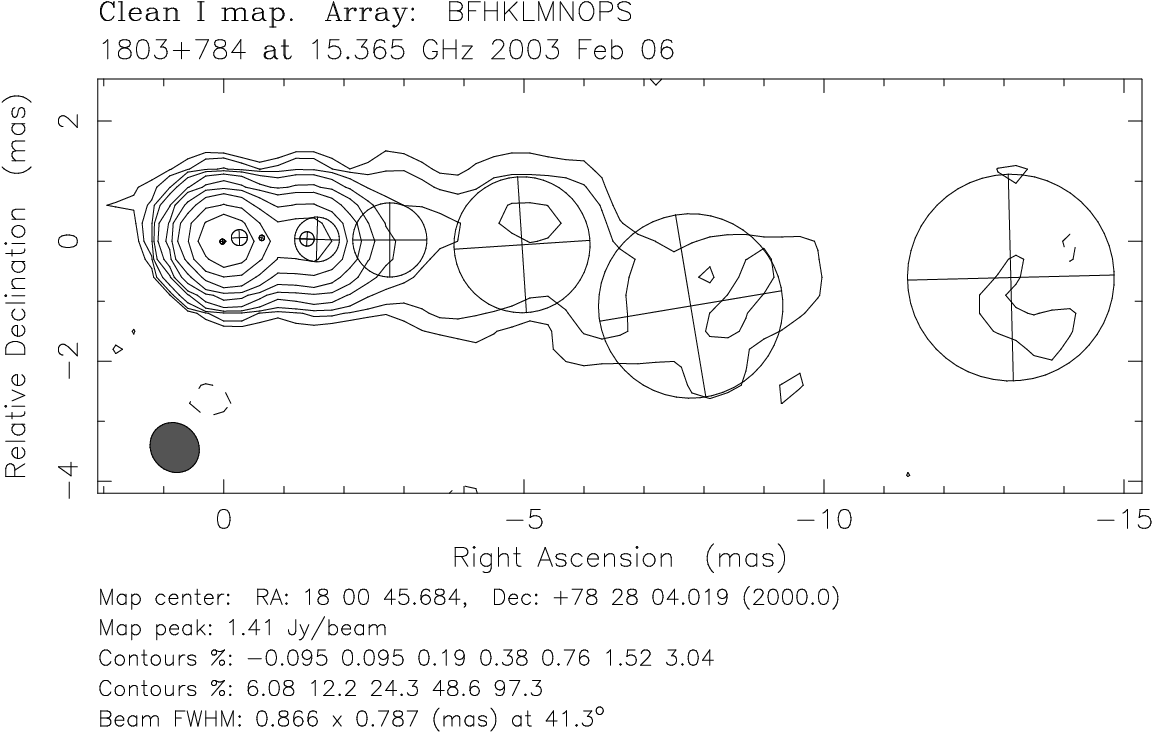,width=7.5cm}}
\subfigure[]{\psfig{figure=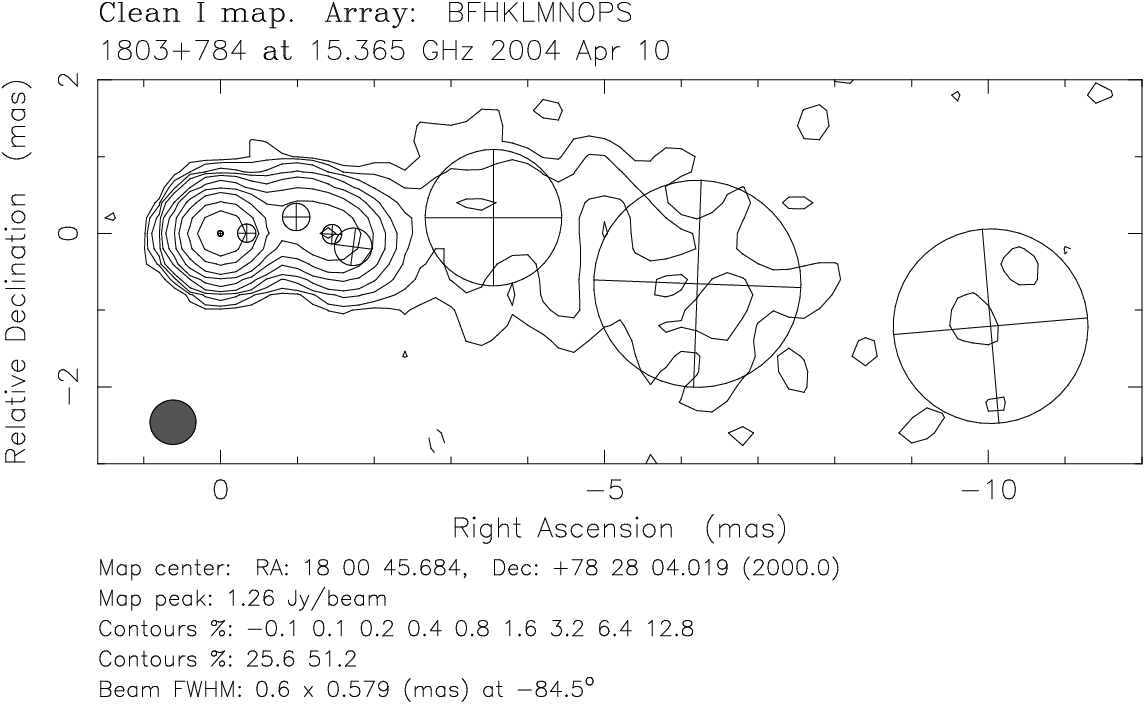,width=7.5cm}}
\hspace*{2cm}\subfigure[]{\psfig{figure=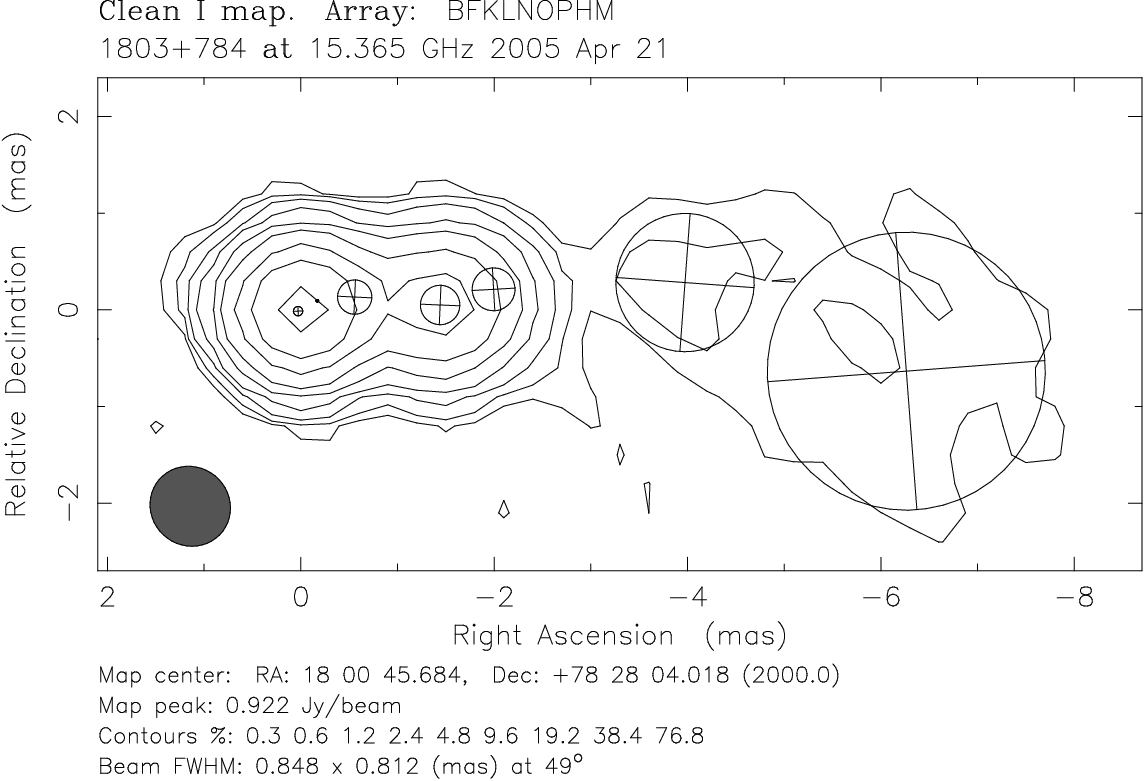,width=7.5cm}}\\
\subfigure[]{\psfig{figure=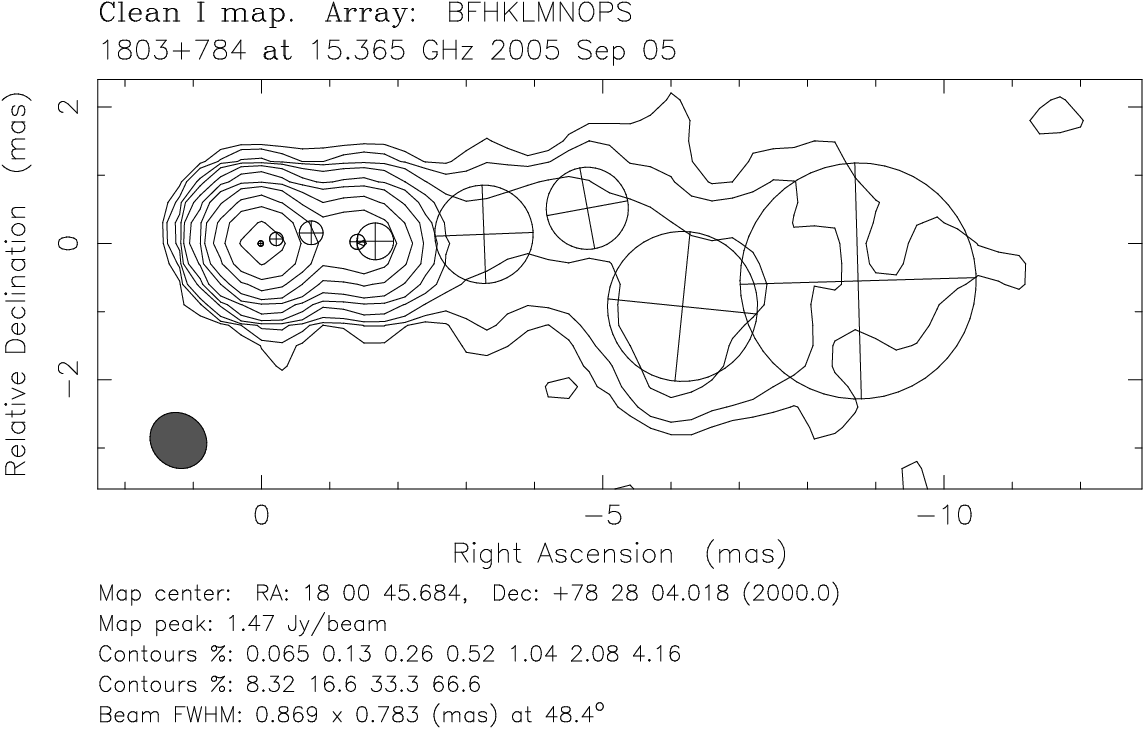,width=7.5cm}}
\caption{Results of hybrid imaging (contours) and circular Gaussian model fitting (circles with crosses) for S5 1803+784 at 15 GHz; continuation of 
figure~\ref{bilder1}.}
\label{bilder2}
\end{figure*}

\begin{figure*}[h]
\subfigure[]{\psfig{figure=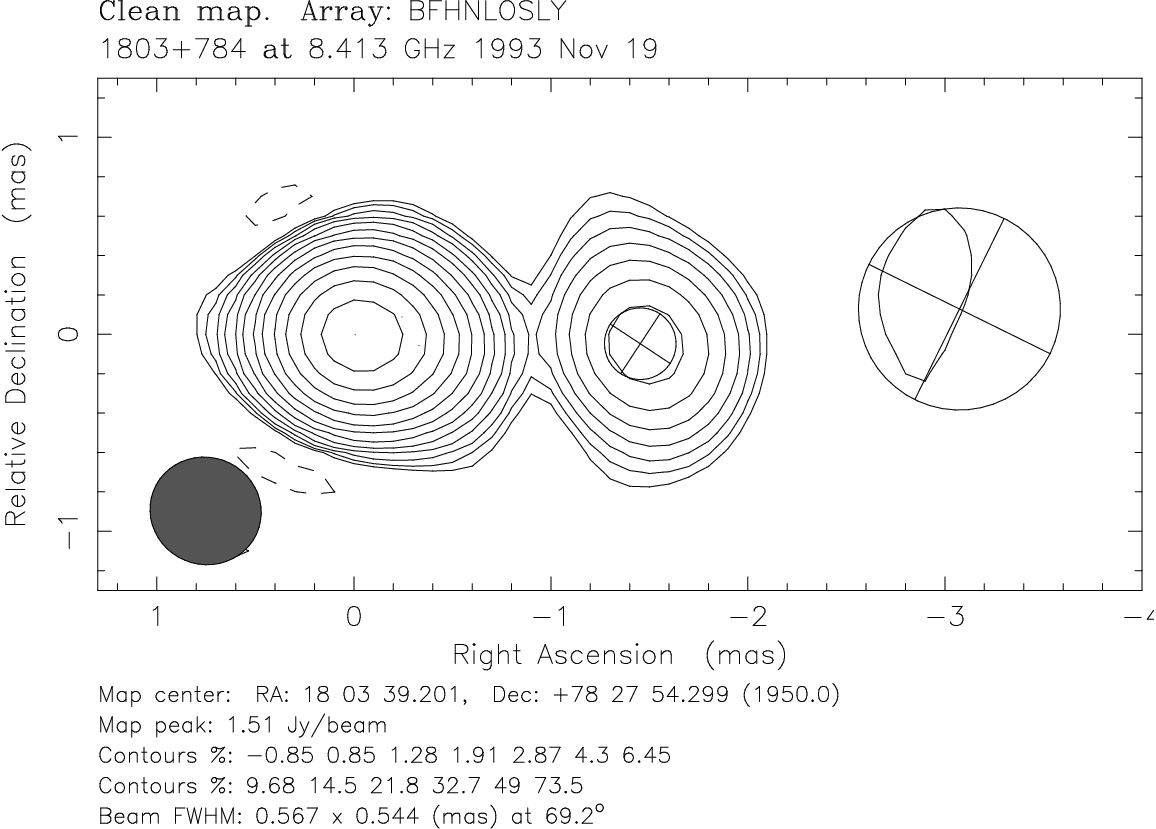,width=7.5cm}}
\hspace*{2cm}\subfigure[]{\psfig{figure=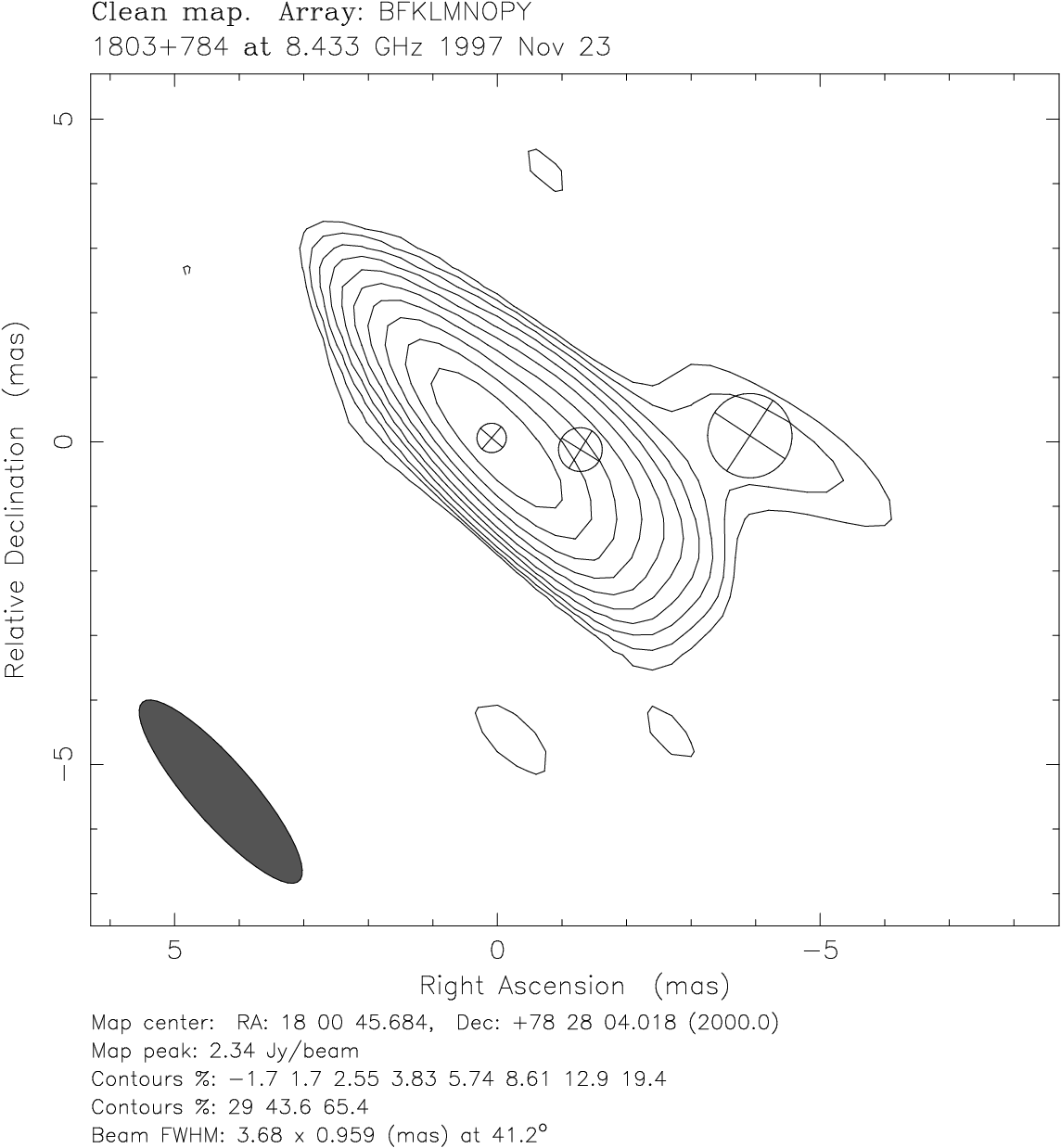,width=7.5cm}}\\
\subfigure[]{\psfig{figure=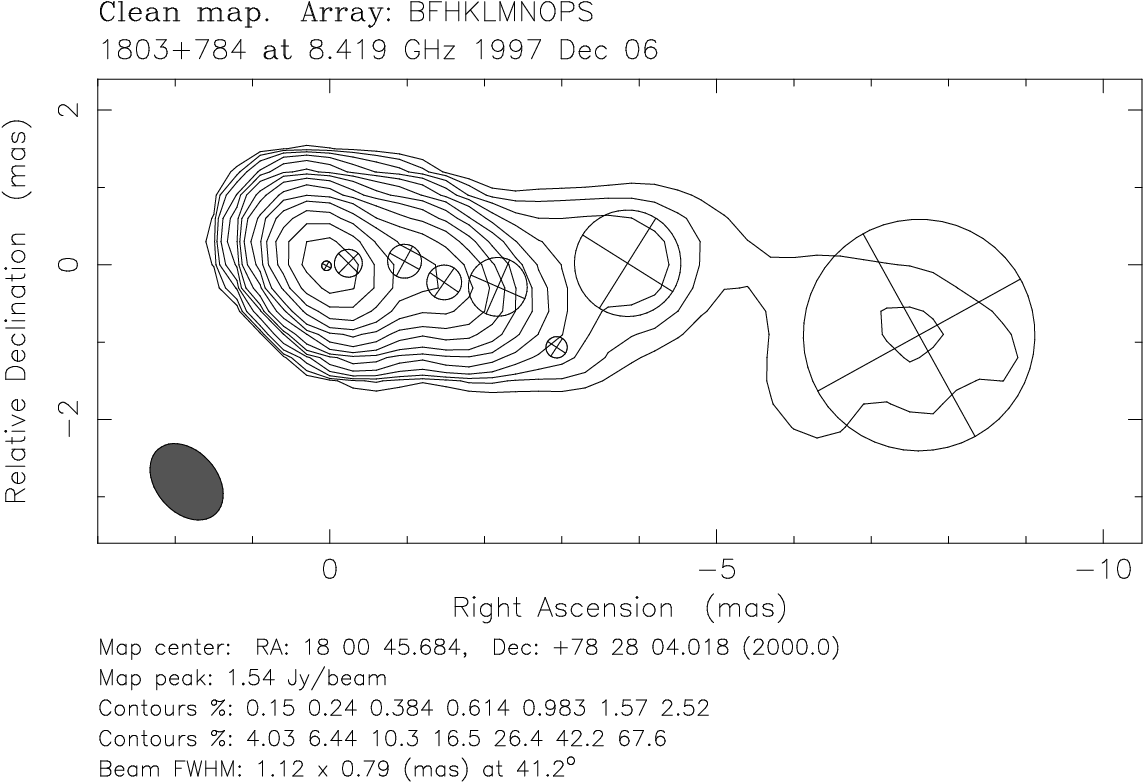,width=7.5cm}}
\hspace*{2cm}\subfigure[]{\psfig{figure=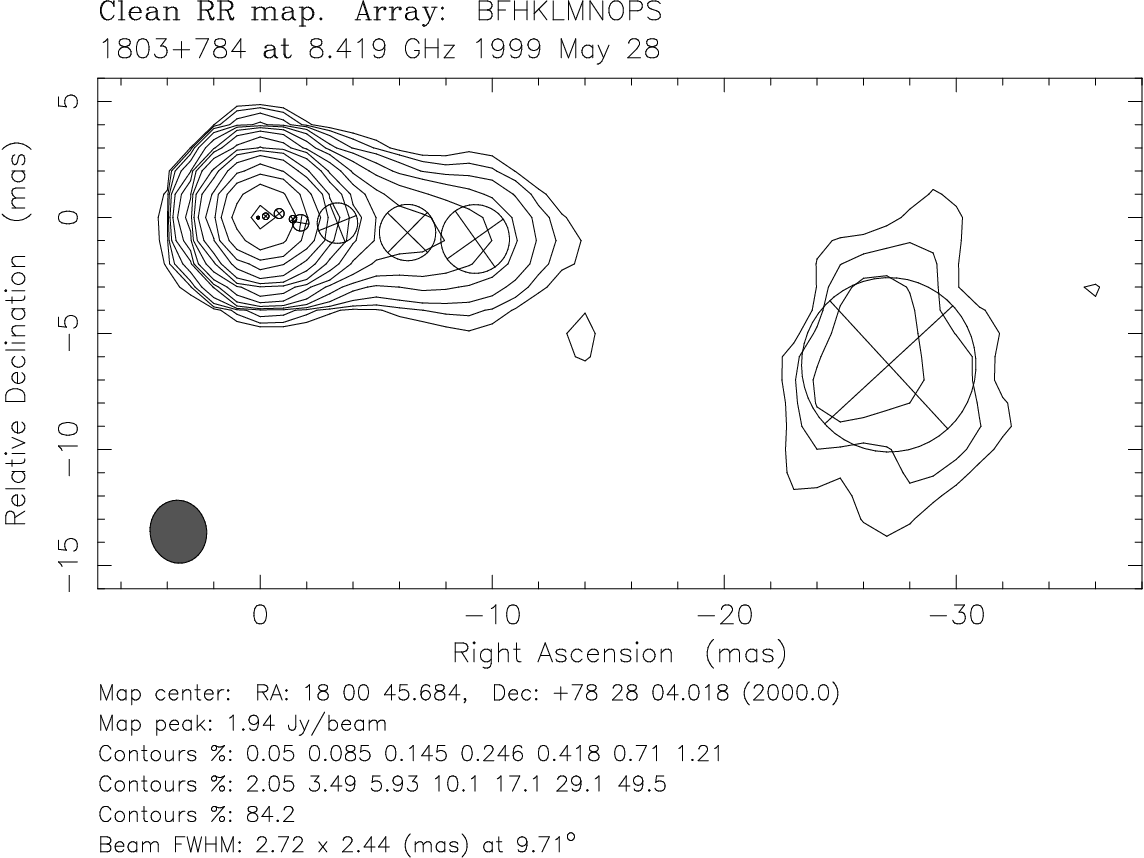,width=7.5cm}}\\
\subfigure[]{\psfig{figure=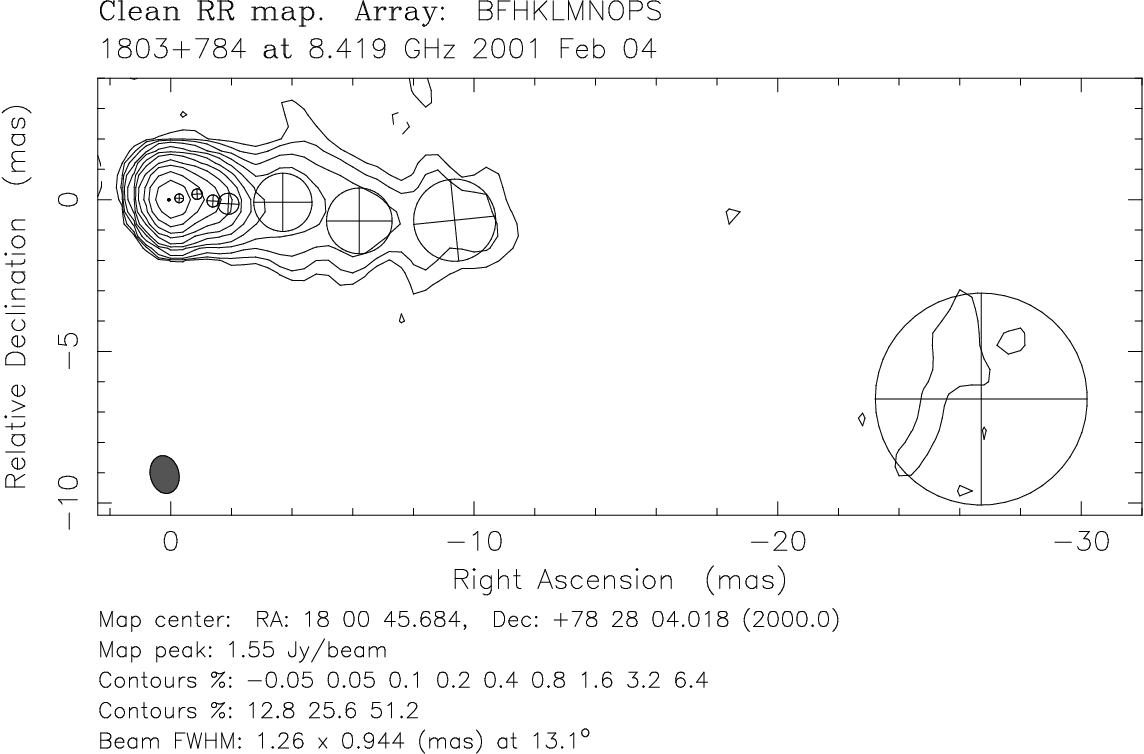,width=7.5cm}}
\caption{Results of hybrid imaging (contours) and circular Gaussian model fitting (circles with crosses) for S5 1803+784 at 8.4 GHz.}
\label{bilder3}
\end{figure*}

\begin{figure*}[htb]
\subfigure[]{\psfig{figure=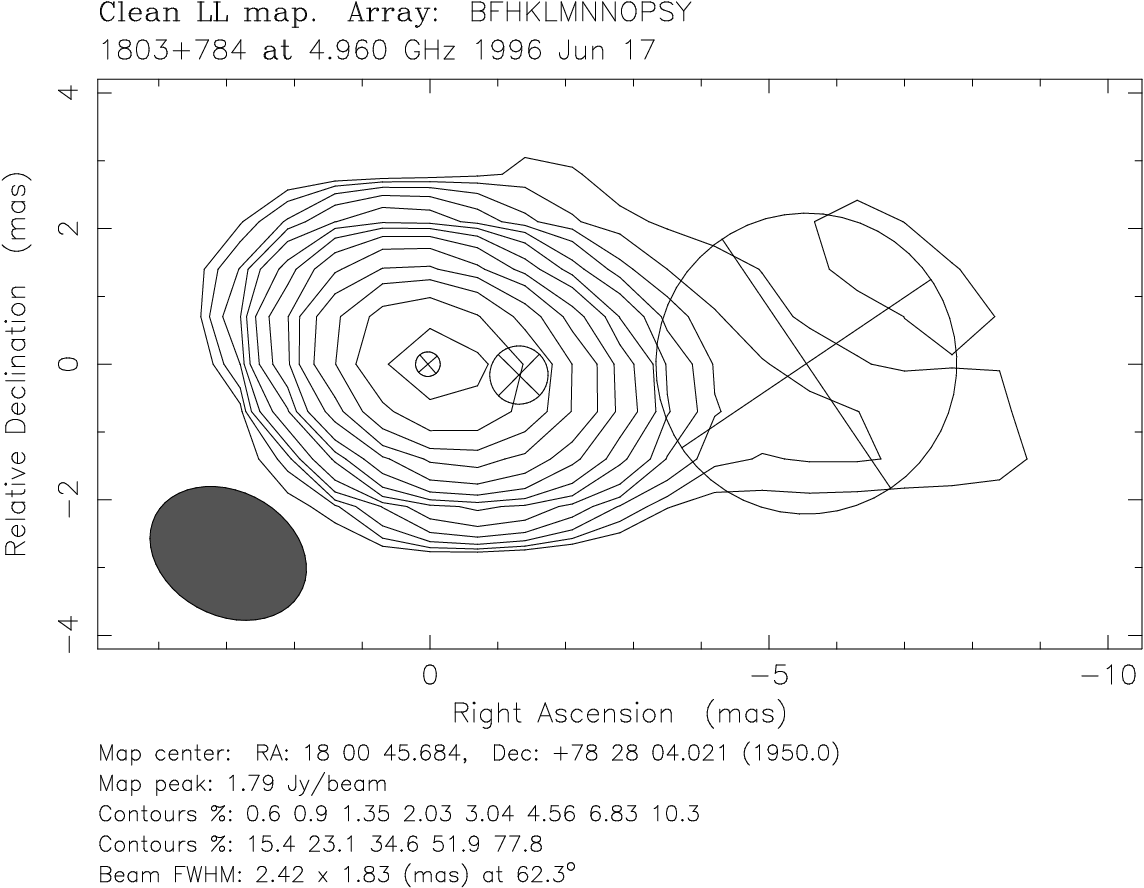,width=7.5cm}}
\hspace*{2cm}\subfigure[]{\psfig{figure=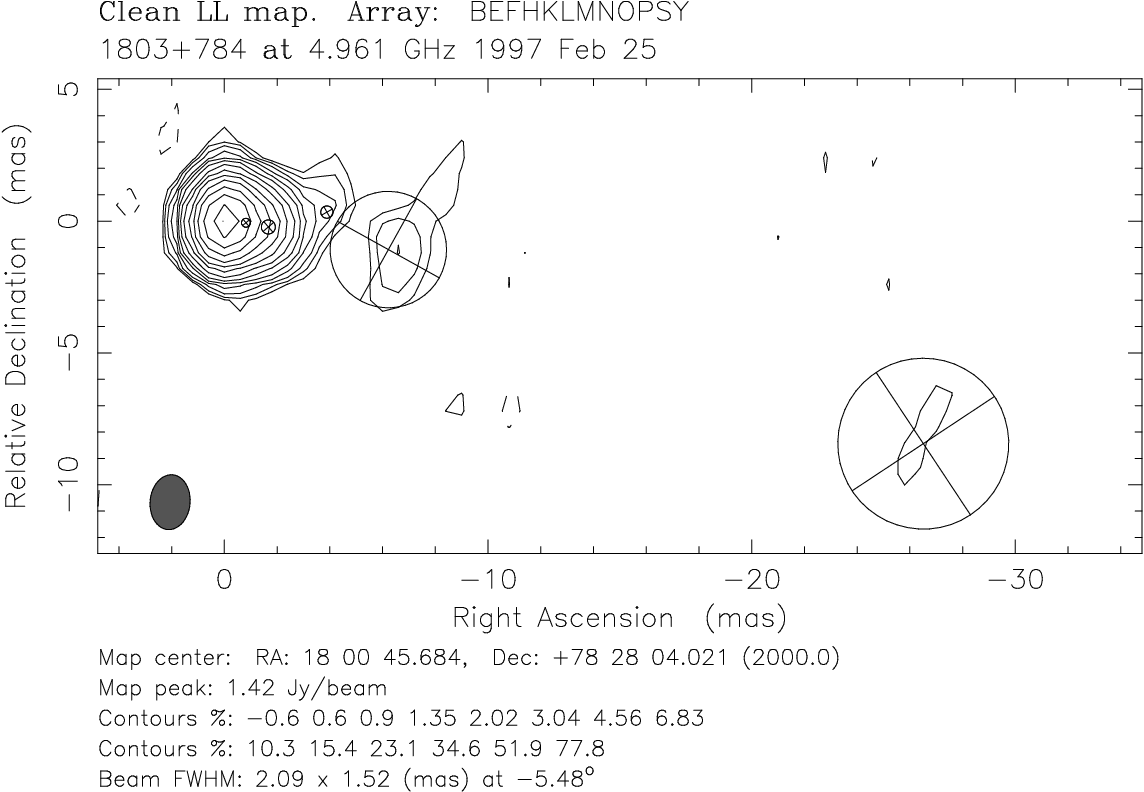,width=7.5cm}}
\subfigure[]{\psfig{figure=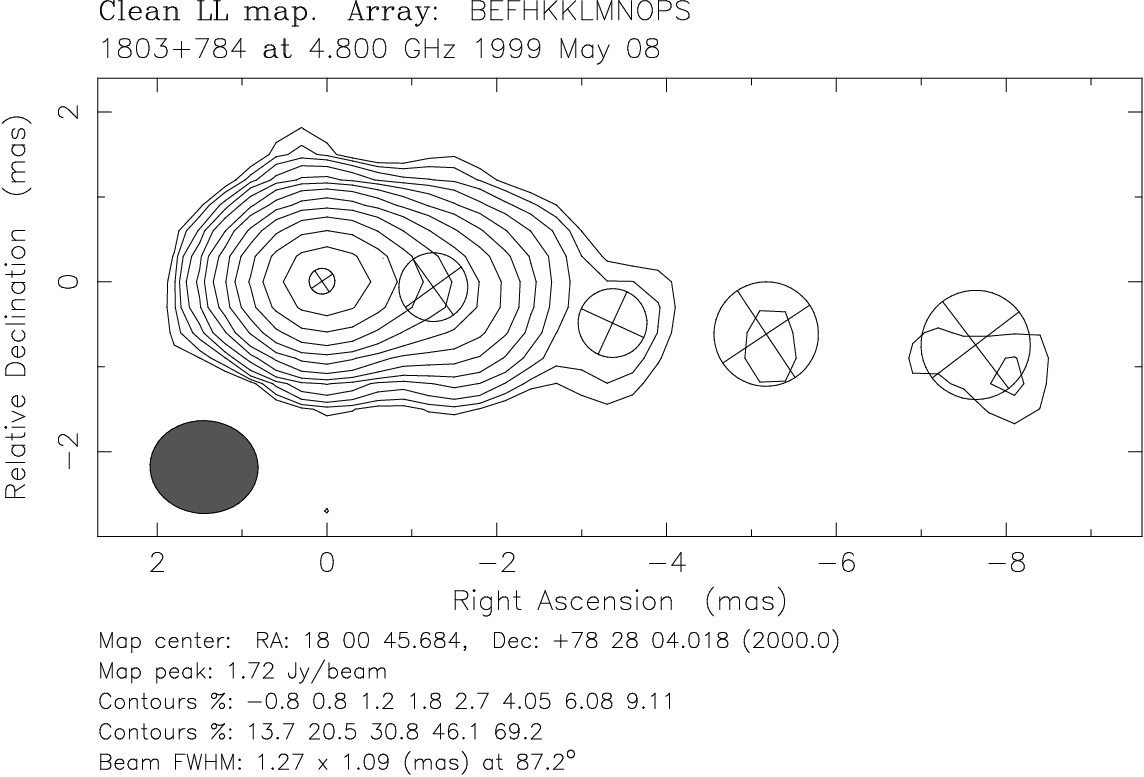,width=7.5cm}}
\hspace*{2cm}\subfigure[]{\psfig{figure=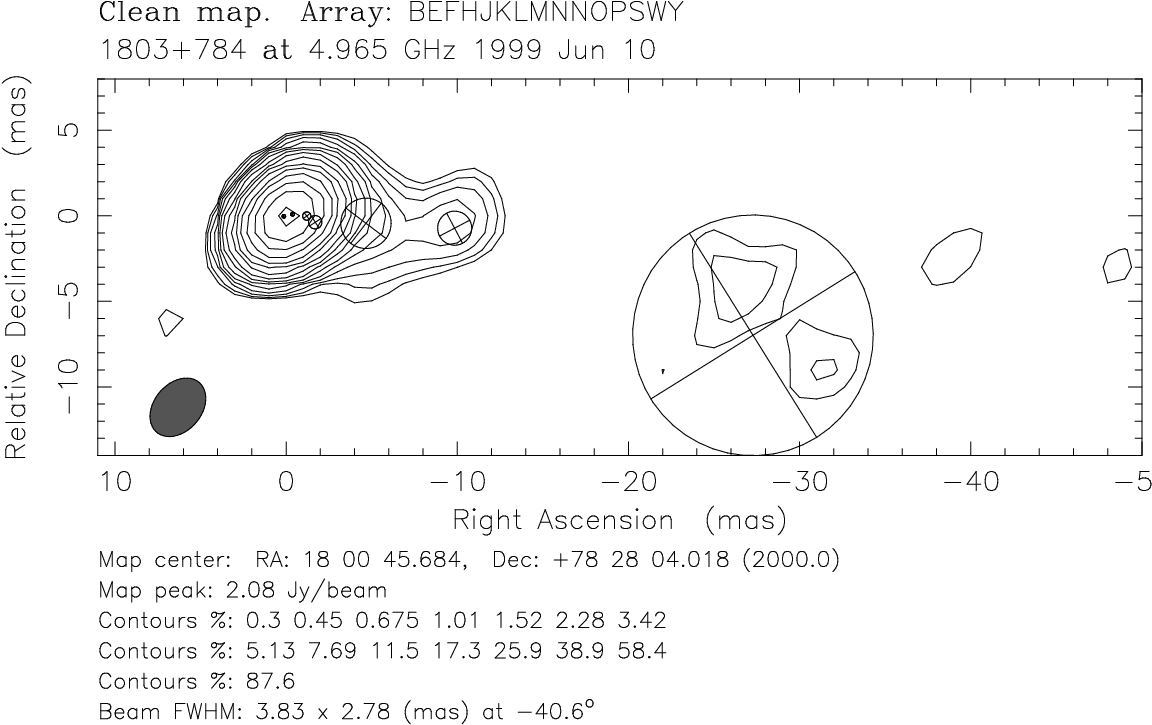,width=7.5cm}}\\
\subfigure[]{\psfig{figure=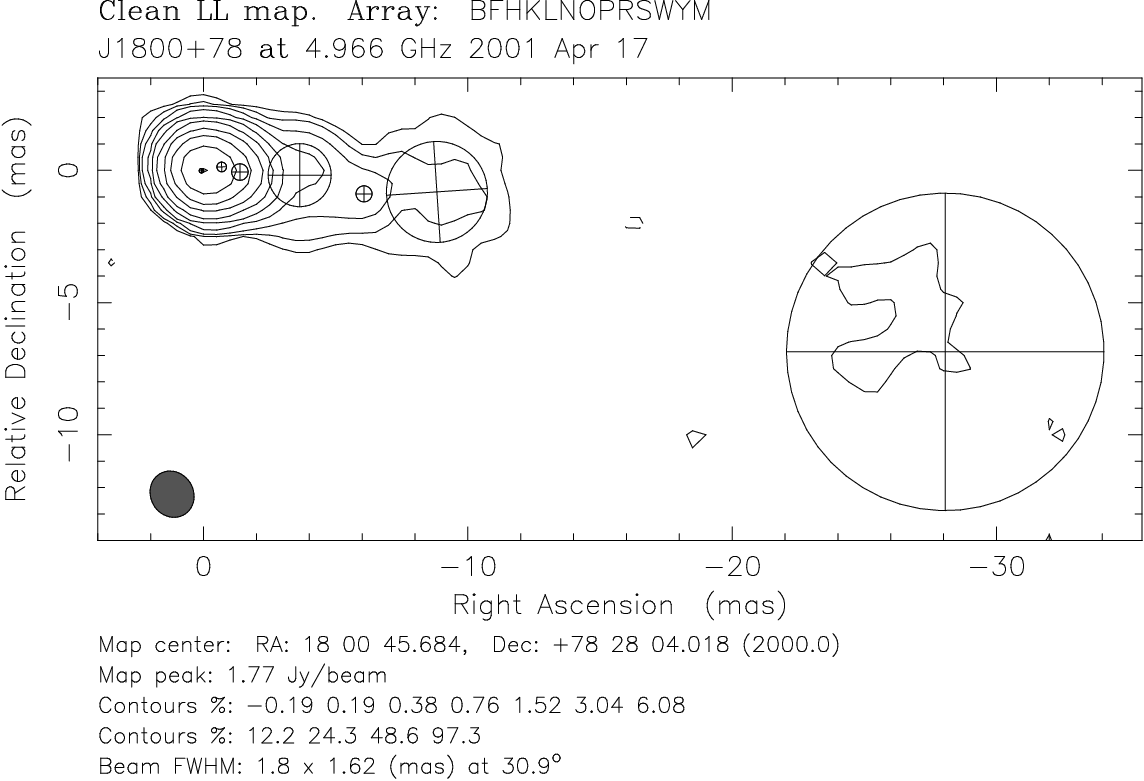,width=7.5cm}}\\ 
\caption{Results of hybrid imaging (contours) and circular Gaussian model fitting (circles with crosses) for S5 1803+784 at 5 GHz.}
\label{bilder4}
\end{figure*}

\begin{figure*}[htb]
\psfig{figure=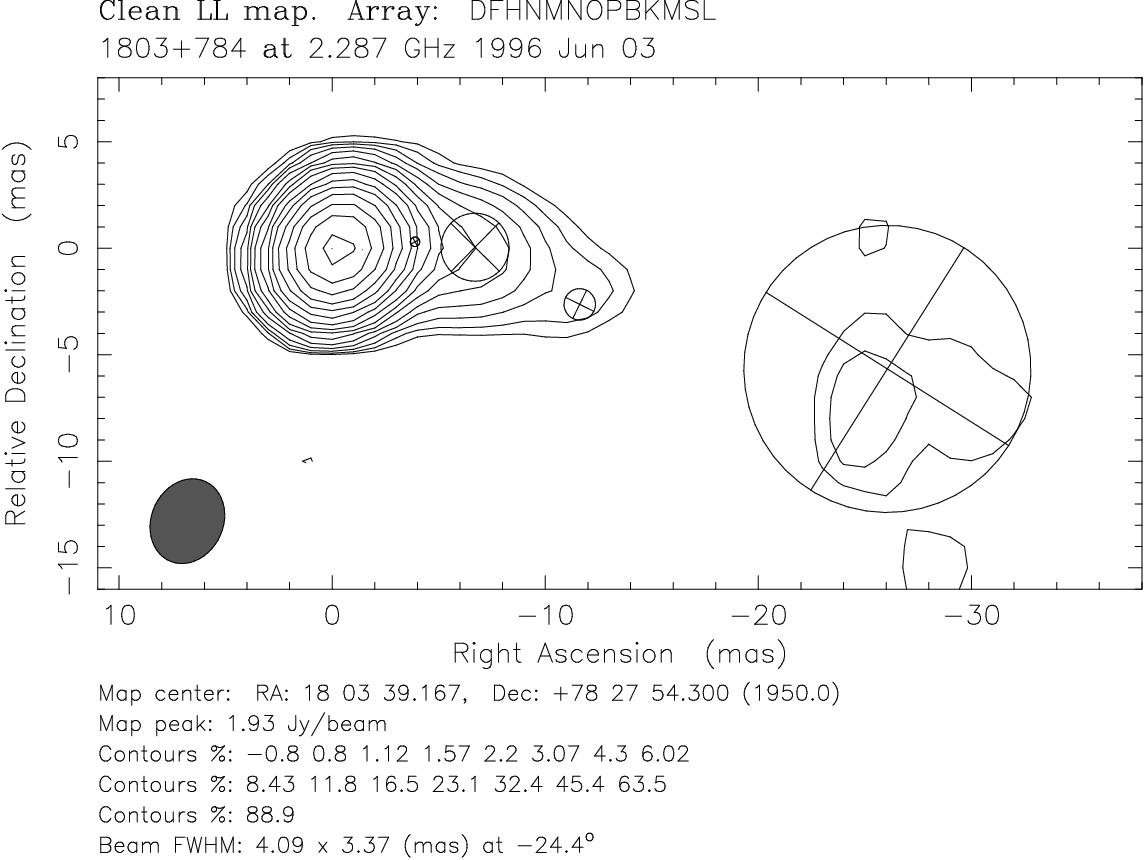,width=7.5cm}
\caption{Results of hybrid imaging (contours) and circular Gaussian model fitting (circles with crosses) for S5 1803+784 at 2.3 GHz.}
\label{bilder5}
\end{figure*}

\begin{figure*}[htb]
\subfigure[]{\psfig{figure=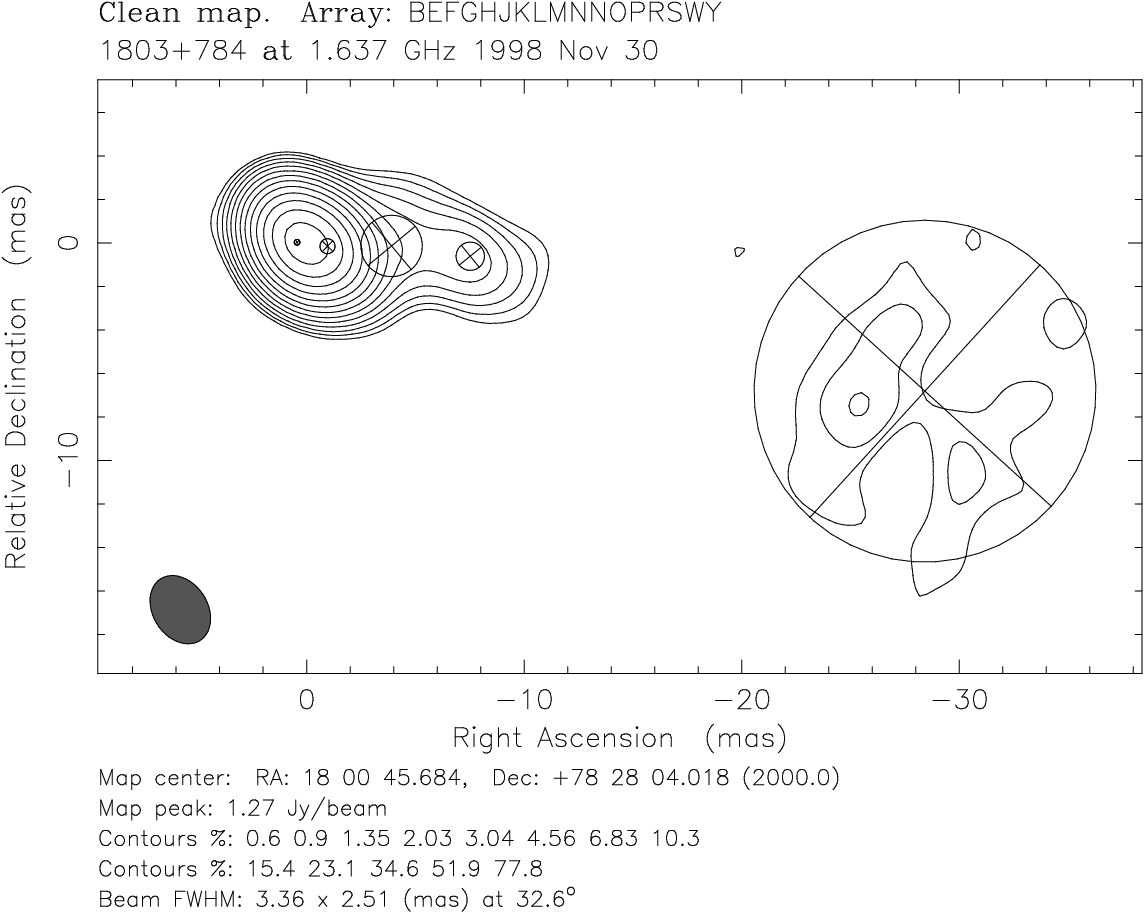,width=7.5cm}}
\hspace*{2cm}\subfigure[]{\psfig{figure=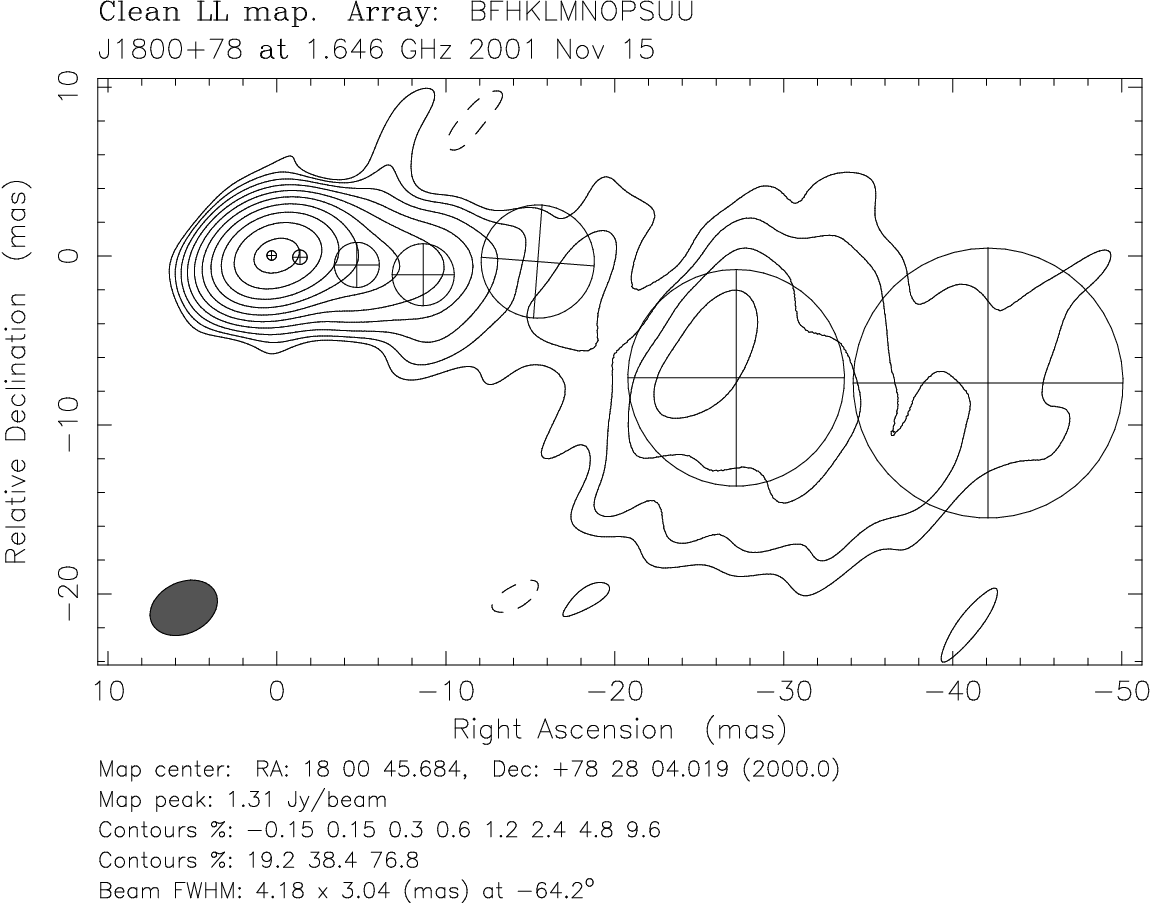,width=7.5cm}}\\ 
\caption{Results of hybrid imaging (contours) and circular Gaussian model fitting (circles with crosses) for S5 1803+784 at 1.6 GHz.}
\label{bilder6}
\end{figure*}

In Table~\ref{1803_table} we list the parameters of the best fits to the data
and their uncertainties. We also give the component identification and the
reference to the original publication of the data sets. In
Figs.~\ref{bilder1}--\ref{bilder6} we show the hybrid images with model-fit
components superimposed. For the images from observations
between February 2001 and September 2005, the beam sizes have been
adjusted for better comparison. The original beams were: epoch 2002 (15 GHz)
0.4207~x~0.6068 mas, epoch 2003 (15 GHz) 0.4764~x~0.4929 mas, epoch 2005/2 (15
GHz) 0.4768~x~0.4962 mas, epoch 2001.09 (8 GHz) 0.697~x~1.036 mas, epoch
2001.29 (5 GHz) 1.137~x~1.651 mas.

\begin{figure*}[htb]
\begin{center}
\subfigure[]{\includegraphics[angle=-90,clip,width=7.5cm]{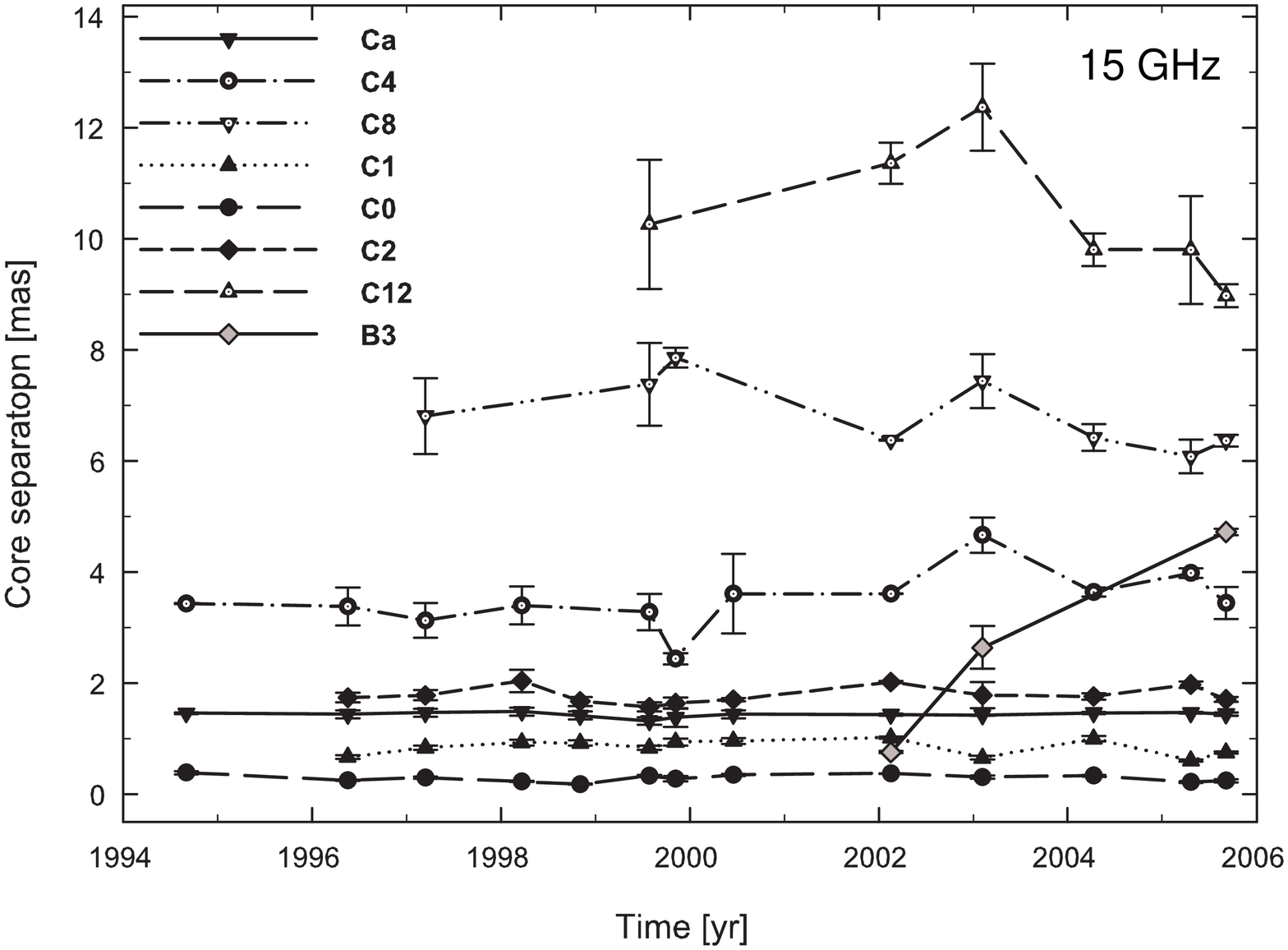}}
\subfigure[]{\includegraphics[angle=-90,clip,width=7.5cm]{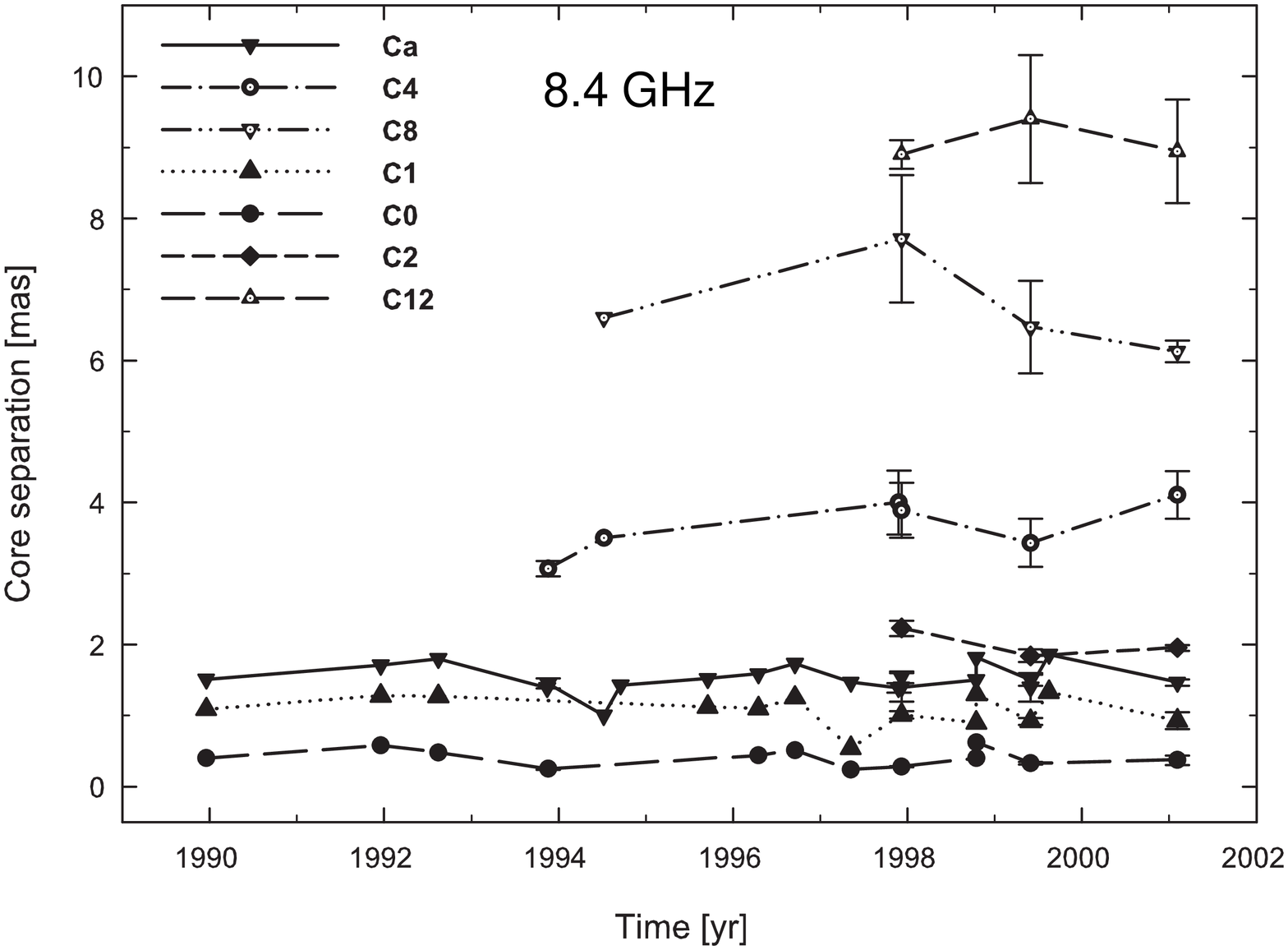}}\\
\subfigure[]{\includegraphics[angle=-90,clip,width=7.5cm]{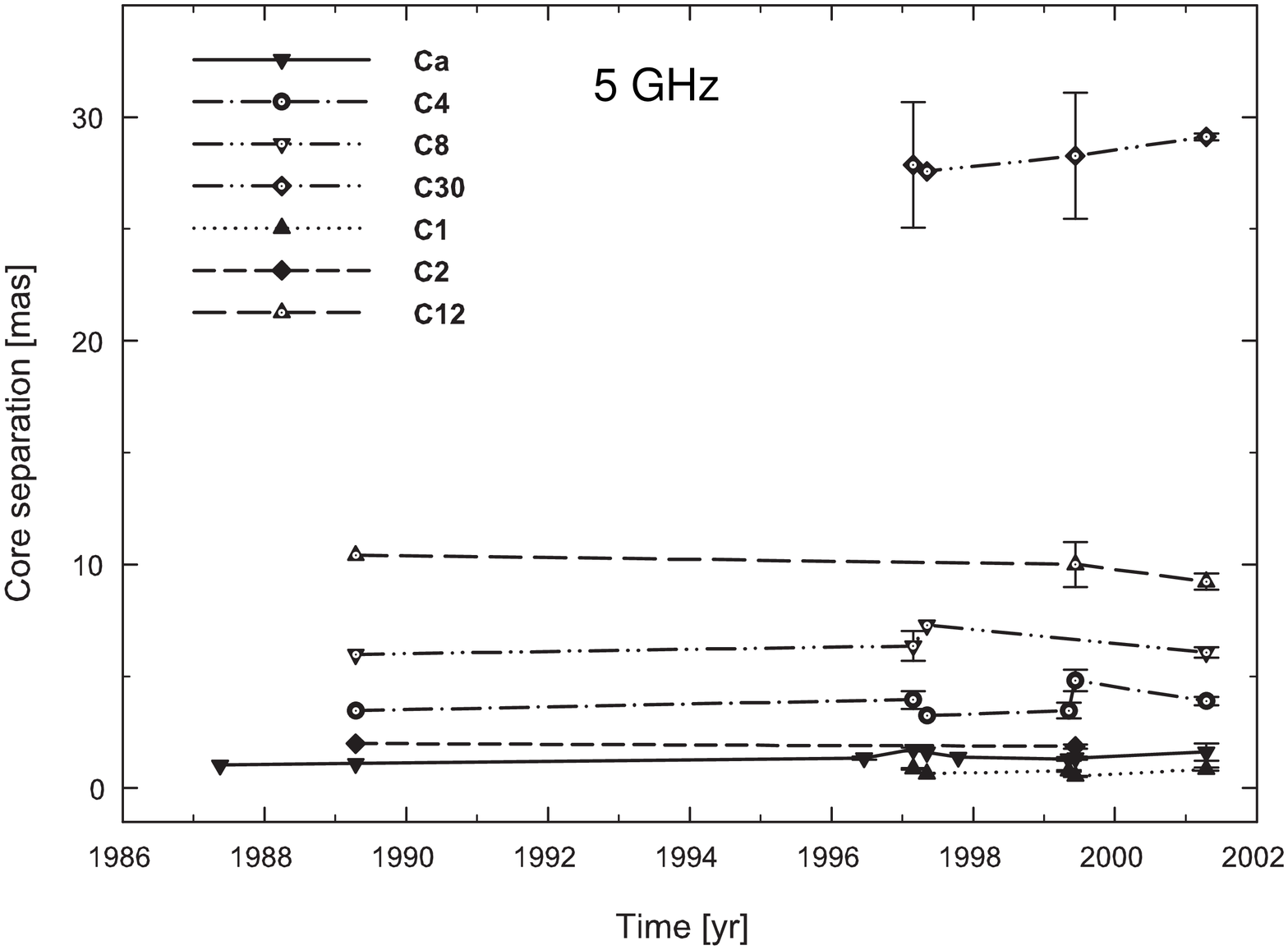}}
\subfigure[]{\includegraphics[angle=-90,clip,width=7.5cm]{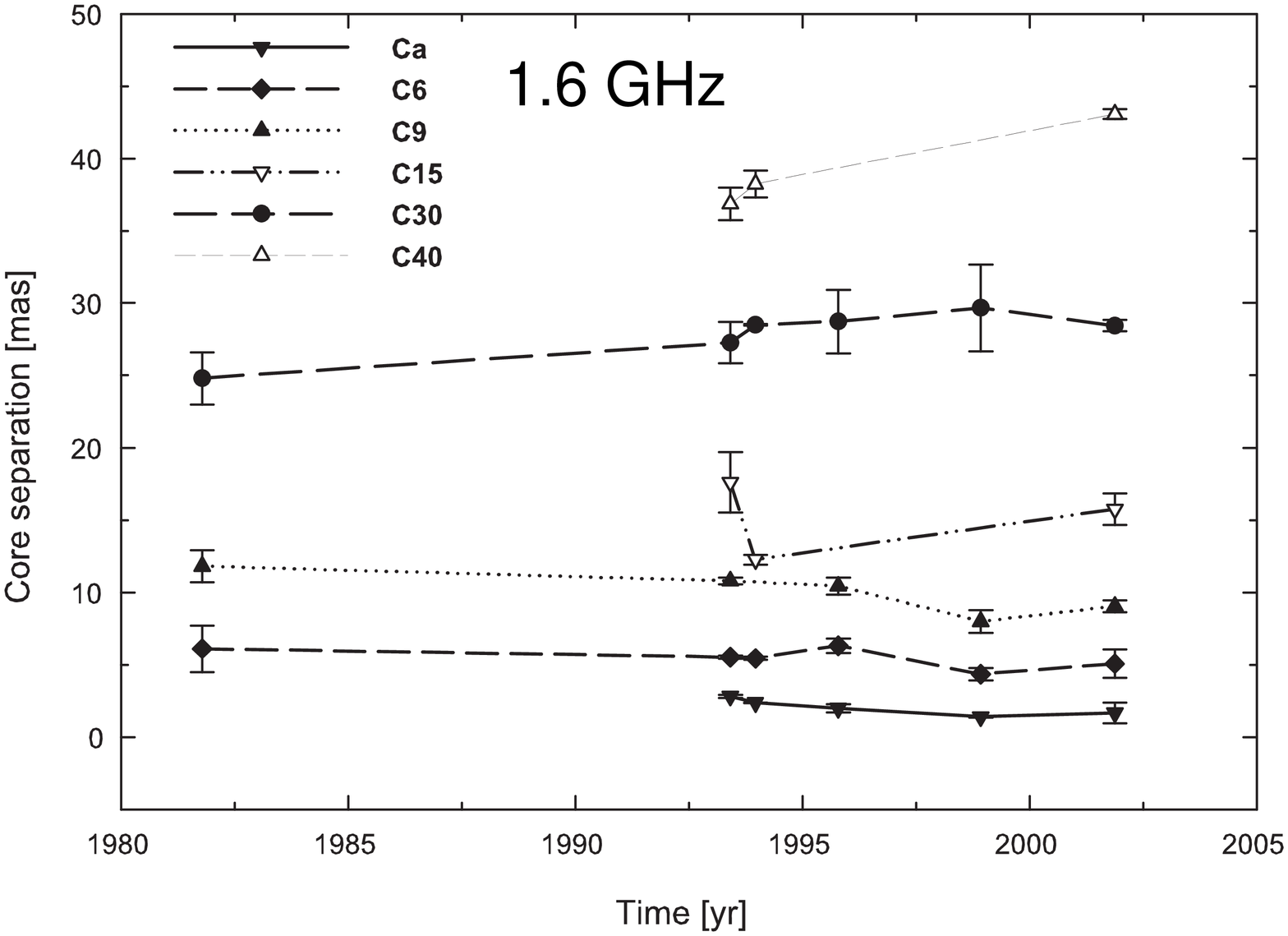}}\\
\subfigure[]{\includegraphics[angle=-90,clip,width=7.5cm]{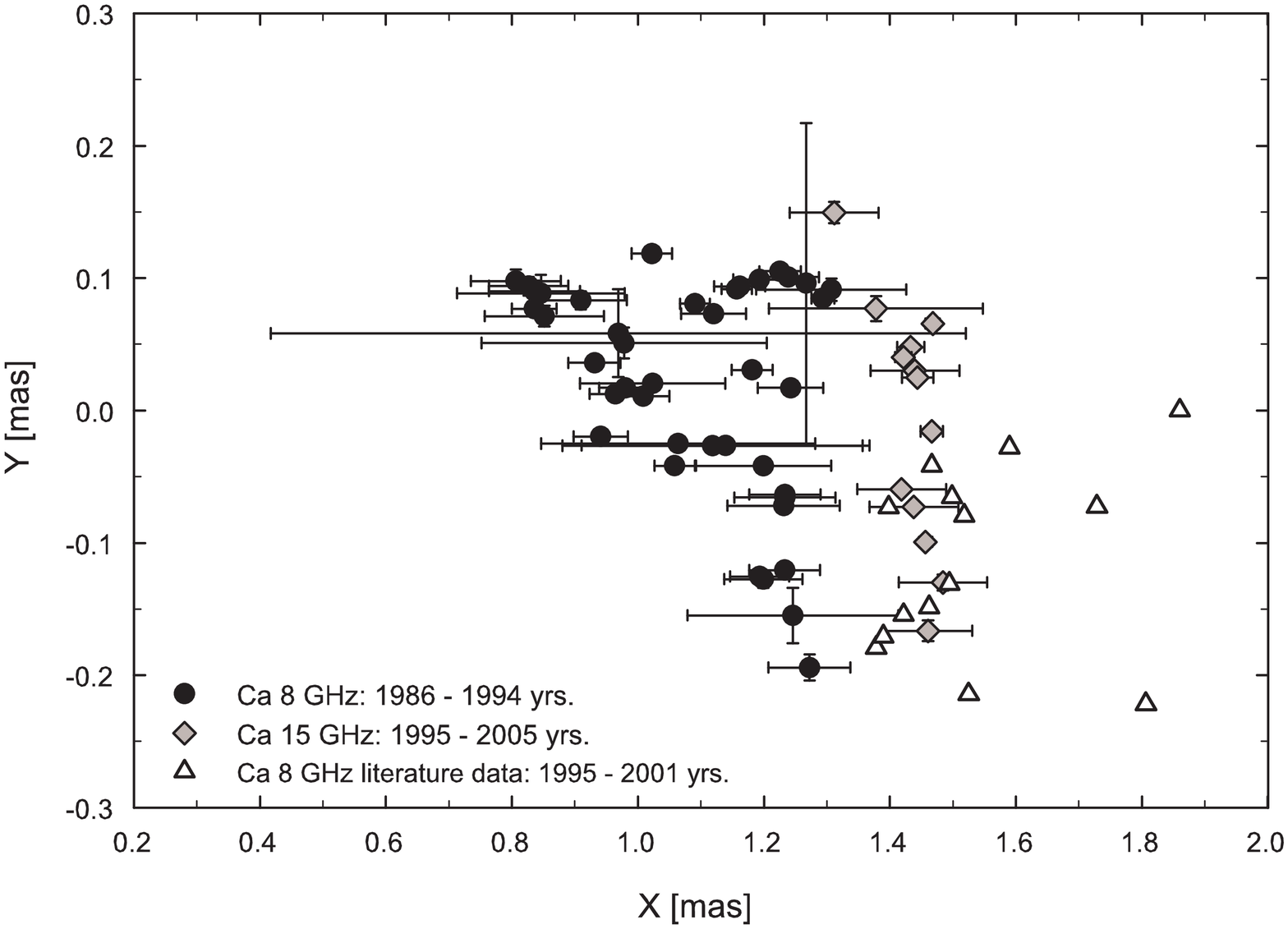}}
\subfigure[]{\includegraphics[angle=-90,clip,width=7.5cm]{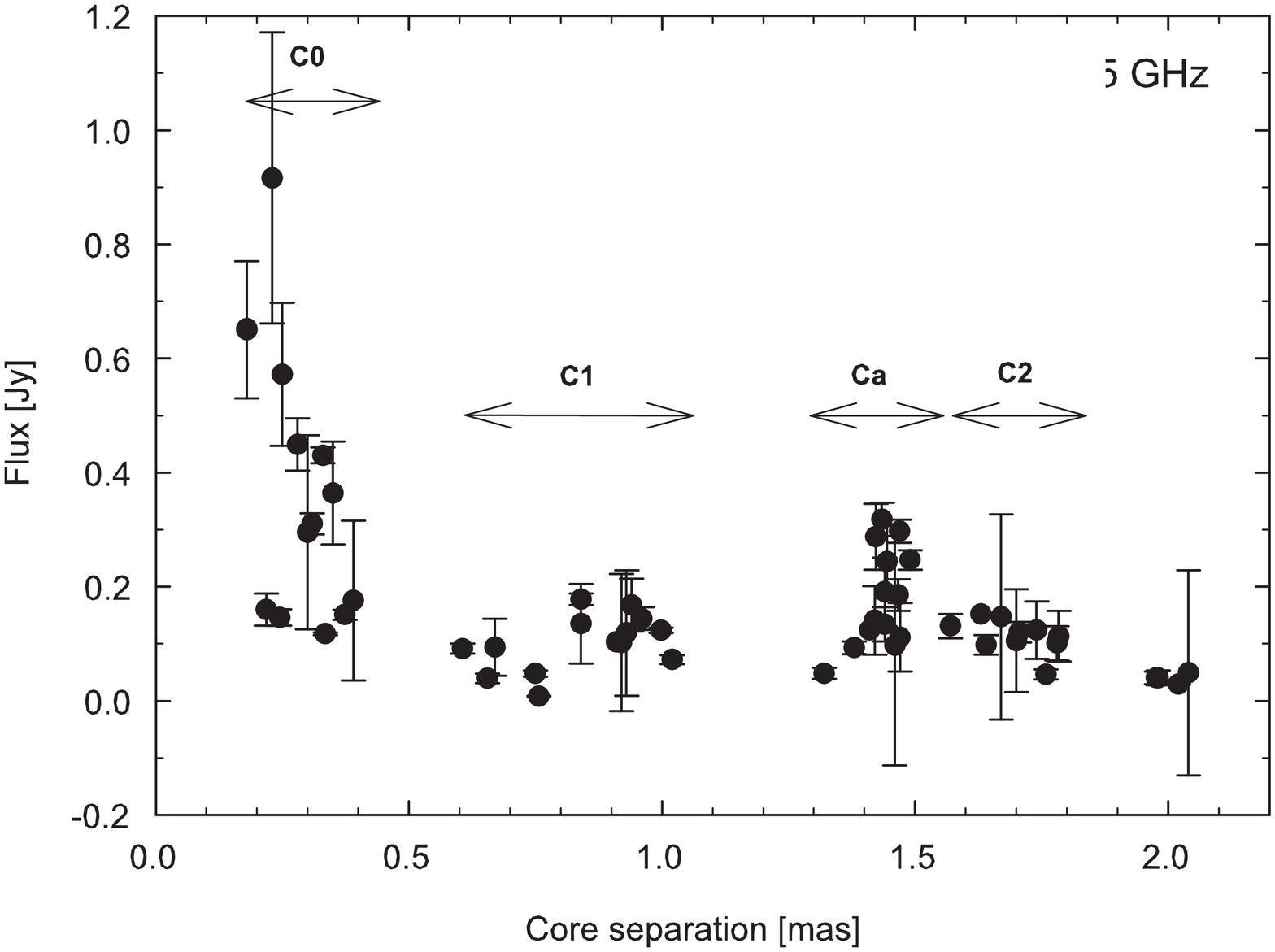}}
\end{center}
\caption{Panel (a) shows the core separation as function of time, $r(t)$, for those jet
components detected at 15 GHz. Individual components are denoted by different
symbols and lines (usage common across all panels). Note
the straight line for component {\bf Ca} and the fast moving component {\bf B3}. Panels 
(b) and (c) show $r(t)$ for 8.4~GHz and 5~GHz observations, respectively, based on data collected from
the literature, with component re-identification following the new
scenario discussed in this
paper. Panel (d)
shows $r(t)$ for components farther out along the jet based on 1.6 GHz
observations (new + from the literature). Panel (e) shows the motion of
component
{\bf Ca} in rectangular coordinates at 8.4 and 15~GHz, in which a displacement 
towards larger $x$ values is apparent between the fits from the earlier
(1986-1994) and newer (1995-2001) 8.4~GHz data with the 15~GHz fits falling in between.
Panel (f) shows the flux-density along the jet based on
all new epochs of the 15~GHz data. The {\it x} and {\it y} axes scales differ
across panels (a)--(d).}
\label{identification}
\end{figure*}

\subsection{Component identification}

Our kinematic analysis relies
critically on the component identification.  Our new scenario is
based on the flux-density, core separation, and position angle
of the individual components. All components have been identified in a way such
that all of these parameters vary as little as possible between adjacent
epochs. For those components that have been identified at several frequencies,
we did not find a significant frequency-dependent offset. The results are
presented and discussed in detail in the following sections.

\section{Results}
\label{results}

\subsection{Lack of long-term $r(t)$ change in most of the model jet components}

In addition to the formerly so-called {\it stationary} component ({\bf Ca}) at
$\sim$1.4~mas, we find components at core separations of $\sim$0.3~mas ({\bf
C0}), $\sim$ 0.8~mas ({\bf C1}), $\sim$2~mas ({\bf C2}), $\sim$3--4~mas ({\bf
C4}), $\sim$6--8~mas ({\bf C8}), and $\sim$10--12~mas ({\bf C12}). Most of
these jet components --- based on the new data presented here --- remain at
similar core separations between 1993.88 and 2005.68. The core separations of
{\bf C12} and {\bf C8} at 15 GHz are less well defined, reflected by their
larger uncertainties. Since the identification of {\bf C12} suffers from a
reduced number of data points, we do not include this component in our further
analysis. Components {\bf C30} and {\bf C40} can only be traced at 1.6 GHz, and
also will not be considered in the detailed analysis. Fig.~\ref{identification} panels (a--d)
present an
overview of the $r(t)$ behavior for all jet components at 15, 8.4, 5, and
1.6~GHz, respectively, out to a core separation of $\sim$50~mas and over a time span of up to twenty years.
Panels (b) and (c)
show data collected from the literature, but with component re-identifications
following the scenario we present in this paper; these previously published
data support this new component identification scenario.

\subsubsection{Comparison with data from the literature}

In order to check whether the components we find showed a similar or different
kinematic behavior in earlier times, we collected all available VLBI
information for S5 1803+784 from the literature. We list the model-fit
parameters derived for the total intensity VLBI observations in Table
\ref{literatur}. We re-identified all the components from these published model
fits according to the scenario proposed in this paper and plot the results in
Fig.~\ref{identification} panels (b) and (c). At both frequencies (5, 8.4~GHz),
the components have remained at similar core separations for almost 20 years;
we find no long-term outward, apparent superluminal motion based on the
aggregate data. Although different kinematic scenarios have been proposed by
different authors, the data are fully consistent with a scenario in which the
components tend to remain at long-term roughly constant core separations.\\

  Panel (d) of Fig.~\ref{identification} shows $r(t)$ from the 1.6~GHz
data. We again find that the inner components (up to ~12 mas) maintain roughly
constant core separations, although
the outer components ({\bf C30} and {\bf C40}) do show some evidence for
outward motion.\\

In Table \ref{regression} we give the parameters of the linear regressions
performed for the jet components' core separation as a function of time,
$r_{\rm mean}$ and $\mu_r$ and calculated apparent speeds, $\beta_{\rm app}$.
Based on the values for the apparent speeds calculated for all frequencies, we
find no superluminal motion for the components within $\sim$4~mas of the core.
{\bf C1 and C2} show no significant motion within the errors. {\bf C0} and {\bf
Ca} show some evidence for subluminal outward motion. Although {\bf C4} shows
apparent superluminal motion at 15GHz, combining data at all frequencies gives
an apparent speed consistent with zero. We find significant apparent
superluminal outward motion of the faint component {\bf B3}. {\bf C8} and {\bf C12} 
show some evidence of apparent inward motion. For these two components
fewer epochs having well-constrained positions are available. In the following 
we concentrate on the inner components.\\

\begin{table*}[htb]
\setcounter{table}{2}
\begin{center}
\caption{Linear regression fits to the motion of listed components. $r_{\rm mean}$ is the mean 
core separation over the time-range of the regression, $\mu$ is the proper motion of 
the component, and $\beta_{\rm app}$ is the apparent
speed. $\mu$ and $\beta_{\rm app}$ are computed twice: once using only 15~GHz 
results, and once with the combined results from all frequencies.  $t_0$ is the
extrapolated time of component ejection from the core in the case of component
{\bf B3}.}
\label{regression}
\medskip
\begin{tabular}{lrrrrrc}
\hline \noalign{\smallskip}
Comp. ID & $r_{\rm mean}$ [mas] & $\mu_{\rm r}^{\rm 15}$ [mas~$\rm yr^{-1}$] &$\beta_{\rm app}^{\rm 15}$ [$c$]& $
\mu^{\rm all}_{\rm r}$ [mas~$\rm yr^{-1}$] & $\beta^{\rm all}_{\rm app}$ [$c$]&$\rm t_{0}$ [yr]  \\
\hline \noalign{\smallskip}
{\bf C0} & $0.30 \pm 0.02$ & $0.017 \pm 0.001$ & $0.39 \pm 0.02$ & $0.006 \pm 0.001$ & $0.14 \pm 0.02$& /\\
{\bf C1} & $0.79 \pm 0.03$ & $-0.037 \pm 0.003$ & $-0.86 \pm 0.07$ & $0.001 \pm 0.001$& $0.02 \pm 0.02$ &/ \\
{\bf Ca} & $1.27 \pm 0.03$ & $0.007 \pm 0.004$ & $0.16 \pm 0.09$ & $0.022 \pm 0.001$ & $0.51 \pm 0.02$ &/ \\
{\bf C2} & $1.93 \pm 0.03$ & $0.017 \pm 0.007$ & $0.39 \pm 0.16$ & $-0.002 \pm 0.001$ & $-0.05 \pm 0.02$ &/ \\
{\bf C4} & $3.70 \pm 0.08$ & $0.145 \pm 0.017$ & $3.37 \pm 0.39$ & $0.001 \pm 0.001$ & $0.02 \pm 0.02$ &/ \\
{\bf C8} & $6.78 \pm 0.18$ & $-0.182 \pm 0.005$ & $-4.23 \pm 0.12$ & $-0.182 \pm 0.005$ & $-4.23 \pm 0.12$ &/ \\
{\bf C12} & $10.02 \pm 0.38$ & $-0.715 \pm 0.113$ & $-16.62 \pm 2.63$ & $-0.197 \pm 0.069$ & $-4.58 \pm 1.60$ &/ \\
{\bf B3} &               /    & $0.807 \pm 0.151$ & $18.76 \pm 3.51$ &/ &/ & $1999.8 \pm 1.1$ \\
\hline
\end{tabular}
\end{center}
\end{table*}

\subsubsection{Comparison with geodetic VLBI data}

Britzen et al. (2005a) presented an analysis of the kinematics in the pc-scale
jet of S5 1803+784 based on geodetic VLBI data. Such geodetic observations are
performed in general more frequently than astronomical observations (please
find details in Britzen et al. 2005). In order to have comparable numbers of
observations per time from astronomical and geodetic observations, we smoothed
the geodetic model-fit results at 8.4~GHz in the period 1986.21--1993.95 with
the following sliding mean algorithm. The new data points were produced,
replacing the set of circular Gaussian parameters $m_{i}$ for each 
time $t_{i}$ with the weighted
mean:
\begin{equation}
m_i^{\prime} = \frac{1}{\Sigma p_j} {\sum\limits_{j=1}^k
p_j m_j}, \label{eq:1}
\end{equation}
\noindent where $k$ is the number of data points within interval $
[t_i-\Delta, t_i+\Delta]$, and the weight of $j$th point is
determined as
\begin{equation}
p_j=\exp\left[-(\delta t_j/\Delta)^2\right], \label{eq:2}
\end{equation}
\noindent where $\delta t_j$ is the time span from the $j$th point
to the center of the window. For our analysis we selected a
window value of $\Delta = 0.3$ years, which is consistent with
the temporal resolution of data points obtained at other frequencies.\\


\begin{figure}[htb]
\begin{center}
\hspace*{-0.5cm}\includegraphics[width=6.4cm,clip]{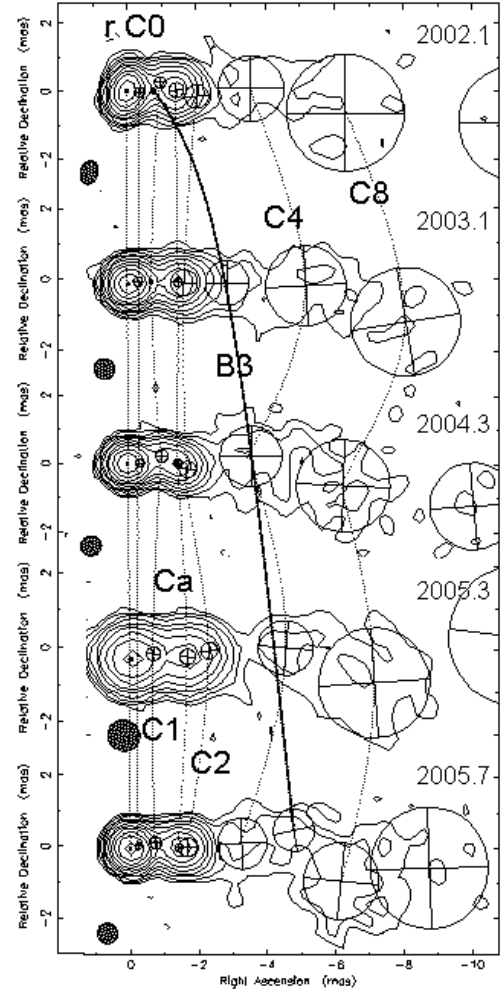}
\end{center}
\caption{Model fits superimposed on the hybrid maps obtained from Cleaning and
self-calibration for the most recently observed five epochs. The component
positions are connected by dotted lines. The positions of the fast component
{\bf B3} are connected by a solid line. Please note that the elapsed time between adjacent epochs differs.}
\label{b3}
\end{figure}

\subsubsection{Flux-density evolution along the jet}

Fig.~\ref{identification} panel (f) shows the 15-GHz flux-density
as function of distance along the jet out to $r=2$~mas, with the various
epochs of observations overplotted. We
find maxima in the observed distribution at $\sim$0.9 mas and 1.4 mas. For the
component at 1.4 mas we find a trend towards a sharp peak in flux-density.

\subsubsection{Displacement of {\bf Ca}}

Fig.~\ref{identification} panel (e) shows the position of component {\bf Ca}
in rectangular coordinates. The different symbols denote frequency (8.4 and
15~GHz), with the 8.4~GHz results further divided into pre- and post-1995.0
sets. The latter 8.4~GHz data are obtained from the literature. These two
8.4~GHz data-sets seem to have a shift in right ascension of $\sim$0.3~mas.
We are comparing here geodetic data with astronomical data.
In the case of the geodetic data (Britzen et al. 2005a), we
assumed that the brightest component is identical with component {\bf Ca}. We
investigate this effect in the scenario of a Binary Black Hole in more detail
(Roland et al. 2008).

\subsection{The "fast" component {\bf B3}}
\label{fast}

In addition to the {\it stationary} components, one fast moving
component ({\bf B3}) has been observed in the time between 2002.1 and 2005.7.
As shown in Fig.~\ref{identification} panel (a) and Fig.~\ref{b3}, this
component shows outward motion with an apparent velocity of $\sim$19$c$ based on 3 epochs of observations.
{\bf B3} is much fainter (between 0.01 and 0.04 Jy) and shows a kinematical
behavior that differs significantly from the
other components, which don't show long-term secular outward motion.
The ejection time of this component computed by extrapolating it's proper 
motion back to $r=0$ is $1999.8 \pm 1.1$. 
Unfortunately, this epoch
is in the middle of the two-year gap in the
observations. \\

\subsection{Further evidence for the coexistence of one "fast" component among "stationary" components}

In order to check our model of the S5~1803+784 jet, in which only one component moves 
with apparent superluminal speed, we investigated in detail the
opacity and spectral evolution of the total flux-density light-curves. The
emergence of a new jet component, according to the shock-in-jet model (e.g.,
Marscher 1996; Gomez et al. 1997), is caused by a shock induced by a primary
excitation at the base of the jet.  This manifests itself in radio light-curves as an
outburst that is delayed at lower frequencies due to the combined effects of
the frequency stratification of the emitting electrons, non-zero opacity, and
light-travel delays. Such time-delayed outbursts are associated observationally
on mas-scales with the brightening of the VLBI core and are accompanied by the
ejection of a new optically thin jet component.\\

\begin{figure} [htb]
\begin{center}
\hspace*{0.5cm}\includegraphics[width=8.5cm,clip]{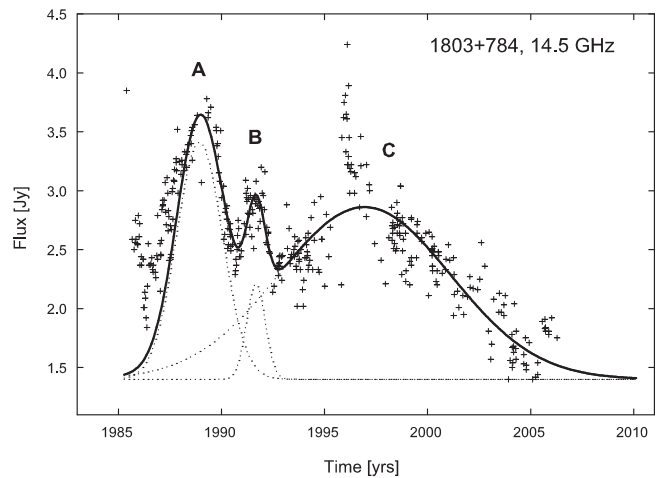}
\end{center}
\caption{Gaussian decomposition of the UMRAO 14.5~GHz light-curve.  The solid line shows the sum of
the Gaussian functions and dotted lines are individual Gaussian functions. Letters indicate the names of the
flares.} \label{gaussians}
\end{figure}

\begin{figure} [htb]
\begin{center}
\hspace*{0.5cm}\includegraphics[width=8.5cm,clip]{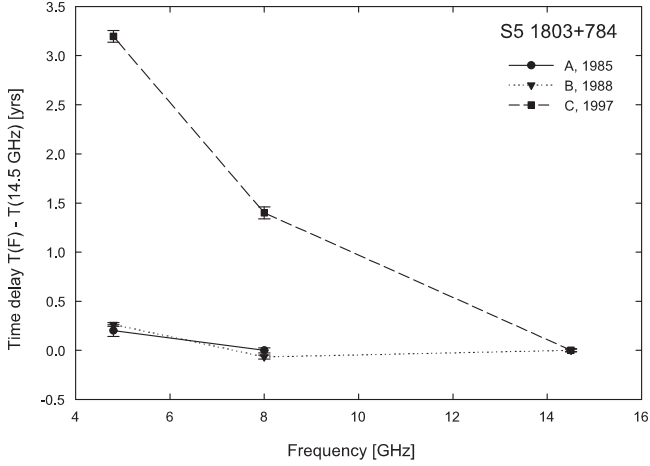}
\end{center}
\caption{Time lags of individual outbursts as functions of frequency.} \label{time_delays}
\end{figure}

\begin{figure} [htb]
\begin{center}
\hspace*{0.5cm}\includegraphics[width=8.5cm,clip]{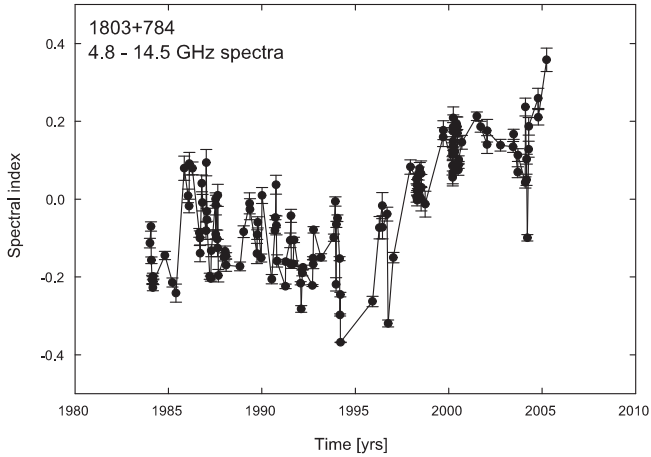}
\end{center}
\caption{Time evolution of the quasi-simultaneous spectral index, fit to the
4.8~GHz, 8~GHz, and 14.5~GHz flux densities.} \label{sed}
\end{figure}

We decomposed the total flux-density light-curves at 4.8~GHz, 8~GHz, and
14.5~GHz from the University of Michigan Radio Observatory (UMRAO) monitoring campaign (Aller et al. 1999, 2003) into
Gaussian components, as was described in Pyatunina et al.  (2006, 2007).
Fig.~\ref{gaussians} shows the light-curve of S5~1803+784 at 14.5 GHz and the
resulting Gaussian decomposition. Three prominent outbursts are visible: 
(A) in 1985; (B) in 1988, and (C), 
a very prolonged outburst starts in $\sim$1992 and ends in
$\sim$2005 with a peak in 1997. The Gaussian parameters, fitted to
the light-curve are listed in Table~\ref{1803outbursts}. Frequency-dependent
time delays for each flare are the time difference between the Gaussian peak at
each frequency with respect to the position of the peak at 14.5~GHz.
These are plotted in Fig.~\ref{time_delays}. Flares A and B show moderate time delays of
0.3 years, whereas the prolonged flare C shows an enormous time delay of 3.2
years between 4.8~GHz and 14.5~GHz, which is an indication of high opacity in
the source during the C outburst and a flat spectrum that becomes gradually
steep. Outburst $C$ has a very broad profile at all
frequencies ($\Theta\sim 8$~yr). Similarly broad outbursts were observed for the
source 1308+326 in Pyatunina et al.\ (2007), where it was mentioned that
such a broad outburst can be associated with the development of a dense region,
some kind of ``cocoon'' that confines the innermost part of the jet that is
excited by the primary perturbation. We calculated quasi-simultaneous spectra
using the measurements at all three frequencies for which time separation is
less than two weeks; Fig.~\ref{sed} shows the spectral evolution fit as a 
power law to each spectrum. The spectral behavior changes significantly after
$\sim$1996, where the spectra change from being flat, and gradually become steeper. The
changes of the spectral evolution occur in the beginning of a bright, prolonged
flare C and can be associated with it. 
We suggest that this can be explained by a jet
component ejection. Small frequency-dependent time delays and steep spectra of
other observed flares of the source suggest that there probably were no jet
component ejections before 1995 which is in good agreement with our model for
the S5~1803+784 jet.

\begin{table*}[htb]
\setcounter{table}{3}
\begin{center} \caption{Gaussian parameters of outbursts in the UMRAO 
light-curves} \label{1803outbursts}
\medskip
\begin{tabular}{lrlllr}
 \hline \noalign{\smallskip}
Comp.& Freq. & Amplitude & $T_{max}$ &$\Theta$ & Time delay \\
      &  GHz  &  Jy         &  yr       &   yr   &   yr     \\
\hline \noalign{\smallskip}

A   & 8.0    & $1.94\pm0.02$ & $1985.19\pm0.02$ & $1.41\pm0.02$ & $0.00\pm0.02$ \\
A   & 4.8    & $0.77\pm0.04$ & $1985.39\pm0.04$ & $1.41\pm0.04$ &  $0.20\pm0.06$ \\
B   & 14.5   & $2.01\pm0.01$ & $1988.93\pm0.01$ & $2.63\pm0.01$ & $0.00\pm0.01$ \\
B   & 8.0    & $2.15\pm0.02$ & $1988.86\pm0.01$ & $4.07\pm0.02$ & $-0.07\pm0.02$ \\
B   & 4.8    & $1.25\pm0.01$ & $1989.19\pm0.01$ & $2.54\pm0.01$ & $0.26\pm0.02$ \\
C   & 14.5   & $1.46\pm0.01$ & $1996.94\pm0.02$ & $9.83\pm0.03$ & $0.00\pm0.02$ \\
C   & 8.0    & $1.57\pm0.02$ & $1998.34\pm0.04$ & $7.89\pm0.04$ & $1.40\pm0.06$ \\
C   & 4.8    & $0.93\pm0.01$ & $2000.13\pm0.04$ & $6.28\pm0.05$ & $3.20\pm0.06$ \\
\hline
\end{tabular}
\end{center}
\end{table*}

\begin{figure}[htb]
\begin{center}
\subfigure[]{\includegraphics[clip,width=7.5cm]{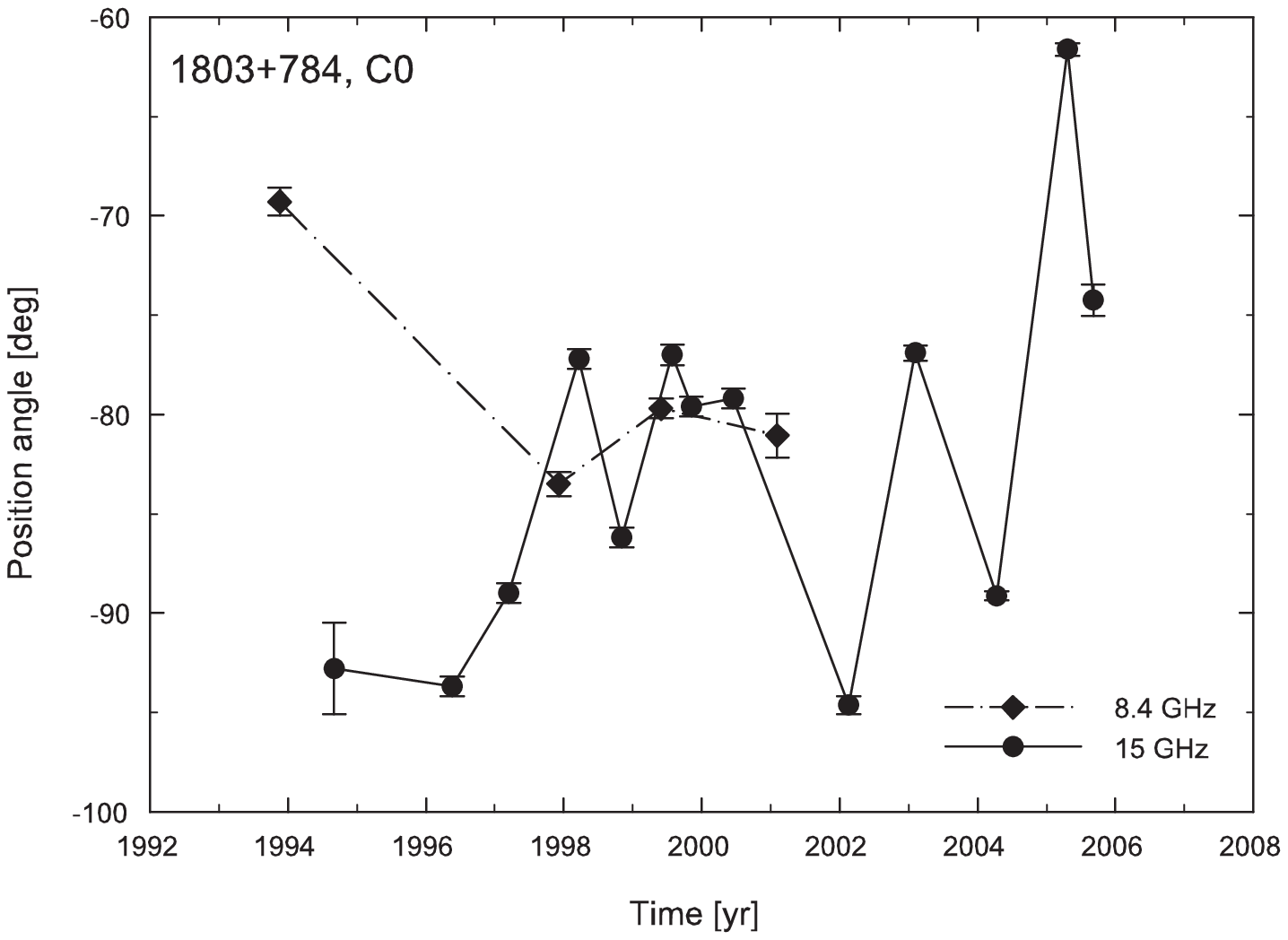}}
\subfigure[]{\includegraphics[clip,width=7.5cm]{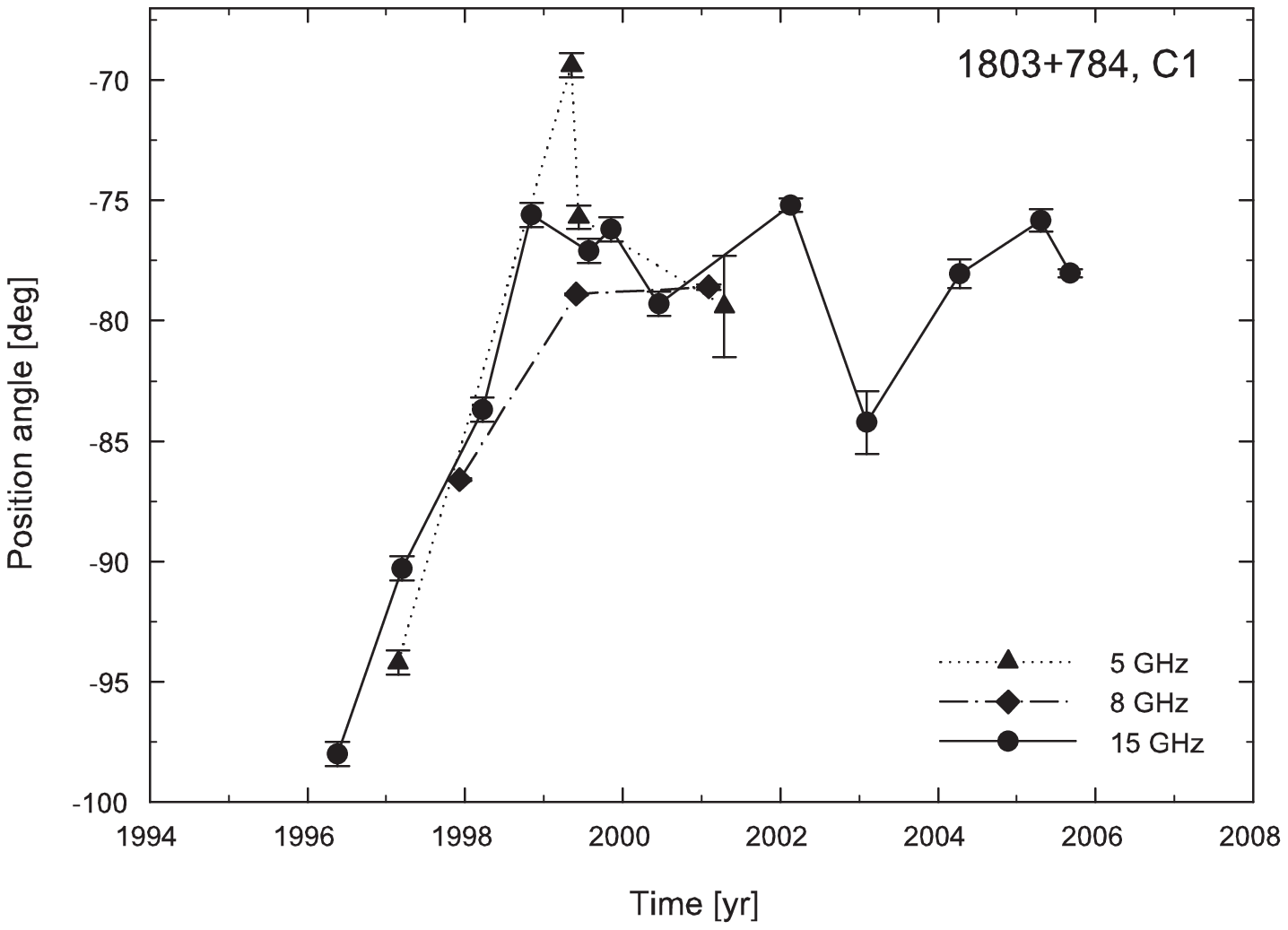}}\\
\subfigure[]{\includegraphics[clip,width=7.5cm]{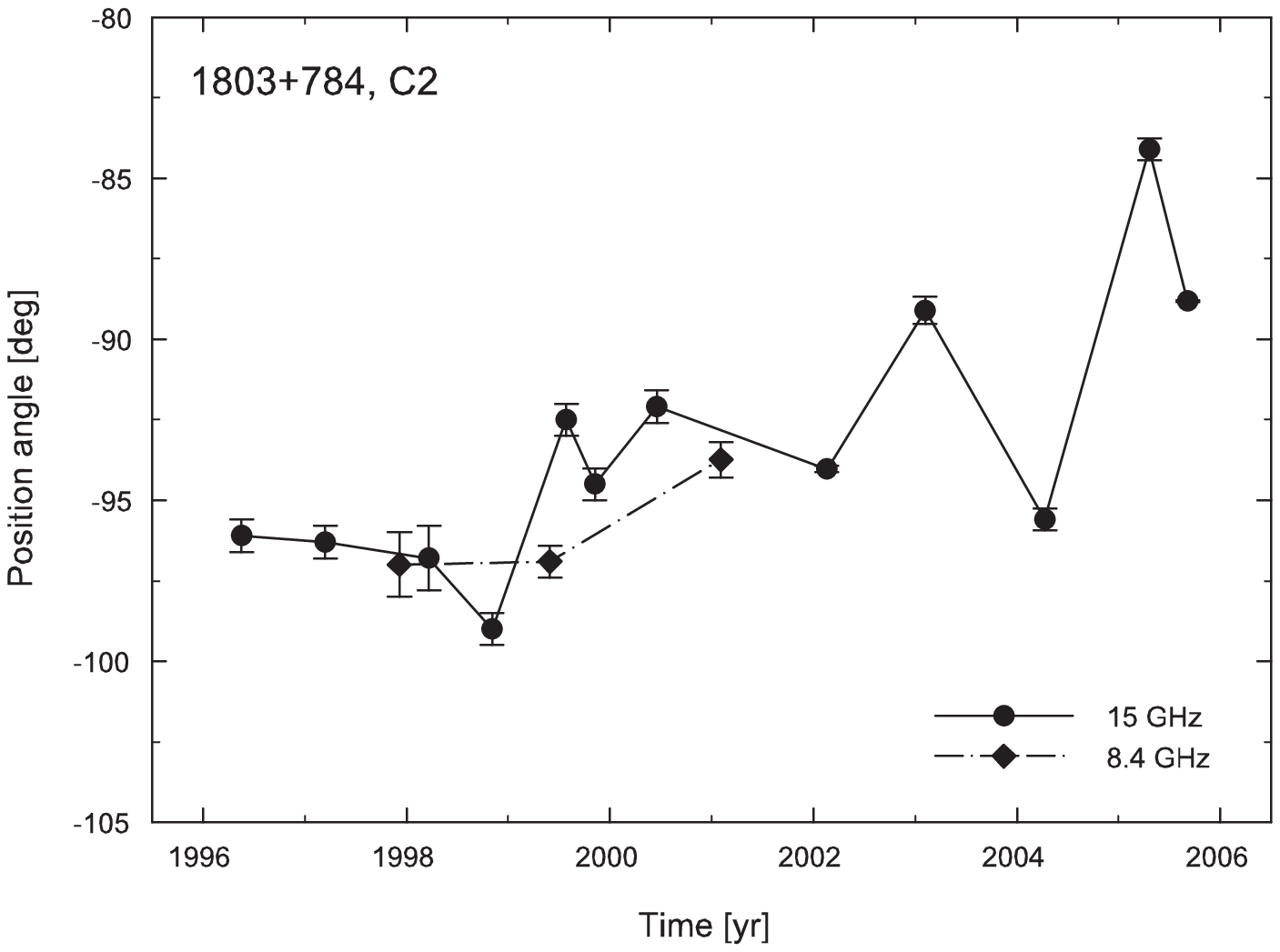}}
\subfigure[]{\includegraphics[clip,width=7.5cm]{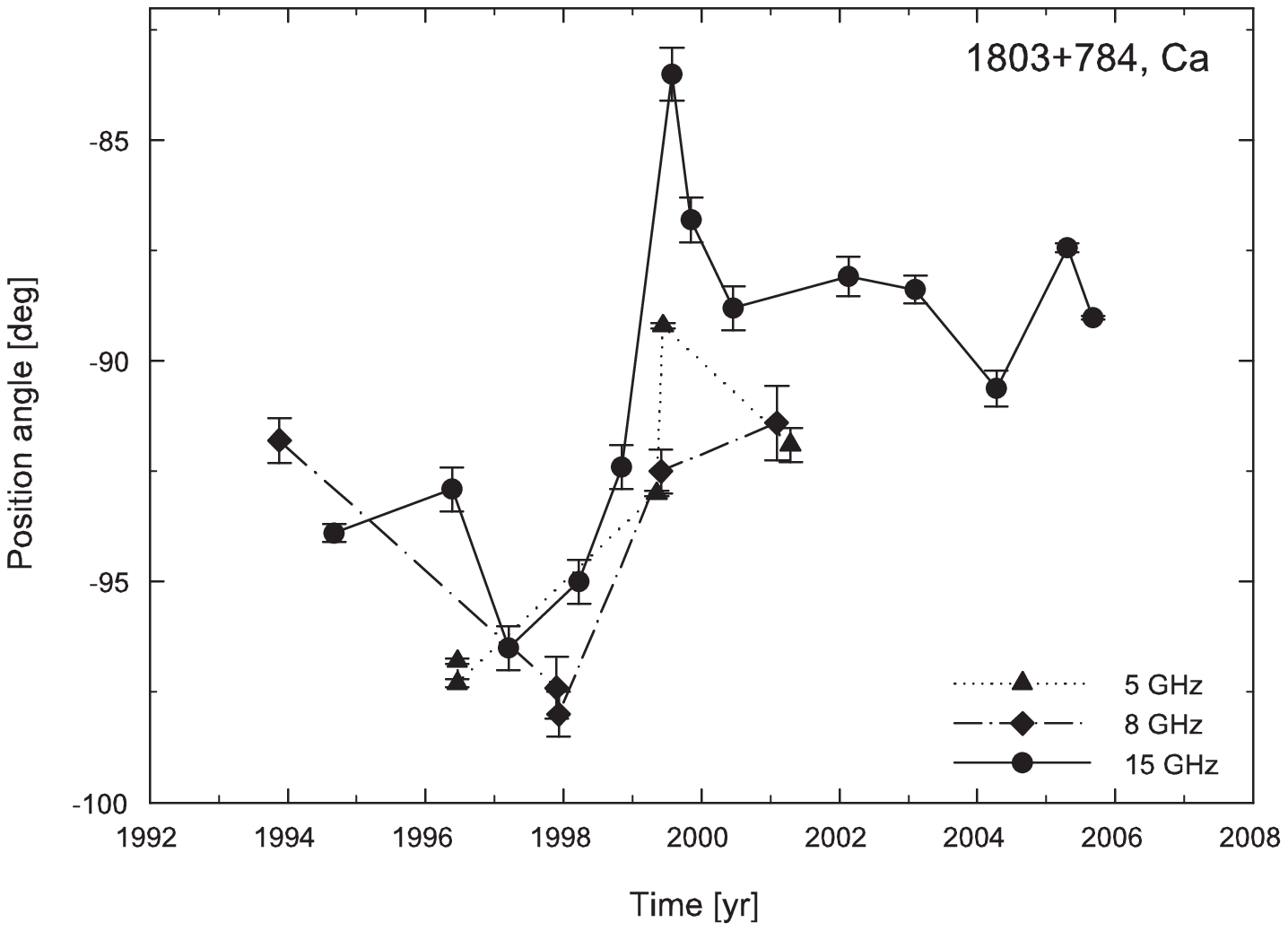}}\\
\end{center}
\caption{Position angle as function of time at different frequencies for four individual components.}
\label{freq_pa}
 \end{figure}

\subsection{Significant position angle changes}

As shown before, most of the components remain at similar core separations for
the time covered by the observations. Although it is obvious from the panels in Fig.~\ref{identification} (a) and (b) that most jet components show 
an additional motion with regard to the core separation superimposed on their 
rather stationary behavior, we concentrate in this paper on the most important 
aspects only. A detailed analysis of the second order motion is complicated by the 
uneven sampling. This effect of ``oscillating'' motion has been analyzed in 
detail for {\bf Ca} in 1803+784 by Britzen et al. (2005a). The phenomenon of  ``oscillations'' 
is clearly seen in other components of 1803+784 and in other BL Lac objects as well and will be analyzed in detail 
in Glueck et al. (in prep). \\
The components show significant position angle changes. 
Fig.~\ref{freq_pa} panels (a)--(d) show the evolution of the position angle for
four jet components ({\bf C0}, {\bf C1}, {\bf C2}, and {\bf Ca}, respectively) 
at different frequencies. The position angles vary over
35$^{\circ}$ in the case of {\bf C0} and 15$^{\circ}$ in the case of {\bf C2}. 
We find very similar relations at 5 and 8 GHz for {\bf C0}, at 5, 8,
and 15 GHz for {\bf C1}, at 8 and 15 GHz for {\bf C2}, and at 5, 8, and 15 GHz
for {\bf Ca}. One outlier for {\bf C0} in Fig.~\ref{freq_pa} panel (a) in 1994 can
probably be explained by a blending effect with the core or a new component due
to the closeness of the core. In the case of the core separation for {\bf C0}
and {\bf C2}, the evolution is similar at the two frequencies. For {\bf C2} we
find some indication that the lower frequency-data can be found at larger core
separations.

\subsection{Evolution of the mean jet ridge line}

\begin{figure*}[htb]
\begin{center}
\subfigure[]{\includegraphics[width=12.0cm,clip]{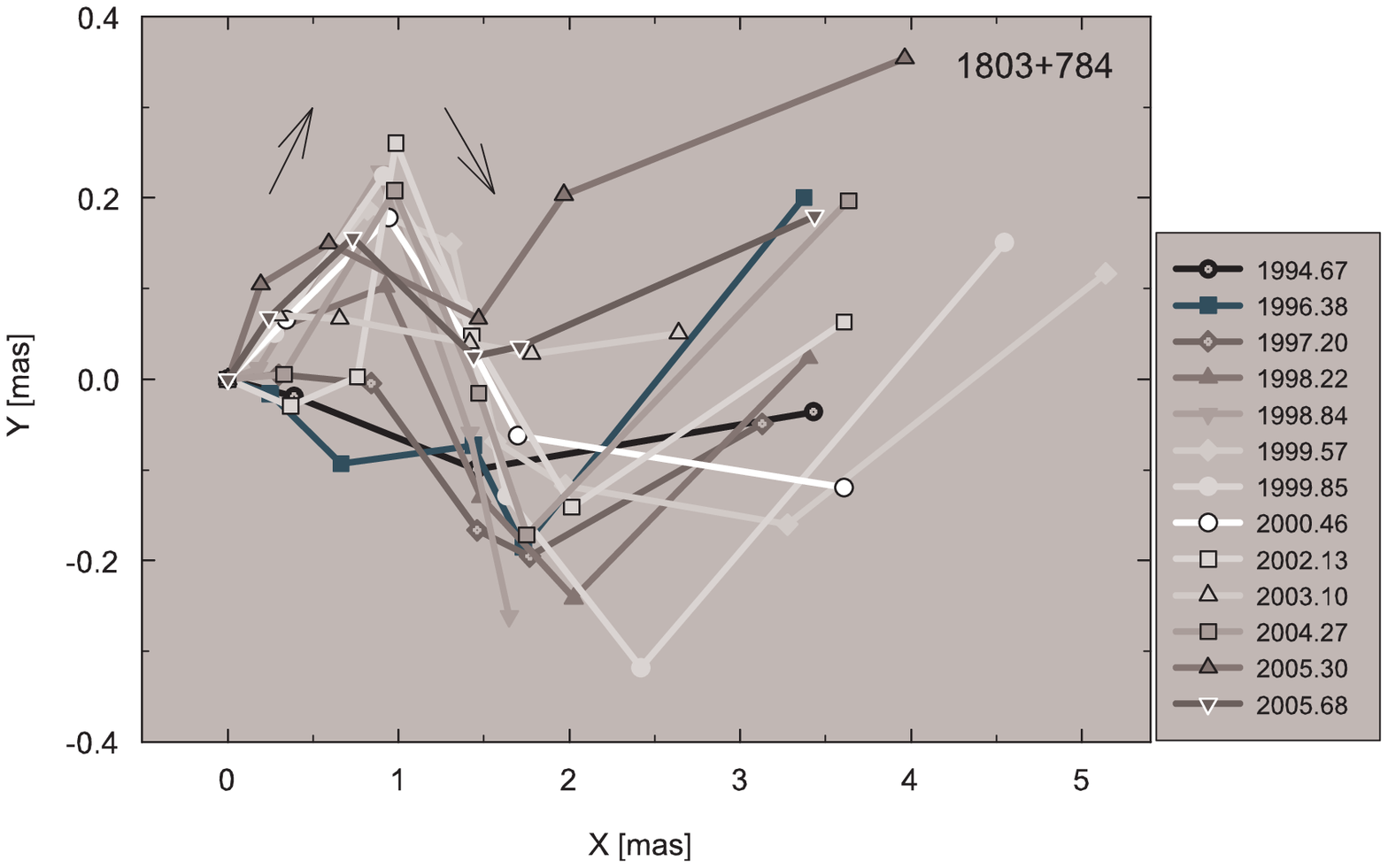}}\\
\end{center}
\vspace*{-0.8cm}\hspace*{0.1cm}\subfigure[]{\includegraphics[width=9.0cm,clip]{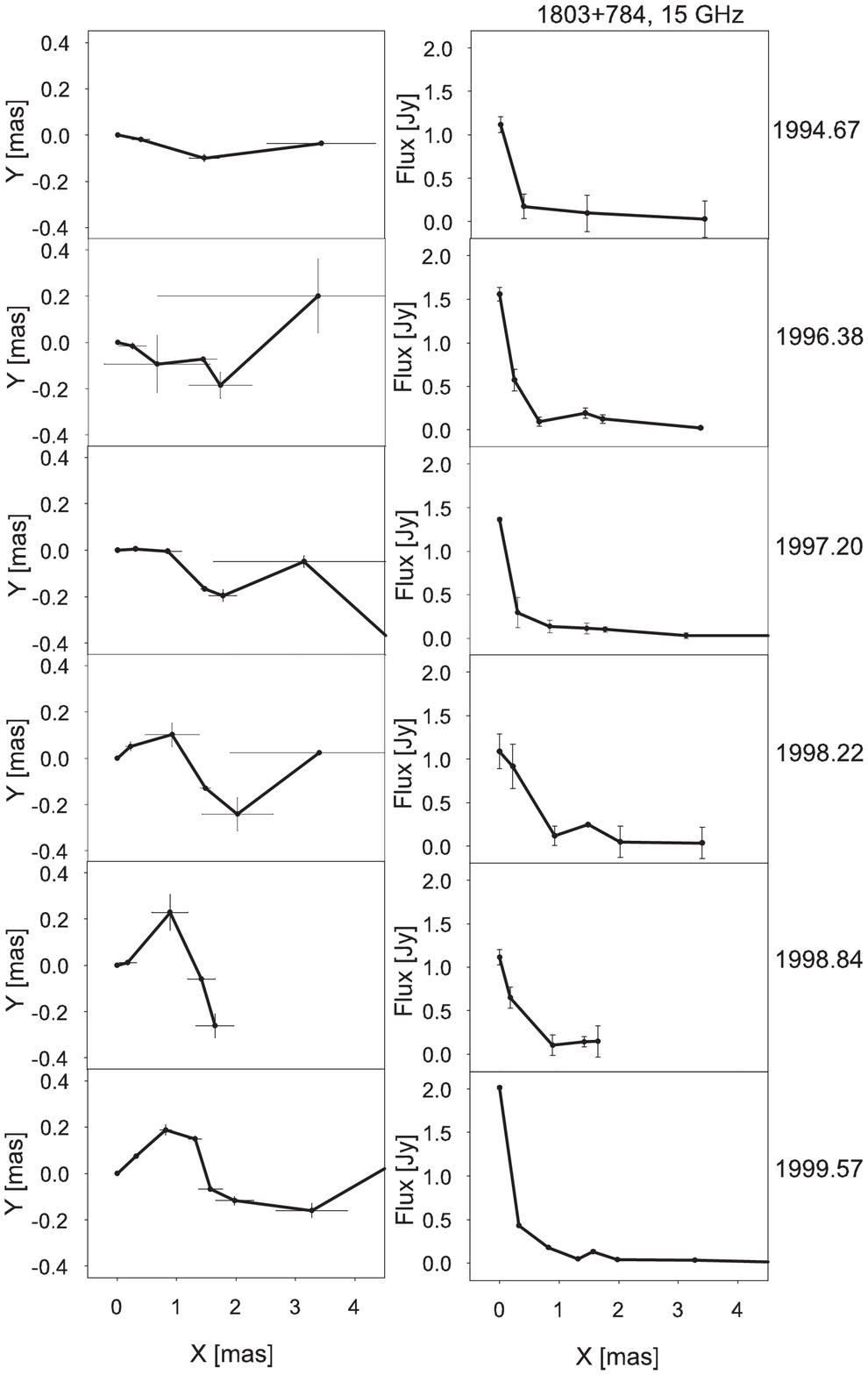}}
\vspace*{-0.8cm}\hspace*{1cm}\subfigure[]{\includegraphics[width=9.0cm,clip]{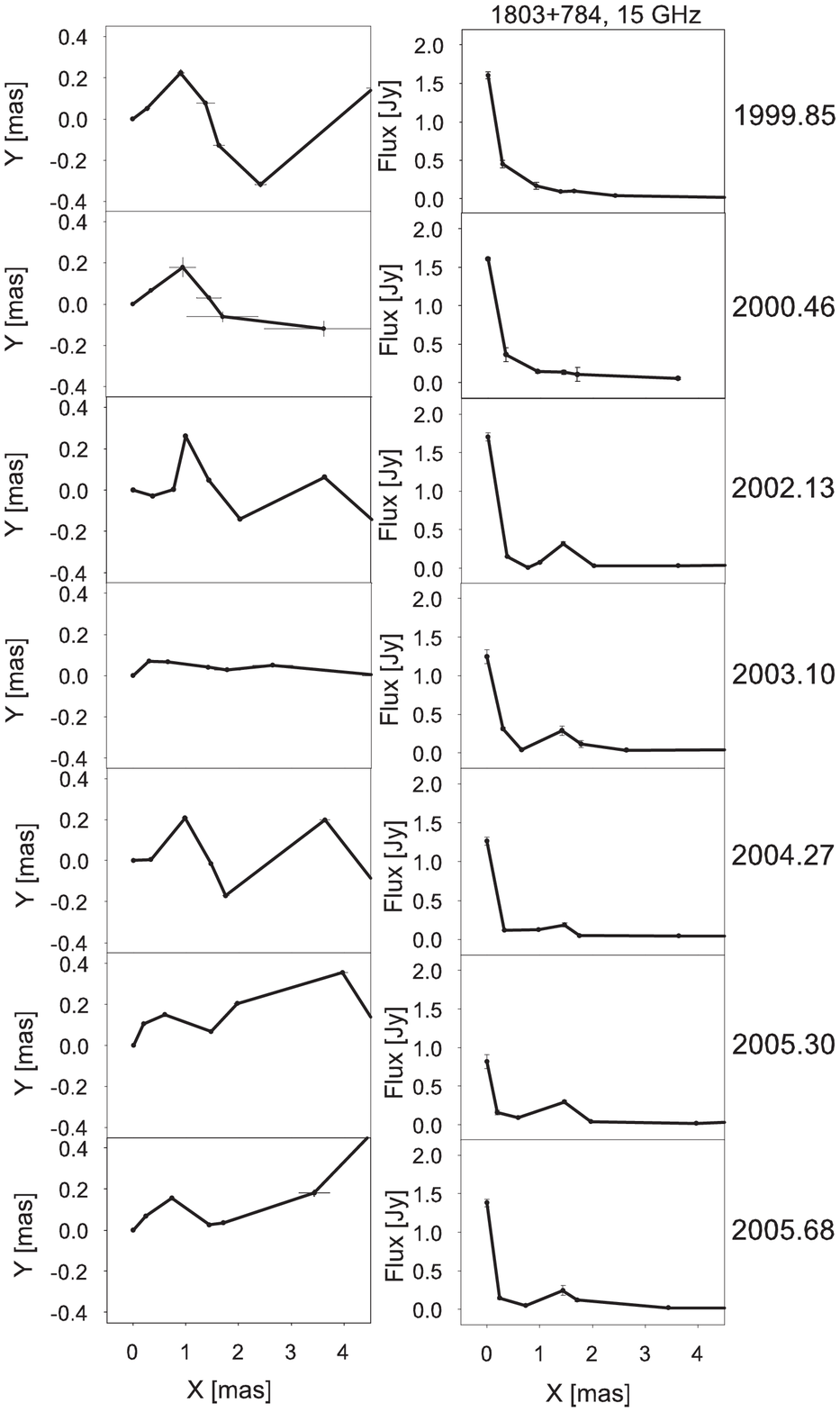}}
\vspace*{1cm}\caption{The jet ridge line for all components detected in 15~GHz
observations within 6 mas of the core.  Panel (a) overplots all epochs, as
denoted by the different lines/points. Panels (b) and (c) show the trace of
the jet in Cartesian coordinates (left-hand columns) and the flux density
along the jet as a function of {\it x} (right-hand columns).}
\label{grey}
\end{figure*}

In Fig.~\ref{grey} we show the jet ridge line for all epochs obtained at 15~GHz
in Cartesian coordinates. The jet ridge line is defined by a line connecting
all component positions at a given epoch and is independent of any component
identification. We find that the jet ridge line
changes its shape between epochs in a periodic way. Fig.~\ref{grey} panel (a) 
shows
the temporal evolution of jet component positions at 15 GHz for the time period
1994--2005.  It overplots the jet ridge lines at each epoch, with
each dot representing the position of one jet component and each line
connecting the jet components for one
particular epoch. Panels (b) and (c) show this
evolution with time in more detail; the left-hand columns showing the trace of
the ridge line at each epoch, and the right-hand columns the flux density of
the components along the jet as a function of {\it x}. From an almost straight line in 1994.67,
the shape of the jet evolves into a sinusoid with the northernmost
value at $x \sim$ 1~mas and the southernmost value at $x \sim$ 2~mas.
The amplitude of the sinusoid reaches its maximal
values at epoch 1998.84 and decreases again, forming an almost straight line 
again in 2003.10. One period of this excursion is completed after $\sim$ 8.5 years and the jet shape
starts to evolve into a sinusoid again. However, the position of the straight
lines in 1994.67 and 2003.10 are different, with an offset in {\it y} of
0.1~mas. We thus see an evolution of the jet ridge line with
time as defined by the positions of the individual components within the jet.

\subsection{Evolution of the jet width}

\begin{figure*}[htb]
\begin{center}
\subfigure[]{\includegraphics[clip,width=8.5cm]{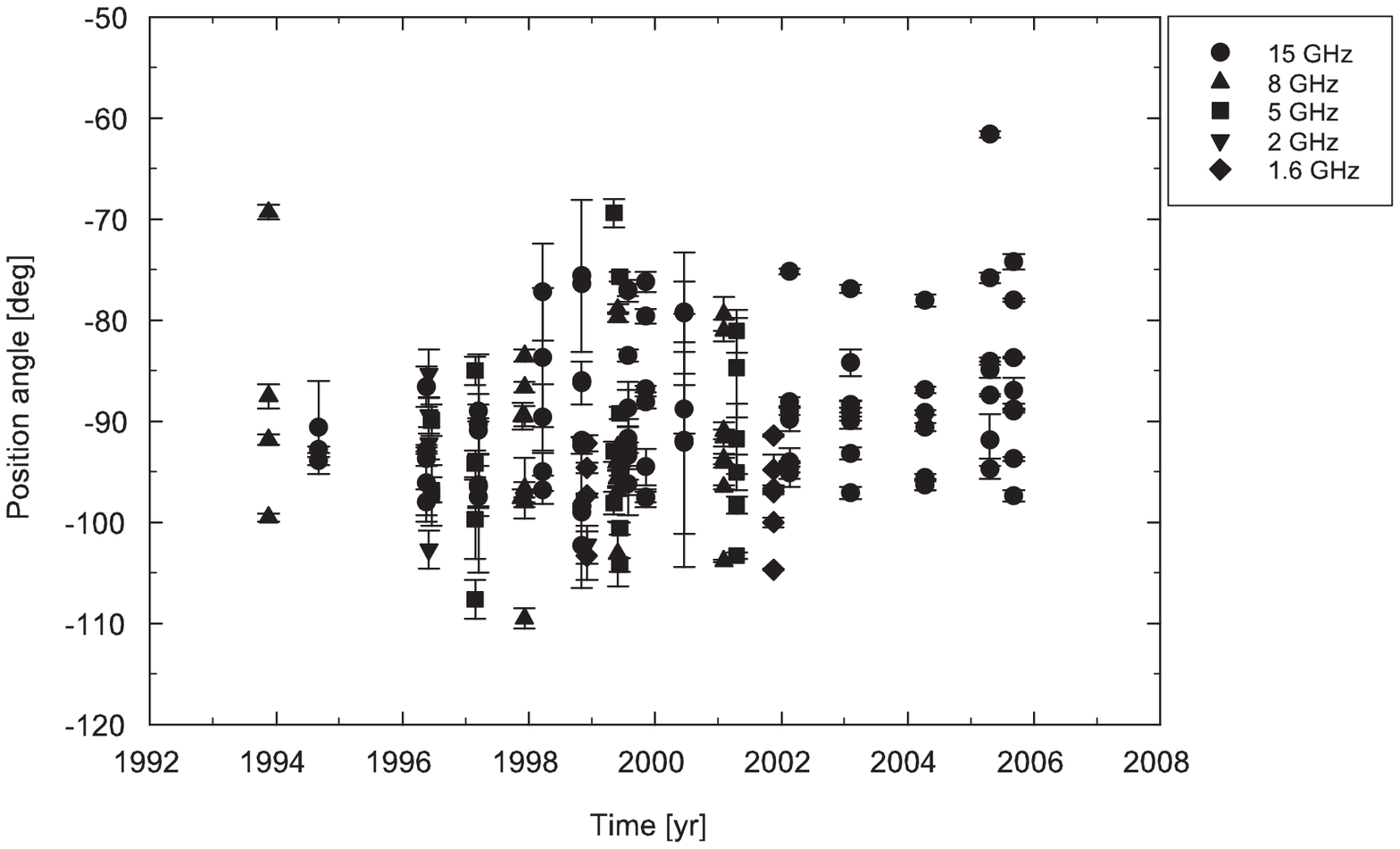}}
\subfigure[]{\includegraphics[clip,width=8.5cm]{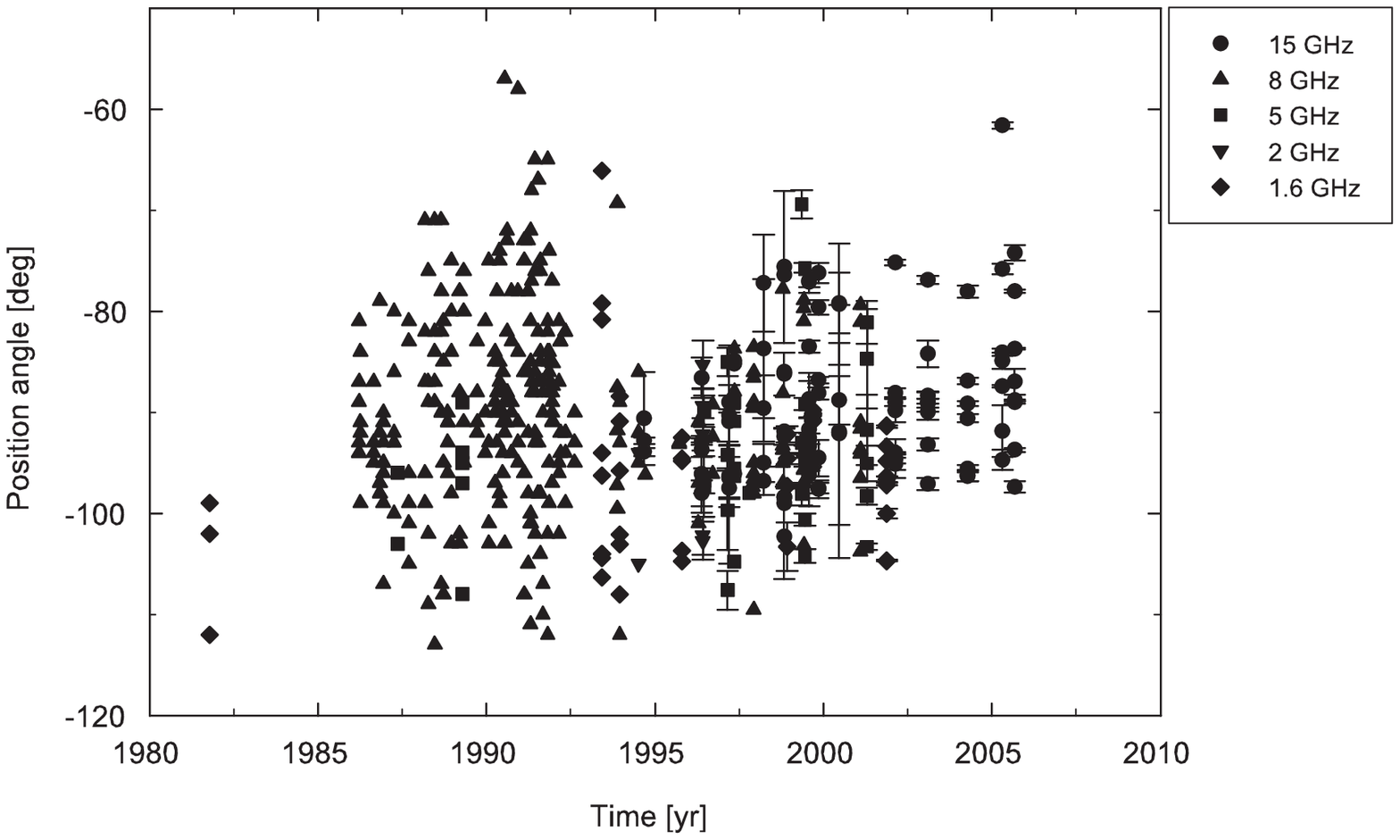}}\\
\hspace*{-1cm}\subfigure[]{\includegraphics[clip,width=7.5cm]{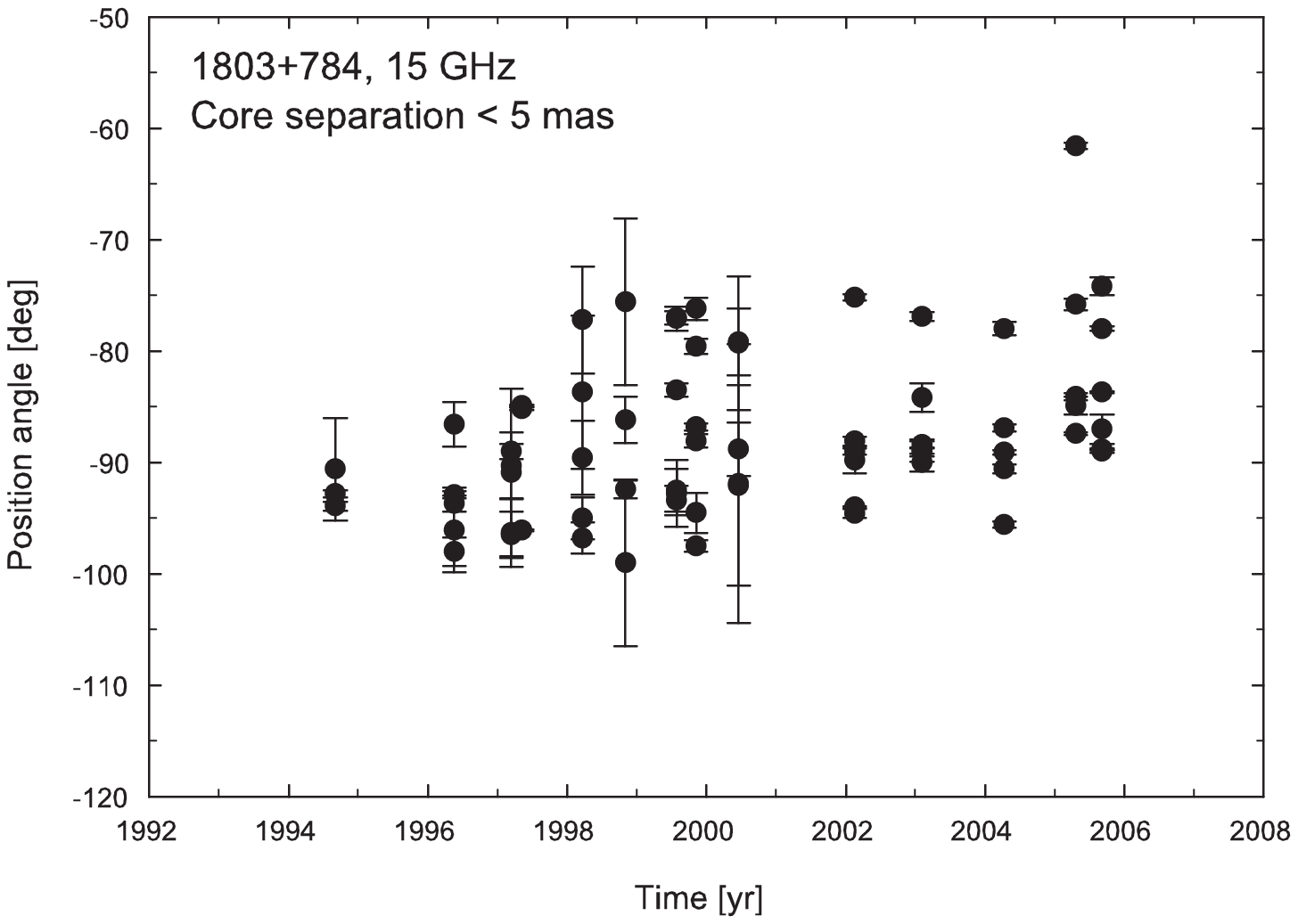}}
\hspace*{1cm}\subfigure[]{\includegraphics[clip,width=7.5cm]{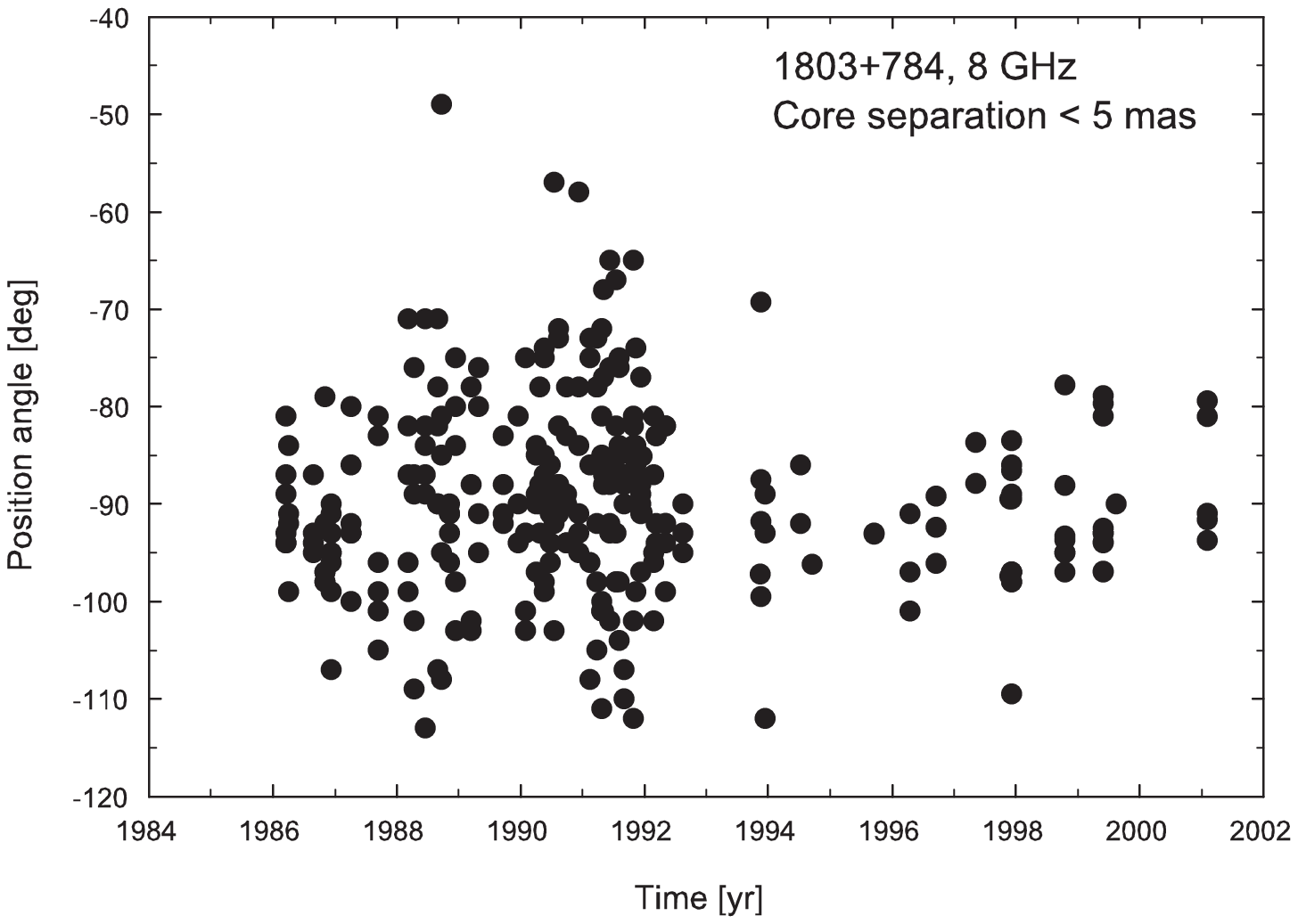}}
\end{center}
\caption{Position angle distribution of individual jet components as a function
of time.  Panels (a) and (b) show the new observations from this paper and
those combined with the observations from the literature, respectively, for
all frequencies.  Panels (c) and (d) show components within 5~mas of the core 
15~GHz and 8.4~GHz, respectively.}
\label{jet_width}
\end{figure*}

In this section we investigate the width of the jet as defined by the position
angle distribution that the jets components span. Fig.~\ref{jet_width}
shows this distribution as function of time, including all jet components seen at all
observed frequencies. This plot shows that the jet changes its width with time:
around 1995 the jet width appears to be quite small ($\sim$5$^{\circ}$), around
1999.5 the jet width appears to be much broader ($\sim$ 40$^{\circ}$) and is
narrower again in 2004. To investigate the longterm evolution of this effect,
we include data taken from the literature in Fig.~\ref{jet_width} panel (b) and find
a similar evolution of the width. The complete dataset (our data +
literature data) seems to cover two cycles of jet-width broadening with a 
period of about 8--9 years. To investigate whether this broadening is mainly a
frequency dependent effect, we show the distribution at 15~GHz in panel (c) and
at 8.4~GHz
in panel (d); both panels confirming the trend of changing jet width. To
investigate whether less well-constrained model components at large core
separations with large position angles produce this effect, we checked whether
the shape of the plot changes when we include only jet components within a
certain range of core separations. We checked the data for the jet components
within the inner 1 mas, 2 mas, etc. up to 7 mas. However, the shape of the
position angle distribution is stable and does not change significantly. Panels
(c) and (d) discussed above 
show the distribution for components within 5~mas of the core.

\begin{figure} [htb]
\begin{center}
\hspace*{0.5cm}\includegraphics[width=6.2cm,clip]{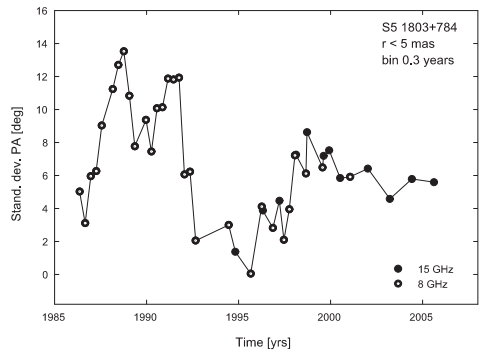}
\end{center}
\caption{Time evolution of standard deviation of position angles, calculated 
for components within 5~mas of the core in 0.3~year bins. Dotted circles
show 8.4~GHz data; circles 15~GHz data.} \label{pa_bin}
\end{figure}

To check whether periods of more frequent observations produce more outliers in
position angle and thus affect the position angle distribution, 
we binned the data in 0.3 year bins and calculated the standard
deviation in position angle within each bin, both for 5~GHz and 8.4~GHz.
This procedure used the components within 5~mas of the core.
Fig.~\ref{pa_bin} shows the
evolution of the standard deviation per bin 
(the outliers were removed from the plot), using the
center of each bin as its epoch for plotting. 
The position angle spread with time shows
a smooth increase starting in 1985, and reaches its
maximum in 1988--1992.  It then decreases again and has the second maximum in about
1998. It is clearly seen that the spread of the position angle evolution has
two cycles with a characteristic timescale of about 8--9 years, similar to the 
period suggested by the evolution of the jet ridge line. The
autocorrelation function gives the value of the timescale: $9.9\pm0.2$ with a
correlation coefficient of 0.79.
Binning the data with different bin length, such as 0.5 or 1.0
years gives the same results. In order to check whether the different number of
jet components for different epochs of observations can affect these results, we
selected only data points within 0.5 mas core separation. We only took the
first three jet components for each particular epoch into account and binned
the position angles into 0.3 year bins. The evolution of the standard
deviation of the position angles in these bins looks similar. We thus conclude
that the shape of the plot and characteristic time scale of the position angle
changes do not depend on the number of the jet components for a particular
epoch of observation.

\clearpage

\section{Summary: kinematic results}
\label{summary}

Before we describe the periodicities and correlations found in S5 1803+784, we
give a brief summary of the results of the kinematic analysis. In Britzen et
al. (2005a) we presented a component identification scenario for the pc-scale 
jet in S5~1803+784 based on the assumption that the brightest component seen in each observing epoch can be
identified with the same component ({\bf Ca}). The most
plausible identification scenario resulting from this assumption was the one
presented in that paper: the coexistence of several superluminally moving and
one {\it oscillating} component.\\

\begin{figure}[h]
\begin{center}
\includegraphics[width=8.5cm,clip]{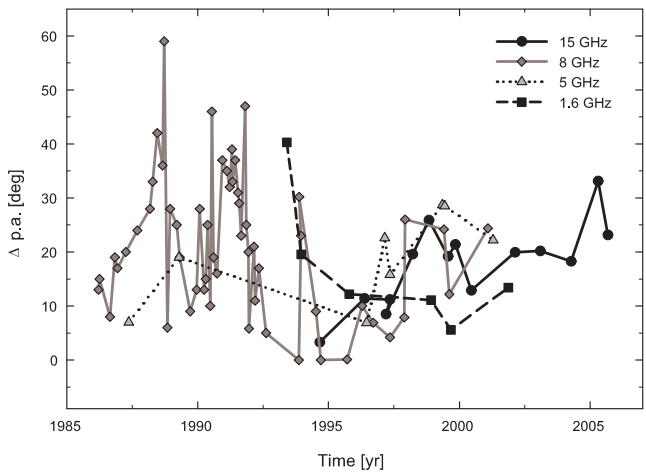}\\
\end{center}
\caption{The spread in the position angle for four different frequencies as a function of time, based on all the data (this paper + literature). We clearly find several maxima that are correlated with the maxima found in the total flux-density data.}
\label{dpa}
\end{figure}

The data presented in this paper cover a broader range in frequency
(1.6--15~GHz
and time (1993.88--2005.68). Thus, a more detailed investigation of the
jet component motion is possible. In addition, we collected kinematic
information from the literature and tested our hypothesis against these data.
In total, 94 data sets have been investigated.  Conclusions derived from
the present analysis include:\\
\begin{itemize}
\item most jet components within the inner $\sim 10$~mas of the core remain 
at roughly constant core separation over the long term (see Fig.~\ref{identification}).
\item the position angles of these components change significantly over 
time (see Fig.~\ref{freq_pa}).
\item component ({\bf B3}) moves with a superluminal apparent speed of $\sim$
19$c$ based on 3 epochs (see Figs.~\ref{identification} panel (a) and \ref{b3}).
\item the jet ridge line evolves with time, with a likely period of 
$\sim$8.5~years (see Fig.~\ref{grey}), independent of any component 
identification.
\item the width of the jet changes periodically with a similar period of $\sim$8--9 years (see Fig.~\ref{jet_width}).\\
\end{itemize}
In the following section we show that the epochs of the maximum widths of the position-angle distribution
correlate well with maxima in the total flux density
observed within the UMRAO monitoring programme (see Figs.~\ref{dpa} and \ref{light-curve}). 
We find convincing correlations between parameters such as the flux-density, core separation, position angle, and describe them below.

\section{Correlations \& Periodicities}
\subsection{Correlations \& Anti correlations}

As shown in previous sections, the jet width changes with time and the jet
ridge line itself shows a quasi-periodic variation with a characteristic time
scale of $\sim$ 8.5 yrs. Both facts could result from a geometric origin. This
would also produce correlation among the core separation, the position angle,
and the flux density changes for each jet component and across the different
components. To investigate these correlations, we first performed a visual
analysis of the time series of the position angle,
core separation, and the flux of each
component at 8.4 and 15 GHz. Maxima and minima at 8.4~GHz are listed in Table
\ref{8ghz}. Major changes in the variations of the jet parameters at 8 GHz
occurred in 1987, 1988.5, 1990 and 1991 for {\bf C0}, {\bf C1}, {\bf Ca} and
the core. The core flux reached a maximum in 1987, accompanied by minima in
flux, position angle, and core separation of {\bf C1}, maxima of the core
separation of {\bf C0} and maxima for flux and position angle of {\bf Ca}. In
1988.5, the values that were maximal became minimal and vice versa, and switched
again in 1990 and 1991. This indicates that changes in flux, position angle and
core separation of the components and the core-flux are correlated and change
with characteristic time scales of a few years. Similar behavior was found as
well at 15~GHz at two other epochs (Table~\ref{15ghz}). 
The major changes in core separation, position angle, and flux variations
of the inner components occurred in 1998 and 2000.\\

To investigate the correlations among these circular Gaussian parameters of
a component and across multiple components more carefully, we calculated the discrete
cross-correlation functions (Edelson \& Krolik 1988) for the pairs of
parameters (core separation, flux density, position angle) for each of the
components ({\bf C0}, {\bf C1}, {\bf Ca}, {\bf C2}, {\bf C4}) at 8.4~GHz.
Component {\bf C2} at 15 GHz possibly comprises 3 components at 8.4~GHz. 
We call these three {\bf C2} components at 8.4~GHz {\bf C2$_{1}$, C2$_{2}$}, and {\bf C2$_{3}$}. 
We find the following results:\\

\begin{table*}[htb]
\begin{center}
\caption{Maxima and minima in flux density, position angle and
core separation at epochs
1987, 1988.5, 1990, and 1991 at 8.4~GHz.}
\label{8ghz}
\medskip
\begin{tabular}{l|ccc|ccc|ccc|ccc}
\hline \multicolumn{1}{c|}{Time} & \multicolumn{3}{c|}{{\bf 1987}}
&
\multicolumn{3}{c|}{{\bf 1988.5}} & \multicolumn{3}{c|}{{\bf 1990}} &
\multicolumn{3}{c}{{\bf 1991}}\\
\hline \hline
Comp. ID & Flux & P.A. & Core sep. & Flux & P.A. & Core sep. & Flux & P.A. &
Core sep.& Flux & P.A. & Core sep.\\
\hline
{\bf Core} & max & & & min& & & max & & & min& & \\
{\bf C0} & & & max& & & min& & & max& min & & min\\
{\bf C1} & min & min & min & max & max & max & min & min & min & max & max
&max  \\
{\bf Ca} & max  & max & min & min  & min & max & max & max & min & & & \\

\hline
\end{tabular}
\end{center}
\end{table*}

\begin{figure}[ht]
\begin{center}
\includegraphics[width=6.0cm,clip]{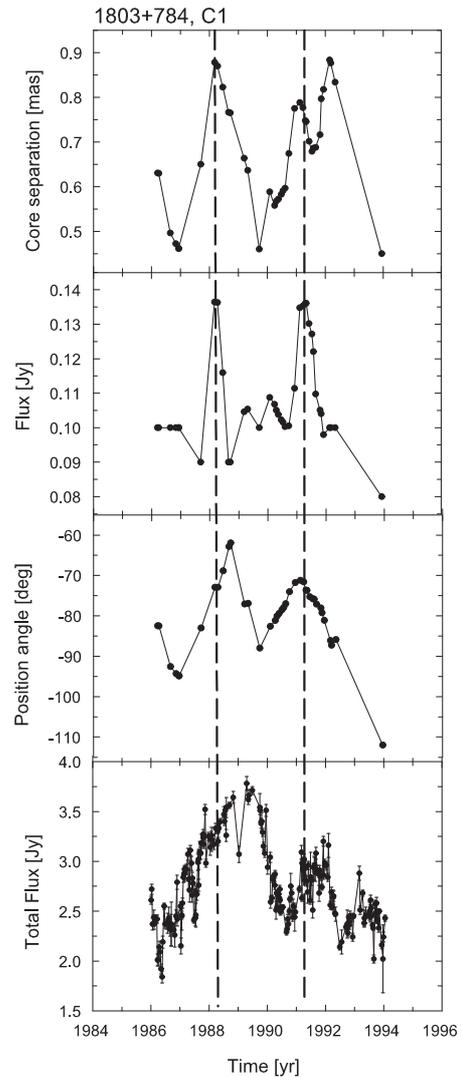}
\end{center}
\caption{Core separation, flux density, position angle of component {\bf C1}, and total flux-density light-curve at 8 GHz in the period 1984--1996. The two dashed lines indicate the nearly simultaneous peaks in all four plots.}
\label{corr_c1}
\end{figure}

\begin{itemize}
\item
For the brightest jet components {\bf Ca} and {\bf C1},
all the parameters are correlated.
As an example,
Fig.~\ref{corr_c1} shows the time series of the core separation, the
position angle, and the flux density of the component {\bf C1} from
1984 to 1996. 
\item
For the jet components {\bf C2$_{1}$}, {\bf C2$_{2}$} and {\bf C4}, we
find a strong correlation or anti-correlation among the variations in
flux density, position
angle, and core separation within each component. Table~\ref{dcf_total} shows
the discrete correlation function peaks for these parameters in jet components.
\item The core flux is correlated with
the core separation, the flux density, and the position angle changes
of all inner jet components {\bf C0}, {\bf C1}, {\bf Ca}, {\bf
C2$_{1}$}, {\bf C2$_{2}$} and {\bf C2$_{3}$} (see
Table~\ref{dcf_comp}).
\item
Variations of individual circular Gaussian parameters also show correlations 
across different
jet components.
Table~\ref{dcf_comp} also shows the results of the discrete
correlation function calculation. 
\end{itemize}

\begin{figure}[ht]
\includegraphics[width=8.0cm,clip]{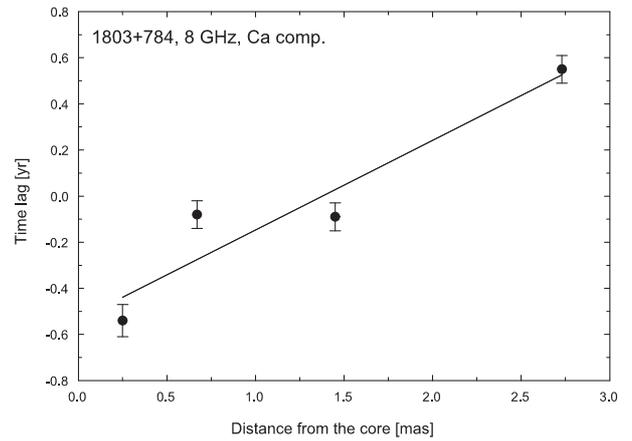}
\caption{Time lag of the discrete cross-correlation function of the core separation of component {\bf Ca} vs. those of components {\bf C0}, {\bf C1}, {\bf C2$_1$}, and {\bf C2$_3$} as a function
of the mean value of distance from the core for these components.
The linear regression fit to these time lags is overplotted.}
\label{ca_lag}
\end{figure}

When cross-correlating core separation changes of {\bf Ca} and those of other
components, significant peaks in the discrete correlation function (e.g., {\bf
C0}, {\bf C1}, {\bf C2$_{1}$} and {\bf C2$_{3}$}) are associated with time lags
that change gradually with distance along the jet.  
The time lags appear to increase
roughly linearly with core separation;
Fig.~\ref{ca_lag} shows
the time lags for these component pairs as a function of the components'
mean core separation
(time lags for components interior to {\bf Ca} have had their sign inverted
from Table~\ref{dcf_comp}), and fit with a linear regression.
Similar behavior is found for the cross-correlation between the core separation
changes of {\bf C1} and {\bf Ca}, {\bf C2$_{1}$}, {\bf C2$_{2}$}, and {\bf
C2$_{3}$}.  This can be explained if we assume that the motion of the
components is mostly due to a geometrical origin. The components follow
almost the same trajectories, but with a certain time delay. The components are
connected with each other and the information is spreading with a proper motion
of about 2.3 mas/yr if the reference point is component {\bf Ca} and 0.79
mas/yr if the reference point is component {\bf C1}, corresponding to
extremely fast speeds of $53c$ and $18.2c$, respectively.\\

\begin{figure}[h]
\begin{center}
      \includegraphics[width=8.5cm,clip]{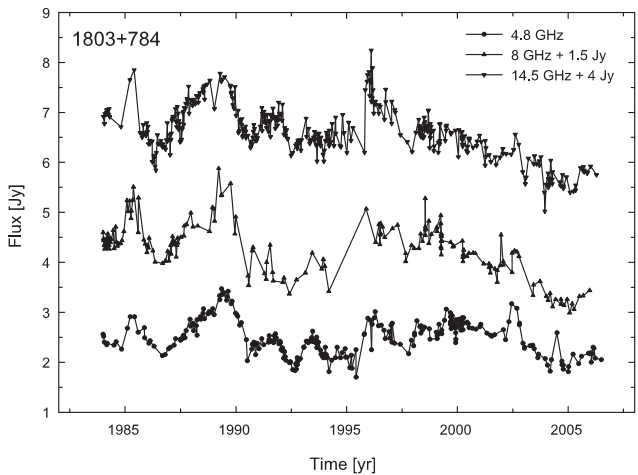}
\end{center}
\caption{Total single-dish flux-density light-curve for S5 1803+784 at 4.8, 8
       and 14.5~GHz. The 8 and 14.5 GHz light-curves are shifted by 1.5 and
       4~Jy, respectively. All the data are
       from the UMRAO monitoring campaign.}
\label{light-curve}
\end{figure}

We investigated possible correlations among the jet-component
parameters and the total flux-density variability. For this purpose
we used the UMRAO data of multi-frequency AGN monitoring (Aller et al. 1999, 2003), spanning almost 30 years
(see Fig.~\ref{light-curve}). Using the 8~GHz UMRAO light-curve over
the time-range spanned by the VLBI data, we find:
\begin{itemize}
\item variations in the core separation and position angle of 
{\bf Ca} and {\bf C2$_{1}$} are correlated with the
total flux-density variability (see Table~\ref{dcf_total}).
\item variation in the core separation and the
position angle changes of the jet component {\bf C1} are correlated
with the total flux density at 14.5~GHz (see Fig.~\ref{corr_c1}),
but do not show any correlation at 8~GHz.
\end{itemize}
\begin{table}[htb]
\begin{center}
\caption{Maxima and minima in flux density, position angle and
core separation at epochs 1998
and 2000 at 15~GHz.}
\label{15ghz}
\medskip
\begin{tabular}{l|rrr|rrr}
\hline \multicolumn{1}{c|}{Time} & \multicolumn{3}{c|}{{\bf 1998}}
&
\multicolumn{3}{c}{{\bf 2000}}\\
\hline \hline
Comp. ID & Flux & P.A. & Core sep. & Flux & P.A. & Core sep.\\
\hline
{\bf Core} & min & & & max & & \\
{\bf C0} & max & max & min& & & \\
{\bf C1} & & & & max & max & max \\
{\bf Ca} & max & min & max & min & max& min\\

\hline
\end{tabular}
\end{center}
\end{table}

\begin{figure}[ht]
\begin{center}
\subfigure[]{\includegraphics[clip,width=7.0cm,clip]{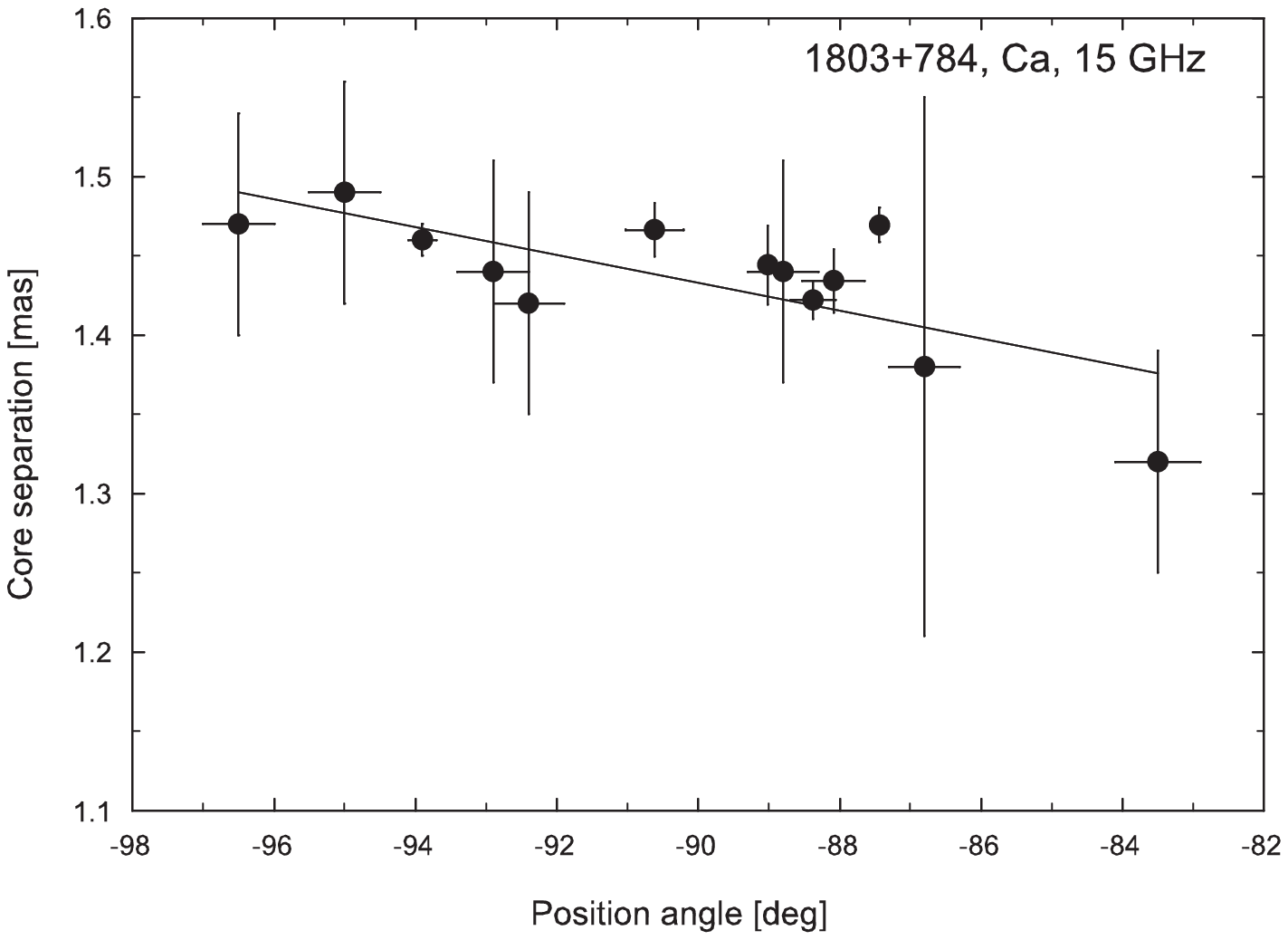}}
\subfigure[]{\includegraphics[clip, width=7.0cm]{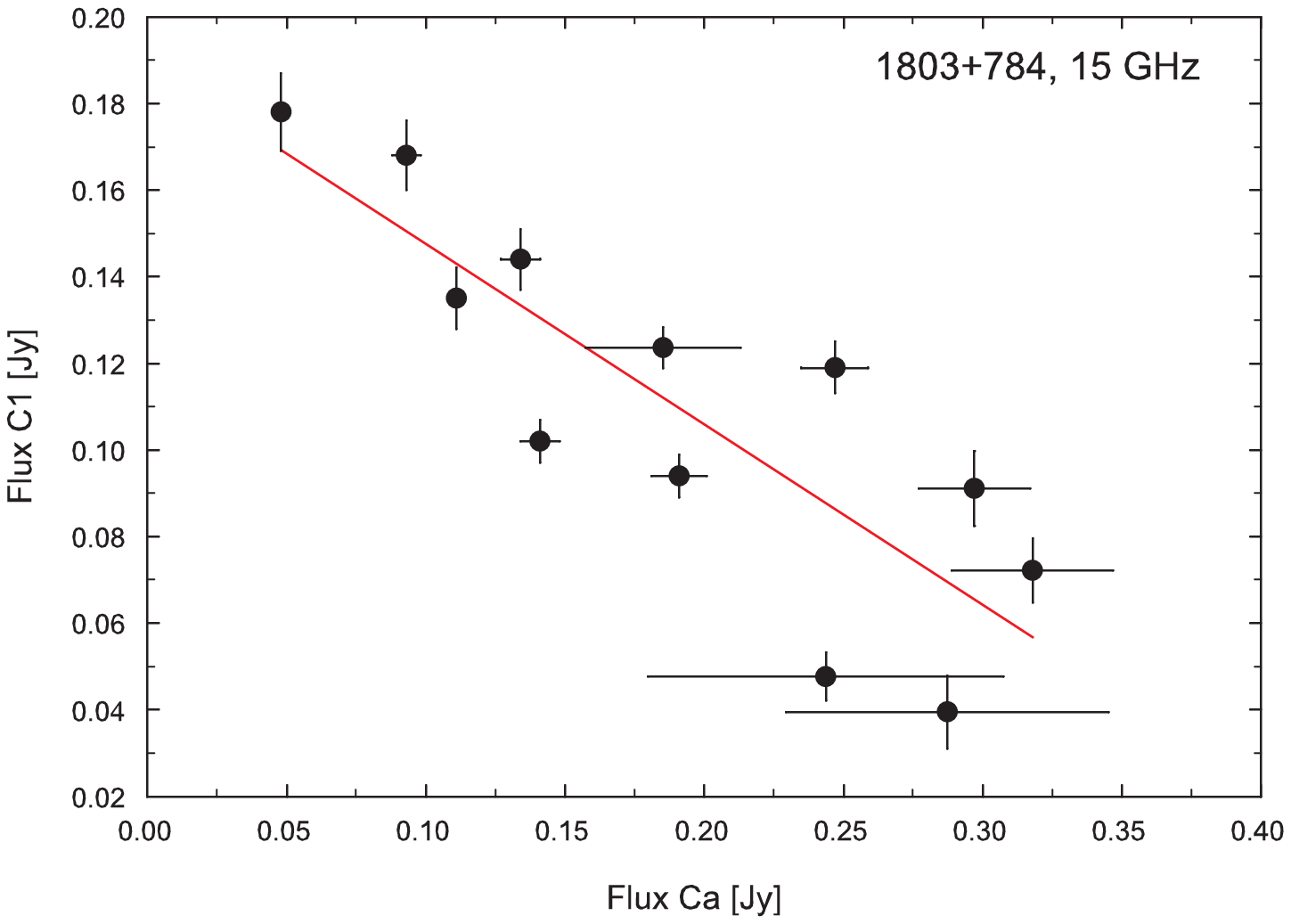}}
\end{center}
\caption{Panel (a) shows core separation as a function of position angle for the brightest component {\bf Ca}.  Panel (b) shows the flux density of component {\bf C1} against that of component {\bf Ca}.  Both panels plot the parameters estimated from 15~GHz observations from all the different epochs.}
\label{corr}
\end{figure}

In order to investigate the same correlations at 15~GHz where the number of
data points was not sufficient for the discrete correlation function analysis,
we calculated the Pearson's correlation coefficient and inspected them
visually.  Tables~\ref{dcf_total} and \ref{dcf_comp} show the 
Pearson correlation coefficients and the probabilities for finding
such a value by chance for various pairings of 
different parameters for a single component and the same parameter across
two components, respectively.
The flux-density
variability is anti-correlated for the pairs of
{\bf C1}/{\bf Ca}, and {\bf Ca}/{\bf C2}.
We only find correlated core separation changes
for the pair {\bf Ca}/{\bf C2}. 
Fig.~\ref{corr} shows notable examples for a correlation
and anti-correlation:
the position angle vs.\ core separation in component {\bf Ca} (panel a) and
the flux densities in components {\bf Ca} vs.\ {\bf C1} (panel b),
respectively.\\





\subsection{Periodicities}
\label{Periodicities}

To test for possible periodic changes in the component parameter variations, we
applied a discrete autocorrelation function (Edelson \& Krolik 1988), a
data-compensated discrete Fourier transform (Ferraz-Mello 1981), and the
Jurkevich method (Jurkevich 1971). All three methods were applied over the
time-range 1981--2005 for all components out to {\bf C12}.  We
find a quasi-periodicity of about 2 and 4 years in the variation of the core
separation, the position angle, and the flux with time for the inner jet
components {\bf C0}, {\bf C1}, {\bf Ca}, and {\bf C2} at 8 and 15 GHz. This
agrees well with the periods of 2 and 3.9 years found by Kelly et al. (2003) in
the total flux density variability. A more detailed description of this
analysis will be presented by N.A. Kudryavtseva et al. (in prep.).

\section{Discussion}
\label{discussion}

In this paper we have presented multi-frequency kinematic analysis for S5 1803+784
based on observations covering roughly twenty years in time. The most important
result is
that most of the features in the pc-scale jet up to a core distance of $\sim$12
mas remain at roughly constant core separations over long time scales. This is in contrast to most
previously presented kinematic scenarios for this source or flat-spectrum AGN in general. While many authors
find that component {\bf Ca} is stationary, most authors claim outward motion for the
remaining components. In contrast, the 2cm/MOJAVE webpage and Lister et al. (2009) show a
core separation/time plot in which several components within 
$\sim$~8~mas of the core also maintain roughly constant core separations. 
This is consistent with our results presented in this
paper. The 2cm data from the 2cm/MOJAVE survey form part of the literature
data sets included in the present analysis.
The fast-moving component {\bf B3} is not seen by the 2cm/MOJAVE.\\ 

The combination of new and re-analyzed data spanning 12--20 years shows that
all bright components of the inner jet remain at roughly constant core
separations, with no significant outward superluminal motion.
Applying our component identification procedure to
available data sets from the literature leads to a consistent kinematic
picture.
Although these components remain at roughly constant core separations
on the long term,
they do exhibit significant changes in
their position angles.
This results in an evolution of the jet
ridge line from a straight line to
a sinusoidal form with a characteristic time-scale of 8--9 years.
A similar evolution of the jet ridge line has
so far only been observed in one other BL Lac object: 0716+714 (Britzen et al.
in prep.). Following such jet ridge-line evolution proceeds independently of any
component-identification-scenario but requires a large number of epochs to be
discovered and traced over several periods of evolution.\\

We find a number of correlations between the parameters of the core and those
of individual components. The best case is seen for {\bf C1} where the total
flux, the component flux, the core separation, and the position angle are
obviously related. Many other pairs of correlations or
anti-correlations are presented in this paper. The set of all these correlations
and periodicities together argues in favor of a geometric origin.
In the following sub-sections, we discuss possible scenarios to explain the observed
evolution in the pc-scale jet of S5 1803+784 and the kinematic behavior of
its components.

\subsection{Curved jet structures / Precession?}

Curved jet structures have been observed in many different sources,
both extragalactic (e.g., 1928+738, OJ 287, 3C 273, 3C 449) and Galactic (e.g.
SS433, HH111). There are prominent differences between stellar and AGN jets,
including the role of magnetic fields in the collimation and dynamical
evolution of their structure and 
the jet-to-environment density ratio that is significantly larger for the
stellar jets.
Based on the later characteristic, the dynamical evolution of a stellar jet can
more easily be treated ballistically, whereas such an assumption would not 
necessarily apply to the case of AGN jets (Hardee et al. 1994).\\

Curving and wiggling of jets is usually attributed to precession of the base of
the jet. Two different scenarios concerning the central engine that feeds the
jets can explain the observed curvature.\\

\subsubsection{Models relying on a single central Black Hole}

In the first scenario, the core that ejects the jet is considered to be a
single object (usually a supermassive black hole, SMBH). Precession can then
occur due to the Bardeen-Petterson effect through Lense-Thirring precession
of the accretion disk (Caproni et al. 2006). However, this model assumes a
rigid body configuration between the accretion disk and the jet, an assumption
that we are not able to test or verify. A magnetic torque mechanism has also
been proposed (Lai 2003), driven by the large-scale magnetic field
threading through the accretion disk, leading to precession of the disk and
consequently of the jet. In the same frame, a group of models attributes the
curving and wiggling of the jet to magnetohydrodynamic (MHD) instabilities,
most prominently Kelvin-Helmholtz (K-H) instabilities (Camenzind \&
Krockenberger 1992; Hardee \& Norman 1988; Birkinshaw 1991; Zhao et al. 1992;
Hardee et al. 1994; Hardee et al. 1997; Meier \& Nakamura 2006; Perucho et al.
2006). Meier \& Nakamura (2006) consider the instability to be driven by
current, related to the strong-field pinch, rather than by K-H instabilities. All
models mentioned above strongly depend on the jet-to-environment
density ratio. However, instabilities have to be excited in order to affect the
structure of the jet. Thus, although the MHD instabilities (and mainly the
helical K-H kink instability) can possibly explain the structure of pc and kpc
scale jets, the problem of how these modes are excited 
still remains.\\

\subsubsection{Models relying on a Binary Black Hole}

In the second scenario, models have been built under the assumption that a
binary system can be found in the centers of AGN. The most popular idea
concerns binary black hole systems (BBH) as a product of galaxy merging. The
problem is complex and each of the models makes a number of assumptions to
explain the observed structures. In such systems, the primary black hole is
considered to have an accretion disk and jets, whereas the secondary black hole
is in orbit around the primary. Gravitational perturbations associated with
this orbiting secondary black hole
can disrupt the accretion disk of the primary black hole and lead to
precession. Lobanov \& Roland (2005) argue that in light of the ``two-fluid"
model, the orbital motion and the disk precession can lead to curving and
wiggling of the jet. Furthermore, they propose that for strong enough magnetic
fields, the K-H instabilities can be considered negligible. Roos et al. (1993)
argue that the orbital motion of the binary is responsible for the change of
the jet-ejection direction and thus leads to a wiggling structure
(the jet is treated ballistically). Kaastra \& Roos (1991) argue that
the precession angle is determined by the angle between the primary black hole
spin axis and the angular momentum of the binary system. As before, the jet is
treated ballistically. Similar models have been proposed by Katz (1997) (a
gravitational torque induced by the secondary black hole onto the
accretion disk) and  Romero et al. (2000) (near-rigid body configuration).\\ 

Roland et al. (2008) apply a BBH model to the data for S5 1803+784 presented
and discussed in this paper. The wiggling of the jet is attributed to a
precession of the accretion disk, the orbital motion of the binary, and
the movement of the whole system around the center of gravity of the AGN host
(the ``two-fluid'' model is used again).\\

 \subsubsection{Garden hose models}

Another group of models, usually identified as ``garden-hose'' models, have
been extensively used for stellar jets.  They assume a ballistic evolution of
the jet, and place specific conditions on the distribution of the density of
the jet and surrounding medium. Most prominent examples for this type of
model are Hjellming \& Johnston (1981) (a kinematic model for SS433 X-ray
binary), Raga et al. (1993) (an analytical model for the expansion of jets that
get ejected with a time-dependent direction), Biro et al. (1995) (a 2D numerical
simulation based on the analytical model of Raga et al. (1993), considering adiabatic and
non-adiabatic expansion of the jet), Cliffe et al. (1996) (a 3D
numerical simulation for precessing stellar jets considering interaction with
the surrounding medium), and Stirling et al. (2002). Along the lines
of Hjellming \& Johnston (1981), Gower et al. (1982) produced a similar model
for radio AGN jets, taking into account projection, relativistic, and
cosmological redshift effects. This family of models again does not answer the
problem of how the original precession arises, but they do offer a
plausible explanation of the evolution of such perturbed jets. As all these
models assume ballistic behavior, it is possible that they might not
describe AGN jets sufficiently well.\\

Rieger (2004) presented a comparison of different configurations for curved
jets, explained by ballistic and non-ballistic models. He proposes that a
non-ballistic description of AGN jets is more appropriate. Geodesic 
(Lens-Thirring) precession
of the accretion disk (Begelman et al. 1980) would
produce a period much longer than observed in the light-curves of AGN (e.g.
OJ 287, S5 1803+784). Instead those authors favor non-relativistic
precession which leads to significantly smaller periods. Finally, the complete
absence of magnetic fields and their effects on the evolution to the jet is
another reason to doubt the applicability of the above ``garden-hose'' models
in their present form to modeling of AGN jets. A model that would take into
account MHD instabilities and the role of large-scale magnetic fields on the
collimation and evolution of AGN jets could provide a good explanation of the
observed structure in S5 1803+784. We are currently in the process of 
preparing a model in this
direction and will apply it to the data for S5 1803+784 presented in this
paper. This model shall explain the position angle changes as well as the secondary ``oscillatory'' motion
of the individual jet components.\\

\subsubsection{External causes of jet curvature}

Finally we mention another group of models that rely on external causes for
explanation of the curving and wiggling of jets. The bending of the jet could
be attributed to interaction with the surrounding medium. This however can not
fully explain the case of S5 1803+784, since such periodic and fine structure
within the jet would imply an equally finely structured surrounding medium, for
which there is no evidence. Blandford \& Icke (1978) attributed the structure
of the jet of 3C 31 to dynamical interaction with a nearby galaxy (again
treating the jet ballistically). J\"agers \& de Grijp
(1985), based on the model of Blandford \& Icke (1978), proposed a similar
model for 3C~10, while studying the movement of discrete blobs along the jet.
Lupton \& Gott (1984), proposed that the curving of the jet of 3C~449 is due to
the orbital motion of the the central engine around the host's center of 
gravity.  But likewise, the 
periodic and finely structured curvature of the S5~1803+784 jet would seem to
preclude such longer-term motions as the full explanation.\\

Yokosawa \& Inoue (1985) combine some of the above models to the case of 3C~75,
which has a double radio source with two pair of jets that seem
to intertwine. Again treating the jets ballistically, the authors take into
account the pressure of the surrounding medium, assume a time-dependent
ejection velocity for the jets, and account for the orbital motion of the
central engines of the jets.\\

A viable model should offer a self-consistent description and physical basis 
for the observed structure and behavior. It is important to include 
the accretion disk as well
as the jet into the model. Moreover, in the case of S5 1803+784, the observed
behavior in which most jet components remain at roughly constant core
separations should also be explained.
In this light, a twisting jet model can 
connect the observed structure and evolution of the jet to the base of the
jet and the accretion disk itself. We follow Steffen et al. (1995) in describing
an expanding synchrotron blob ejected from a rotating hot spot in an accretion
disk around a supermassive black hole. This model has been introduced to
capture the kinematic aspects of the jet model of Camenzind \& Krockenberger
(1992) in the case of the apparent superluminal motion of jet
components in 3C\,345.  In our scenario the jet is
generated by collimated, relativistic outflow from an accretion disk, which
is assumed to be in Keplerian rotation. Although our knowledge of the nozzle region
between the disk and jet is not at all concrete, we can nonetheless make some viable
assumptions that lead to a fairly good fit of the model to the data
available. We are currently in the process of applying a twisted jet model to the data presented in this
paper. Further studies on how
such a hot spot in the accretion disk could come into being are however needed.

\subsection{Jet components at constant core separation}

As illustrated by the kinematic results for S5 1803+784 presented here,
components that show no significant apparent superluminal (outward) motion over long time-scales can exist
along the jet. Such components are also observed in a growing
number of AGN (see e.g., Jorstad et al. 2001; G$\acute{\rm o}$mez et al.;
Britzen et al. 2007). Different physical models and scenarios have been invoked
to explain this phenomenon.  A prominent example is the superluminal
radio source 4C39.25, in which a stationary component was observed by Alberdi et
al. (2000). They explain this in geometrical terms, proposing that
the stationary component is at a 90$^{\circ}$ bend in the jet away from the
observer. They disfavor the case of a recollimation shock based on polarization
data. G$\acute{\rm o}$mez et al. (2001) find three quasi-stationary components in
the jet of 0735+178 in a quiescent state of the source, and attribute them to
recollimation shocks and differential Doppler boosting. Britzen et al. (2009) show that
0735+178 is a ``hybrid'' in the sense that the kinematic properties change simultaneously with morphological changes.
At times when the pc-scale jet is straight, the jet components appear to be stationary.
At times when the pc-scale jet appears to be strongly bent, the jet components show apparent superluminal motion.
Britzen et al. (2007) and Kellermann et al. (2004) find stationary components in their 6 and 2cm AGN
surveys. We also note the case of the quasar 3C 395 (Simon et al. 1988; Lara et
al. 1994; Lara et al. 1999), whose jet exhibits a possible stationary component
due to possible bending of the jet. Mart\'i \& M\"uller (2003) model
AGN jets and find that pressure mismatches exist between the jet and the surrounding
medium. These lead to the production of reconfinement shocks and energy density
enhancements downstream of these shocks, which in turn give rise to stationary
radio knots. Numerical simulations by Agudo et al. (2001) of an expanding
relativistic jet show that the ejection of a superluminal component into the
jet can trigger instabilities that can give rise to stationary
components (shocks) in the wake of the superluminal component, especially in
the near vicinity of the core.\\

In the case of S5 1803+784 we observe components that show no long-term change
in their core separations, but do have position angles that vary in what
appears to be a (quasi-)periodic fashion.
Correlations among circular Gaussian parameters such as flux density,
position angle, and core separation argue in favor
of a geometric origin for the observed phenomena. We conclude that the
kinematics in S5 1803+784 can not be explained easily with recourse to published
models. With curved jets, there will be some cases where the jet crosses the
observer's line of sight, creating inward and stationary features. However,
it is difficult to apply such a scenario here, in which
a number of components show roughly constant core separations and 
varying position angles.  As mentioned in the previous sub-section,
we are currently working on a jet model to explain these characteristics of S5~1803+784.

\begin{acknowledgements}
We thank the anonymous referee for helpful comments and suggestions.
We are especially grateful to K.I. Kellermann, M.A. P\'erez-Torres,
A. Alberdi and J.M. Marcaide for sharing data, in part prior to publication.
N.A. Kudryavtseva and M. Karouzos were supported for this research through a stipend
from the International Max Planck Research School (IMPRS) for Radio
and Infrared Astronomy. This research has made use of data from the MOJAVE database that is maintained by the MOJAVE team (Lister et al., 2009, AJ, 137, 3718). This research has made use of data from the
University of Michigan Radio Astronomy Observatory which is supported
by the National Science Foundation and by funds from the
University of Michigan.
S. Britzen acknowledges support by the Claussen-Simon-Stiftung.
This research has made use of the NASA/IPAC Extragalactic Database (NED) which
is operated by the Jet Propulsion Laboratory, California Institute of Technology,
under contract with the National Aeronautics and Space Administration.
The National Radio Astronomy Observatory is a facility of the National Science Foundation operated under cooperative agreement by Associated Universities, Inc.
The European VLBI Network is a joint facility of European, Chinese, South African and other radio astronomy institutes funded by their national research councils.
Based on observations with the 100-m telescope of the MPIfR (Max-Planck-Institut f\"ur Radioastronomie) at Effelsberg.
\end{acknowledgements}

\clearpage
\clearpage

\onecolumn
\tabcolsep1.0mm
\setcounter{table}{0}
\tablecaption{Circular Gaussian model-fit results for S5 1803+784.  The columns list the epoch of observation, the jet-component identification, the flux density, the radial distance of the component center from the core, the position angle of the center of the component, the FWHM major axis, and the chi-squared of the fit.}
{\smallskip}
\hspace*{-6cm}
\tablehead{\noalign{\smallskip} \hline \noalign{\smallskip}
\label{1803_table}
(1)&(2)&(3)&(4)&(5)&(6)&(7)\\
\hline
{\small Epoch} & {\small Id.} & {\small S [Jy]}& {\small r [mas]} & {\small $\theta$ [deg]} & {\small M.A. [mas]} & $\chi^{2}$\\
\noalign{\smallskip} \hline \noalign{\smallskip}}
\tabletail{\hline\multicolumn{5}{r}{\small continued on next  page}\\}
\tablelasttail{\hline}
\begin{supertabular}{ccrrrrcccc}
\label{proper}
       &      &          &   {\small \bf 1.6 GHz}   &        &      &\\
\hline \noalign{\smallskip}
1998.92&{\bf r} &  0.946$\pm$0.142 & 0.00 &   0.0  & 0.26$\pm$0.02 &38.232803\\
1998.92&{\bf Ca}&  0.558$\pm$0.084 & 1.41$\pm$0.05 & 262.7$\pm$0.8 & 0.69$\pm$0.03  &\\
1998.92&{\bf C4}&  0.091$\pm$0.014 & 4.35$\pm$0.44 & 267.8$\pm$1.0 & 2.82$\pm$0.30  & \\
1998.92&{\bf C8}&  0.059$\pm$0.009 & 7.99$\pm$0.80 & 265.4$\pm$2.0 & 1.30$\pm$0.35  &\\
1998.92&{\bf C30}&  0.259$\pm$0.052& 29.67$\pm$3.00&  256.7$\pm$4.0& 15.72$\pm$4.00& \\
\hline
2001.87&{\bf r}   & 1.233$\pm$0.169&  0.00$\pm$0.01 &  0.0$\pm$0.5& 0.60$\pm$0.08&0.0923\\
2001.87&{\bf Ca}  & 0.266$\pm$0.047&  2.33$\pm$0.71 &  -94.8$\pm$1.5& 1.12$\pm$0.29& \\
2001.87&{\bf C8}  & 0.134$\pm$0.019&  6.15$\pm$0.98 &  -96.7$\pm$1.2  & 3.17$\pm$0.57& \\
2001.87&{\bf C12}  & 0.058$\pm$0.010& 9.45$\pm$0.42&   -97.0$\pm$0.9    & 3.70$\pm$0.11&\\
2001.87&     & 0.010$\pm$0.001&16.85$\pm$1.10&   -91.4$\pm$0.9      & 5.10$\pm$1.60&\\
2001.87&{\bf C30} & 0.241$\pm$0.034&  28.94$\pm$0.49&   -104.7$\pm$1.0  & 14.18$\pm$0.79&\\
2001.87&     & 0.045$\pm$0.007&43.07$\pm$0.61&   -100.0$\pm$0.8       & 15.98$\pm$3.20&\\
\hline \noalign{\smallskip}
             &      &      &{\small \bf 2.3 GHz}&           &      &\\
\hline \noalign{\smallskip}
1996.42&{\bf r} &1.642$\pm$0.246&   0.00      &     0.0     &   0.00$\pm$0.01 & 86.988178\\
1996.42&{\bf Ca} &0.481$\pm$0.072&   1.43$\pm$0.07&   -92.2$\pm$0.5&   0.01$\pm$0.01 & \\
1996.42&{\bf C4} &0.041$\pm$0.006&   3.91$\pm$0.39&   -85.3$\pm$1.0&   0.44$\pm$0.02&  \\
1996.42&{\bf C8} &0.124$\pm$0.025&   6.71$\pm$0.80&   -89.4$\pm$1.8&   3.19$\pm$0.32&  \\
1996.42&{\bf C12} &0.028$\pm$0.010&  11.92$\pm$2.38&  -102.7$\pm$1.9&   1.47$\pm$0.15&  \\
1996.42&{\bf C30}&0.204$\pm$0.069&  26.67$\pm$3.00&  -102.2$\pm$3.9&  13.47$\pm$2.70&  \\
\hline \noalign{\smallskip}
&&&{\small \bf 5.0 GHz} &&&\\
\hline \noalign{\smallskip}
1996.46&{\bf r}&1.673$\pm$0.251&   0.00&     0.0&   0.36$\pm$0.02&52.027293  \\
1996.46&{\bf Ca} &0.422$\pm$0.063&   1.35$\pm$0.07&   -96.8$\pm$1.1&   0.86$\pm$0.04& \\
1996.46&{\bf } &0.090$\pm$0.014&   5.58$\pm$0.60&   -89.9$\pm$1.54&   4.44$\pm$0.44& \\
\hline
1997.15&{\bf r} &1.302$\pm$0.195  &0.00   & 0.0  &0.01$\pm$0.01 &27.458208\\
1997.15&{\bf C1} &0.274$\pm$0.041 & 0.87$\pm$0.04  &-94.2$\pm$0.9 & 0.33$\pm$0.07& \\
1996.15&{\bf Ca} &0.218$\pm$0.033 & 1.73$\pm$0.09  &-97.3$\pm$0.7 & 0.53$\pm$0.08& \\
1997.15&{\bf C4} &0.017$\pm$0.002 & 3.95$\pm$0.40 & -85.0$\pm$1.4 & 0.47$\pm$0.10& \\
1997.15&{\bf C8} &0.081$\pm$0.016 & 6.36$\pm$0.66 & -99.7$\pm$1.9 & 4.42$\pm$0.44& \\
1997.15&{\bf C30} &0.067$\pm$0.013 &27.87$\pm$2.81 &-107.6$\pm$3.9 & 6.48$\pm$0.84&\\
\hline
1999.35&{\bf r}&  1.774$\pm$0.266 & 0.00  &  0.00&  0.30$\pm$0.02&15.738181 \\
1999.35&{\bf C1}&0.033$\pm$0.005 & 0.77$\pm$0.04 & -69.4$\pm$0.5 & 0.00$\pm$& \\
1999.35&{\bf Ca} &0.507$\pm$0.189 & 1.31$\pm$0.07 & -93.0$\pm$1.5 & 0.81$\pm$0.12& \\
1999.35&{\bf C4} &0.031$\pm$0.025 & 3.46$\pm$0.35 & -98.1$\pm$1.1 & 0.81$\pm$0.41& \\
\hline
1999.44&{\bf r} &1.423$\pm$0.213   &0.00     &0.0   &0.21$\pm$0.01  &\\
1999.44&{\bf C1} &0.450$\pm$0.067  &0.53$\pm$0.03   &-75.7$\pm$1.0   &0.22$\pm$0.01  & \\
1999.44&{\bf Ca} &0.274$\pm$0.041   &1.34$\pm$0.07   &-89.2$\pm$0.7   &0.49$\pm$0.03  & \\
1999.44&{\bf C2} &0.101$\pm$0.015   &1.86$\pm$0.09  &-100.6$\pm$0.8   &0.78$\pm$0.08  & \\
1999.44&{\bf C4}   &0.071$\pm$0.011   &4.81$\pm$0.48   &-94.8$\pm$1.1   &2.94$\pm$0.30  &\\
1999.44&{\bf C12}   &0.033$\pm$0.006  &10.00$\pm$1.00   &-93.9$\pm$1.6   &1.99$\pm$0.20  &\\
1999.44&{\bf C30}   &0.099$\pm$0.020  &28.27$\pm$2.83  &-104.2$\pm$2.7  &14.06$\pm$2.58  &\\
\hline
2001.29& {\bf r} &   1.602$\pm$0.246 &  0.00$\pm$0.01 &   0.0$\pm$0.5& 0.21$\pm$0.05& 0.0477   \\
2001.29& {\bf C1}   &   0.511$\pm$0.071 &  0.85$\pm$0.08 &   -81.1$\pm$2.1& 0.43$\pm$0.11&   \\
2001.29& {\bf Ca}   &   0.228$\pm$0.069 &  1.62$\pm$0.38 &   -91.7$\pm$3.5& 0.72$\pm$0.22&  \\
2001.29& {\bf C4}   &   0.045$\pm$0.023 &  3.89$\pm$0.18 &   -84.7$\pm$4.9& 1.38$\pm$0.61&    \\
2001.29& {\bf C8}   &   0.052$\pm$0.022 &  6.07$\pm$0.24 &   -98.3$\pm$0.8& 2.05$\pm$0.85&    \\
2001.29& {\bf C12}   &   0.039$\pm$0.018 &  9.23$\pm$0.35 &   -95.1$\pm$1.7& 3.22$\pm$0.68&    \\
2001.29& {\bf C30}  &   0.112$\pm$0.018 &  29.12$\pm$0.59 &   -103.3$\pm$1.2& 11.83$\pm$0.19&   \\
\hline \noalign{\smallskip}
&&&{\small \bf 8.4 GHz} &&&\\
\hline \noalign{\smallskip}
1993.88&{\bf r}&  1.390$\pm$0.208&  0.00 &   0.00&  0.01$\pm$ &2.8677791\\
1993.88&{\bf C0}& 0.014$\pm$0.002&  0.25$\pm$0.01 & -69.3$\pm$0.7&  0.00$\pm$0.01 & \\
1993.88&{\bf Cx}& 0.262$\pm$0.039&  0.33$\pm$0.02 & -99.5$\pm$0.4&  0.00$\pm$0.01 & \\
1993.88&{\bf Ca}& 0.263$\pm$0.039&1.45$\pm$0.07   & -91.8$\pm$0.8&  0.36$\pm$0.02 & \\
1993.88&{\bf C4}& 0.043$\pm$0.016&3.07$\pm$0.11 & -87.5$\pm$1.2&  1.03$\pm$0.17 &\\
\hline
1997.90&{\bf r}&  2.467$\pm$0.370&  0.00 &   0.00&  0.45$\pm$0.02 &6.8563583\\
1997.90&{\bf Ca}& 0.719$\pm$0.108&  1.39$\pm$0.07 & -97.4$\pm$1.3&  0.68$\pm$0.03 & \\
1997.90&{\bf C4}& 0.127$\pm$0.019&  4.00$\pm$0.45 & -89.5$\pm$1.3&  1.31$\pm$0.13 & \\
\hline
1997.93&{\bf r}&  1.353$\pm$0.203&  0.00 &   0.00&  0.12$\pm$0.01 &1.7453468\\
1997.93&{\bf C0}& 0.270$\pm$0.041&  0.28$\pm$0.01 & -83.5$\pm$1.9&  0.36$\pm$0.02 & \\
1997.93&{\bf C1}& 0.176$\pm$0.026&  1.01$\pm$0.05 & -86.6$\pm$0.6&  0.44$\pm$0.02 & \\
1997.93&{\bf Ca}& 0.212$\pm$0.032&  1.54$\pm$0.08 & -98.0$\pm$0.5&  0.45$\pm$0.02 & \\
1997.93&{\bf C2}& 0.035$\pm$0.005&  2.23$\pm$0.11 & -97.0$\pm$1.0&  0.76$\pm$0.04 &\\
1997.93&{\bf C4}& 0.004$\pm$0.001&  3.16$\pm$0.32 &-109.5$\pm$1.0&  0.28$\pm$0.03 &\\
1997.93&{\bf C4}& 0.028$\pm$0.004&  3.89$\pm$0.39 & -89.5$\pm$1.0&  1.38$\pm$0.14 &\\
1997.93&{\bf C8}& 0.039$\pm$0.008&  7.71$\pm$0.90 & -96.6$\pm$3.0&  3.00$\pm$0.80 &\\
\hline
1999.41&{\bf r}&1.257$\pm$0.188&   0.00  &  0.0 & 0.12$\pm$&1.5628510  \\
1999.41&{\bf C0}&0.496$\pm$0.074&   0.33$\pm$0.02&   -79.7$\pm$1.1&   0.28$\pm$0.01&  \\
1999.41&{\bf C1}&0.171$\pm$0.026&   0.92$\pm$0.05&   -78.9$\pm$0.5&   0.43$\pm$0.02&  \\
1999.41&{\bf Ca}&0.140$\pm$0.021&   1.50$\pm$0.08&   -92.5$\pm$0.5&   0.32$\pm$0.02&  \\
1999.41&{\bf C2}&0.086$\pm$0.013&   1.84$\pm$0.09&   -96.9$\pm$0.5&   0.71$\pm$0.04&  \\
1999.41&{\bf C4}&0.046$\pm$0.007&   3.43$\pm$0.34&   -93.9$\pm$1.0&   1.74$\pm$0.09&  \\
1999.41&{\bf C8}&0.025$\pm$0.004&   6.47$\pm$0.65&   -95.7$\pm$0.9&   2.43$\pm$0.24&  \\
1999.41&{\bf C12}&0.020$\pm$0.003&   9.40$\pm$0.90&   -95.6$\pm$1.0&   2.96$\pm$0.30&  \\
1999.41&{\bf C30}&0.036$\pm$0.006&  27.91$\pm$2.40&  -103.1$\pm$3.2&   7.51$\pm$0.75& \\
\hline
2001.09&{\bf r}&   1.317$\pm$0.146 &  0.00$\pm$0.01 &0.0$\pm$0.5  & 0.13$\pm$0.09 & 1.1756 \\
2001.09&{\bf C0}  &   0.437$\pm$0.068 &  0.37$\pm$0.07 &-81.0$\pm$1.1 & 0.30$\pm$0.03  & \\
2001.09&{\bf C1}  &   0.259$\pm$0.094 &  0.93$\pm$0.12 &-79.4$\pm$1.7 & 0.42$\pm$0.19   & \\
2001.09&{\bf Ca}  &   0.179$\pm$0.032 &  1.47$\pm$0.04 & -91.6$\pm$0.8& 0.45$\pm$0.11  &  \\
2001.09&{\bf C2}  &   0.059$\pm$0.008 & 1.95$\pm$0.04& -93.7$\pm$0.6& 0.72$\pm$0.04   & \\
2001.09&{\bf C4}  &   0.055$\pm$0.010 &  4.11$\pm$0.33 & -91.0$\pm$0.9 & 2.44$\pm$0.64 &  \\
2001.09&{\bf C8}  &   0.027$\pm$0.003 & 6.12$\pm$0.15& -96.5$\pm$0.3& 2.29$\pm$0.13   & \\
2001.09&{\bf C12}  &   0.024$\pm$0.008 & 8.95$\pm$0.73& -94.1$\pm$0.5& 2.90$\pm$0.37  &  \\
2001.09&{\bf C30} &   0.018$\pm$0.003 &27.61$\pm$0.12& -103.8$\pm$0.3& 7.18$\pm$0.19&    \\
\hline \noalign{\smallskip}
&&&{\small \bf 15 GHz} &&&\\
\hline \noalign{\smallskip}
1994.67&{\bf r}&1.115$\pm$0.167 & 0.00  &  0.00 & 0.13$\pm$0.01 &0.43219569\\
1994.67&{\bf C0}&0.176$\pm$0.030 & 0.39$\pm$0.03  &-92.8$\pm$3.9 & 0.40$\pm$0.05 & \\
1994.67&{\bf Ca}&0.097$\pm$0.015 & 1.46$\pm$0.04  &-93.9$\pm$1.5 & 0.61$\pm$0.05 & \\
1994.67&{\bf C4}&0.030$\pm$0.010 & 3.43$\pm$0.10  &-90.6$\pm$1.7 & 1.57$\pm$0.12 &\\
\hline
1996.38& {\bf r} & 1.559$\pm$0.234 &  0.00&     0.00 &  0.12$\pm$0.01&0.43219569  \\
1996.38& {\bf C0} & 0.572$\pm$0.086 &  0.25$\pm$0.01&   -93.7$\pm$2.1 &  0.23$\pm$0.01&  \\
1996.38& {\bf C1} & 0.094$\pm$0.014 &  0.67$\pm$0.03&   -98.0$\pm$1.1 &  0.35$\pm$0.02&  \\
1996.38& {\bf Ca} & 0.191$\pm$0.029 &  1.44$\pm$0.07&   -92.9$\pm$0.6 &  0.45$\pm0.02$&  \\
1996.38& {\bf C2} & 0.124$\pm$0.019 &  1.74$\pm$0.09&   -96.1$\pm$0.5 &  0.45$\pm$0.02&  \\
1996.38& {\bf C4} & 0.022$\pm$0.003 &  3.38$\pm$0.34&   -86.6$\pm$1.0 &  1.43$\pm$0.14&  \\
\hline
1997.20& {\bf r}  & 1.362$\pm$0.204&  0.00         &     0.00     &  0.11$\pm$0.01&1.4601196 \\
1997.20& {\bf C0} & 0.295$\pm$0.044&  0.30$\pm$0.02&-89.0$\pm$1.6 &  0.31$\pm$0.02&  \\
1997.20& {\bf C1} & 0.135$\pm$0.020 &  0.84$\pm$0.04&-90.3$\pm$0.9 &  0.55$\pm$0.03&  \\
1997.20& {\bf Ca} & 0.111$\pm$0.017 &  1.47$\pm$0.07&-96.5$\pm$0.5 &  0.39$\pm$0.02&  \\
1997.20& {\bf C2} & 0.101$\pm$0.015 &  1.78$\pm$0.09&-96.3$\pm$0.5 &  0.59$\pm$0.03&  \\
1997.20& {\bf C4} & 0.030$\pm$0.005 &  3.13$\pm$0.31&   -90.9$\pm$1.0 &  1.64$\pm$0.16&  \\
1997.20& {\bf C8} & 0.034$\pm$0.005 &  6.81$\pm$0.68&   -97.5$\pm$1.0 &  3.40$\pm$0.34&  \\
\hline
1998.22& {\bf r} & 1.089$\pm$0.163 &  0.00&     0.00 &  0.25$\pm$0.01&0.44147467   \\
1998.22& {\bf C0}&0.916$\pm$0.137&  0.23$\pm$0.01    &   -77.2$\pm$5.3 &  0.30$\pm$0.02&  \\
1998.22& {\bf C1} & 0.119$\pm$0.018 &  0.93$\pm$0.05 &   -83.7$\pm$1.2 &  0.47$\pm$0.02&  \\
1998.22& {\bf Ca} & 0.247$\pm$0.037 &  1.49$\pm$0.07 &   -95.0$\pm$0.7 &  0.46$\pm$0.02&  \\
1998.22& {\bf C2} & 0.049$\pm$0.007 &  2.04$\pm$0.20      &   -96.8$\pm$1.0 &  0.62$\pm$0.06&  \\
1998.22& {\bf C4} & 0.038$\pm$0.008 &  3.40$\pm$0.34      &   -89.6$\pm$1.4 &  2.07$\pm$0.21&  \\
\hline
1998.84&{\bf r}& 1.113$\pm$0.167 &  0.00 &    0.00 &  0.08$\pm$0.01&1.4787256  \\
1998.84& {\bf C0}& 0.650$\pm$0.098 &  0.18$\pm$0.02 &  -86.2$\pm$9.4 &  0.35$\pm$0.02&  \\
1998.84& {\bf C1}& 0.102$\pm$0.015 &  0.92$\pm$0.05 &  -75.6$\pm$2.8 &  0.60$\pm$0.03& \\
1998.84& {\bf Ca}& 0.141$\pm$0.021 &  1.42$\pm$0.07 &  -92.4$\pm$0.6 &  0.22$\pm$0.01& \\
1998.84& {\bf C2}& 0.147$\pm$0.022 &  1.67$\pm$0.08 &  -99.0$\pm$1.3 &  0.53$\pm$0.03& \\
\hline
1999.57& {\bf r} &2.015$\pm$0.302&   0.00      &     0.0     &   0.12$\pm$0.01&1.1925775 \\
1999.57& {\bf C0} &0.430$\pm$0.064&   0.33$\pm$0.02&   -77.0$\pm$2.1&   0.26$\pm$0.01& \\
1999.57& {\bf C1} &0.178$\pm$0.027&   0.84$\pm$0.04&   -77.1$\pm$0.9&   0.31$\pm$0.02& \\
1999.57& {\bf Ca} &0.048$\pm$0.007&   1.32$\pm$0.07&   -83.5$\pm$0.6&   0.01$\pm$&  \\
1999.57& {\bf C2} &0.131$\pm$0.020&   1.57$\pm$0.08&   -92.5$\pm$0.6&   0.42$\pm$0.07&  \\
1999.57& {\bf   } &0.040$\pm$0.006&   1.98$\pm$0.10&   -93.4$\pm$1.0&   0.68$\pm$0.11&  \\
1999.57& {\bf C4} &0.032$\pm$0.005&   3.28$\pm$0.33&   -92.8$\pm$1.1&   1.57$\pm$0.14&  \\
1999.57& {\bf   } &0.007$\pm$0.002&   5.14$\pm$0.51&   -88.7$\pm$2.0&   1.01$\pm$0.18&  \\
1999.57& {\bf C8} &0.020$\pm$0.004&   7.38$\pm$0.74&   -96.2$\pm$2.0&   2.47$\pm$0.34&  \\
1999.57& {\bf C12} &0.006$\pm$0.001&  10.26$\pm$1.16&   -91.7$\pm$2.0&   3.04$\pm$0.71&  \\
\hline
1999.85& {\bf r} &1.603$\pm$0.241 & 0.00  &  0.0 & 0.090$\pm$0.01 &1.1036355\\
1999.85& {\bf C0} &0.449$\pm$0.067 & 0.28$\pm$0.05 & -79.6$\pm$2.9&  0.24$\pm$0.01& \\
1999.85& {\bf C1} &0.168$\pm$0.024 & 0.94$\pm$0.06 & -76.2$\pm$1.8&  0.39$\pm$0.02& \\
1999.85& {\bf Ca} &0.093$\pm$0.015 & 1.38$\pm$0.17 & -86.8$\pm$0.6&  0.17$\pm$0.01& \\
1999.85& {\bf C2} &0.098$\pm$0.015 & 1.64$\pm$0.10 & -94.5$\pm$0.5&  0.42$\pm$0.02& \\
1999.85& {\bf C4} &0.039$\pm$0.007 & 2.44$\pm$0.13 & -97.5$\pm$2.8&  1.27$\pm$0.13& \\
1999.85& {\bf } &0.021$\pm$0.004 & 4.55$\pm$0.12 & -88.1$\pm$1.0  &1.61$\pm$0.16& \\
1999.85& {\bf C8} &0.013$\pm$0.003 & 7.86$\pm$0.18 & -97.6$\pm$2.0  &2.05$\pm$0.21& \\
\hline
2000.46& {\bf r} & 1.606$\pm$0.241&  0.00&     0.00 &  0.13$\pm$0.01&0.46644555  \\
2000.46& {\bf C0} & 0.364$\pm$0.055&  0.35$\pm$0.02 & -79.2$\pm$1.8&0.27$\pm$0.01& \\
2000.46& {\bf C1} & 0.144$\pm$0.022 &  0.96$\pm$0.05& -79.3$\pm$0.7&0.36$\pm$0.02& \\
2000.46& {\bf Ca} & 0.134$\pm$0.020 &  1.44$\pm$0.07& -88.8$\pm$0.5&0.29$\pm$0.02&\\
2000.46& {\bf C2} & 0.105$\pm$0.016 &  1.70$\pm$0.03& -92.1$\pm$0.7&0.62$\pm$0.03&\\
2000.46& {\bf C4} & 0.055$\pm$0.008 &  3.61$\pm$0.72& -91.9$\pm$1.1&1.95$\pm$0.20&\\
\hline
2002.13&{\bf r}&   1.702$\pm$  0.251&0.00  $\pm$   0.01&     0.0$\pm$  0.3& 0.11$\pm$0.02&1.3686\\
2002.13&{\bf C0}&  0.151$\pm$  0.023&0.37$\pm$ 0.01&  -94.6$\pm$  1.2& 0.29$\pm$0.01& \\
2002.13&{\bf B3}&  0.008$\pm$  0.001&0.76$\pm$ 0.02&   -89.8$\pm$  1.2& 0.14$\pm$0.03& \\
2002.13&{\bf C1}&  0.072$\pm$  0.012&1.02$\pm$ 0.02&   -75.2$\pm$  0.5& 0.35$\pm$0.02&\\
2002.13&{\bf Ca}&  0.318$\pm$  0.043&1.43$\pm$ 0.02&   -88.1$\pm$  0.5& 0.48$\pm$0.05&\\
2002.13&{\bf C2}&  0.029$\pm$  0.004&2.02$\pm$ 0.02&   -94.0$\pm$  0.5& 0.68$\pm$0.01&\\
2002.13&{\bf C4}&  0.031$\pm$  0.005&3.61$\pm$ 0.05&   -89.0$\pm$  0.8& 1.94$\pm$0.02&\\
2002.13&{\bf C8}&  0.044$\pm$  0.007&6.37$\pm$ 0.09&   -95.1$\pm$  0.8& 3.40$\pm$0.02&\\
2002.13&{\bf C12}&  0.009$\pm$  0.001&11.36$\pm$    0.37&   -94.6$\pm$  1.9& 3.37$\pm$0.67&\\
\hline
2003.10 &{\bf r}&   1.245$\pm$  0.179&0.00$\pm$ 0.02&    0.0$\pm$    0.2& 0.10$\pm$0.03&1.5375\\
2003.10&{\bf C0}&  0.310$\pm$  0.048&0.31$\pm$ 0.03&   -76.9$\pm$  1.7& 0.27$\pm$0.01& \\
2003.10&{\bf C1}&  0.039$\pm$  0.008&0.66$\pm$ 0.03&   -84.2$\pm$  1.3& 0.09$\pm$0.04& \\
2003.10&{\bf Ca}&  0.287$\pm$  0.058&1.42$\pm$ 0.01&   -88.4$\pm$  0.3& 0.35$\pm$0.09&\\
2003.10&{\bf C2}&  0.113$\pm$  0.045&1.78$\pm$ 0.24&   -89.1$\pm$  0.8& 1.00$\pm$0.26&\\
2003.10&{\bf B3}&  0.035$\pm$  0.016&2.64$\pm$ 0.38&   -88.9$\pm$  1.2& 1.17$\pm$0.19&\\
2003.10&{\bf C4}&  0.039$\pm$  0.006&4.67$\pm$ 0.32&   -90.0$\pm$  0.8& 2.01$\pm$0.35&\\
2003.10&{\bf C8}&  0.031$\pm$  0.004&7.44$\pm$ 0.49&   -97.1$\pm$  0.7& 3.00$\pm$0.16&\\
2003.10&{\bf C12}&  0.015$\pm$  0.004&12.37$\pm$    0.78&   -93.2$\pm$  0.6& 4.31$\pm$0.64&\\
\hline
2004.27&{\bf r}&   1.264$\pm$  0.185&0.00$\pm$ 0.01&   0.0$\pm$    0.1& 0.09$\pm$0.03&1.4844\\
2004.27&{\bf C0}&  0.117$\pm$  0.017&0.33$\pm$ 0.01&   -89.1$\pm$  1.3& 0.25$\pm$0.01& \\
2004.27&{\bf C1}&  0.124$\pm$  0.018&1.00$\pm$ 0.05&   -78.0$\pm$  0.6& 0.38$\pm$0.05& \\
2004.27&{\bf Ca}&  0.185$\pm$  0.028&1.47$\pm$ 0.02&   -90.6$\pm$  0.4& 0.31$\pm$0.06&\\
2004.27&{\bf C2}&  0.046$\pm$  0.009&1.76$\pm$ 0.06&   -95.6$\pm$  0.5& 0.49$\pm$0.04&\\
2004.27&{\bf C4}&  0.044$\pm$  0.007&3.64$\pm$ 0.09&   -86.9$\pm$  0.9& 1.83$\pm$0.05&\\
2004.27&{\bf C8}&  0.046$\pm$  0.007&6.42$\pm$ 0.24&   -96.3$\pm$  0.8& 2.71$\pm$0.08&\\
2004.27&{\bf C12}&  0.014$\pm$  0.003&9.81$\pm$ 0.30&   -96.3$\pm$  0.6& 3.09$\pm$0.55&\\
\hline
2005.30&{\bf r}&   0.819$\pm$  0.119&0.00$\pm$ 0.01&   0.0$\pm$    0.2& 0.12$\pm$0.04&0.9761\\
2005.30&{\bf C0}&  0.160$\pm$  0.028&0.22$\pm$ 0.02&   -61.6$\pm$  0.3 & 0.05$\pm$0.03& \\
2005.30&{\bf C1}&  0.091$\pm$  0.014&0.61$\pm$ 0.02&   -75.8$\pm$  1.3 & 0.35$\pm$0.02& \\
2005.30&{\bf Ca}&  0.297$\pm$  0.045&1.47$\pm$ 0.01&   -87.4$\pm$  0.6 & 0.42$\pm$0.03&\\
2005.30&{\bf C2}&  0.040$\pm$  0.011&1.98$\pm$ 0.06&   -84.1$\pm$  0.5 & 0.51$\pm$0.07&\\
2005.30&{\bf C4}&  0.018$\pm$  0.005&3.98$\pm$ 0.09&   -84.9$\pm$  0.8& 1.12$\pm$0.34&\\
2005.30&{\bf C8}&  0.064$\pm$  0.009&6.08$\pm$ 0.30&   -94.7$\pm$  1.0& 3.04$\pm$0.23&\\
2005.30&{\bf C12}&  0.010$\pm$  0.005&9.80$\pm$ 0.97&   -91.9$\pm$  2.5& 2.90$\pm$0.70&\\
\hline
2005.68&{\bf r}&   1.381$\pm$  0.203&0.00$\pm$ 0.01&   0.0$\pm$    0.3& 0.09$\pm$0.02&0.8685\\
2005.68&{\bf C0}&  0.146$\pm$  0.022&0.25$\pm$ 0.03&   -74.2$\pm$  1.5 & 0.21$\pm$0.02& \\
2005.68&{\bf C1}&  0.048$\pm$  0.008&0.75$\pm$ 0.02&   -78.0$\pm$  0.8 & 0.32$\pm$0.03& \\
2005.68&{\bf Ca}&  0.244$\pm$  0.064&1.44$\pm$ 0.02&   -89.0$\pm$  0.3 & 0.31$\pm$0.09&\\
2005.68&{\bf C2}&  0.120$\pm$  0.019&1.71$\pm$ 0.04&   -88.8$\pm$  0.5 & 0.55$\pm$0.02&\\
2005.68&{\bf C4}&  0.021$\pm$  0.007&3.44$\pm$ 0.29&   -87.0$\pm$  1.3 & 1.50$\pm$0.20&\\
2005.68&{\bf B3}&  0.013$\pm$  0.001&4.72$\pm$ 0.06&   -83.7$\pm$  0.4 & 1.16$\pm$0.03&\\
2005.68&{\bf C8}&  0.058$\pm$  0.006&6.37$\pm$ 0.11&   -97.4$\pm$  0.6 & 2.49$\pm$0.18&\\
2005.68&{\bf C12}&  0.022$\pm$  0.001&8.98$\pm$ 0.20&   -93.7$\pm$  0.7& 3.68$\pm$0.22&\\
\hline
\end{supertabular}
\twocolumn
\clearpage

\onecolumn
\tabcolsep1.0mm
\tablecaption{Gaussian model-fit results for S5 1803+784 obtained from the literature. The columns list the epoch of observation, the jet-component identification, the flux density, the radial distance of the component center from the core,  the position angle of the center of the component, the FWHM major axis, and
for elliptical Gaussian fits, the axial ratio and the position angle of the major axis.  The final column provides a reference to the original publication.}
\setcounter{table}{2}
\label{literatur}
{\smallskip}
\hspace*{-12cm}
\tablehead{\noalign{\smallskip} \hline \noalign{\smallskip}
(1)   &(2)&  (3) &(4) & (5) &(6)&(7)&(8)&(9)\\
\hline
{\small Epoch} & {\small Id.} & {\small S [Jy]}      & {\small r [mas]}     & {\small $\theta$ [deg]} & {\small Ma.A. [mas]}  &  {\small Axial ratio}&{\small $\Phi$ [deg]}&{\small Reference}\\
\noalign{\smallskip} \hline \noalign{\smallskip}}
\tabletail{\hline\multicolumn{5}{r}{\small continued on next  page}\\}
\tablelasttail{\hline}
\begin{supertabular}{ccrrrrrrccc}
&&&&{\small 2.32 GHz} &&&&\\
\hline
1994.52&{\bf 1}&  1.63 & 0.00  & 0.00 & 1.37 &  0.00 &  87   & Fey et al. 1996\\
1994.52&{\bf 2}&  0.34 & 4.0  & -94 & 7.24 &  0.28 & 81  & \\
1994.52&{\bf 3}&  0.28 & 27.6  & -105 & 12.11 &  1.00 &   & \\
\hline \noalign{\smallskip}
&&&&{\small 5 GHz} &&&&\\
\hline \noalign{\smallskip}
1997.79&{\bf C1}&  0.413 & 1.38  & -98 & 0.65 &  0.82 & 50  & Lister et al. 2001\\
\hline \noalign{\smallskip}
       &      &      &      &{\small 8 GHz}&      &     &      &\\
\hline \noalign{\smallskip}
1988.79&{\bf Core} &  2.45 & 0.00 &   0.0  & 0.1 & & &Tateyama et al. 2002\\
1988.79&{\bf C1}&  0.23 & 1.50 & -95 & 0.1 & & &\\
1988.79&{\bf C2}&  0.21 & 0.90 &  -95 & 0.1 & &  &\\
\hline
1989.96&{\bf Core}&2.77 &  0.00 &  0.0 &  0.1 &  & &Tateyama et al. 2002\\
1989.96&{\bf C1}&  0.22 & 1.51 &  -94 & 0.1 & &  &\\
1989.96&{\bf C2}&  0.20 & 1.09 &  -90 & 0.3 & &  &\\
1989.96&{\bf E}&  0.65 & 0.40 &  -81 & 0.2 & &  &\\
\hline
1991.96&{\bf Core} & 1.38 &  0.00 &  0.0 &  0.1&  & &Tateyama et al. 2002\\
1991.96&{\bf C1}&  0.14 & 1.71 &  -90.9 & 0.1 & &  &\\
1991.96&{\bf C2}&  0.27 & 1.28 &  -87.4 & 0.1 & &  &\\
1991.96&{\bf E}&  0.06 & 0.58 &  -85.1 & 0.1 & &  &\\
\hline
1992.62&{\bf Core} & 1.68 &  0.00 &  0.0 &  0.2 &  & &Tateyama et al. 2002\\
1992.62&{\bf C1}&  0.11 & 1.80 &  -90 & 0.1 & &  &\\
1992.62&{\bf C2}&  0.40 & 1.27 &  -95 & 0.1 & &  &\\
1992.62&{\bf E}&  0.32 & 0.48 &  -93 & 0.01 & &  &\\
\hline
1993.87&{\bf Core} & 1.25 &  0.00 &  0.0 &  0.2&  & &Tateyama et al. 2002\\
1993.87&{\bf C2}&  0.33 & 1.39 &  -97.2 & 0.1 & &  &\\
\hline
1994.71&{\bf Core} & 1.49 &  0.00 &  0.0 &  0.2&  & &Tateyama et al. 2002\\
1994.71&{\bf C2}&  0.17 & 1.43 &  -96.2 & 0.1 & &  &\\
\hline
1995.71&{\bf Core} & 2.10 &  0.00 &  0.0 &  0.25&  & &Tateyama et al. 2002\\
1995.71&{\bf C2}&  0.28 & 1.52 &  -93.0 & 0.1 & &  &\\
1995.71&{\bf C3}&  0.18 & 1.12 &  -93.1 & 0.1 & &  &\\
\hline
1996.29&{\bf Core} & 1.57 &  0.00 &  0.0 &  0.2&  & &Tateyama et al. 2002\\
1996.29&{\bf C2}&  0.10 & 1.59 &  -91 & 0.1 & &  &\\
1996.29&{\bf C3}&  0.04 & 1.10 &  -97 & 0.1 & &  &\\
1996.29&{\bf E}&  0.18 & 0.44 &  -101 & 0.1 & &  &\\
\hline
1996.71&{\bf Core} & 1.56 &  0.00 &  0.0 &  0.2&  & &Tateyama et al. 2002\\
1996.71&{\bf C2}&  0.15 & 1.73 &  -92.4 & 0.1 & &  &\\
1996.71&{\bf C3}&  0.10 & 1.25 &  -96.1 & 0.1 & &  &\\
1996.71&{\bf E}&  0.17 & 0.51 &  -89.2 & 0.1 & &  &\\
\hline
1998.79&{\bf Core} & 1.60 &  0.00 &  0.0 &  0.2&  & &Tateyama et al. 2002\\
1998.79&{\bf C2}&  0.13 & 1.82 &  -97.0 & 0.1 & &  &\\
1998.79&{\bf C3}&  0.26 & 1.30 &  -93.7 & 0.1 & &  &\\
1998.79&{\bf E}&  0.41 & 0.40 &  -93.3 & 0.1 & &  &\\
\hline
1999.62&{\bf Core} & 1.89 &  0.00 &  0.0 &  0.2&  & &Tateyama et al. 2002\\
1999.62&{\bf C2}&  0.09 & 1.86 &  -90.0 & 0.1 & &  &\\
1999.62&{\bf C3}&  0.19 & 1.33 &  -88.1 & 0.1 & &  &\\
1999.62&{\bf E}&  0.28 & 0.62 &  -77.8 & 0.1 & &  &\\
\hline \noalign{\smallskip}
&&&&{\small 8.4 GHz} &&&&\\
\hline \noalign{\smallskip}
1997.93&{\bf XA}&  1.493$\pm$0.001 & 0.00  &  & 0.2$\pm$0.1 &  0.8$\pm$0.1 & +119$\pm$1  & Ros et al. 2001\\
1997.93&{\bf XB}&  0.231$\pm$0.001 & 0.6$\pm$0.2  & -86$\pm$19 & 0.7$\pm$0.1 & 0.6$\pm$0.1 & +102$\pm$1  &\\
1997.93&{\bf XC}&  0.274$\pm$0.001 & 1.4$\pm$0.2  & -97$\pm$8 & 0.6$\pm$0.1 &  0.7$\pm$0.1 & +41$\pm$1  &\\
1997.93&{\bf XD}&  0.036$\pm$0.001 & 2.0$\pm$0.2  & -98$\pm$5 & 0.5$\pm$0.1 &  0.2$\pm$0.1 & -12$\pm$4  &\\
1997.93&{\bf XE}&  0.024$\pm$0.001 & 2.7$\pm$0.2  & -97$\pm$4 & 1.4$\pm$0.1 &  0.4$\pm$0.1 & +1$\pm$3  &\\
1997.93&{\bf XF}&  0.018$\pm$0.001 & 3.9$\pm$0.2  & -89$\pm$3 & 1.1$\pm$0.1 &  0.9$\pm$0.1 & -30$\pm$16  &\\
1997.93&{\bf XG}&  0.019$\pm$0.001 & 6.3$\pm$0.2  & -96$\pm$3 & 2.8$\pm$0.2 &  0.5$\pm$0.1 & +31$\pm$4  &\\
1997.93&{\bf XH}&  0.025$\pm$0.001 & 8.9$\pm$0.2  & -95$\pm$1 & 4.1$\pm$0.2 &  0.5$\pm$0.1 & +99$\pm$4  &\\
\hline
1999.41&{\bf XA}&  1.517$\pm$0.001 & 0.00  &  & 0.2$\pm$0.1 &  0.6$\pm$0.1 & +107$\pm$1  & Ros et al. 2001\\
1999.41&{\bf XB}&  0.374$\pm$0.001 & 0.5$\pm$0.2  & -81$\pm$21 & 0.5$\pm$0.1 &  0.7$\pm$0.1 & +117$\pm$1  &\\
1999.41&{\bf XC}&  0.192$\pm$0.001 & 1.4$\pm$0.2  & -93$\pm$8 & 0.5$\pm$0.1 &  0.3$\pm$0.1 & +22$\pm$1  &\\
1999.41&{\bf XD}&  0.058$\pm$0.001 & 1.9$\pm$0.2  & -94$\pm$6 & 0.6$\pm$0.1 &  0.2$\pm$0.1 & +3$\pm$2  &\\
1999.41&{\bf XE}&  0.030$\pm$0.001 & 2.7$\pm$0.2  & -93$\pm$4 & 1.9$\pm$0.1 &  0.4$\pm$0.1 & +126$\pm$2  &\\
1999.41&{\bf XF}&  0.030$\pm$0.001 & 4.0$\pm$0.2  & -97$\pm$3 & 2.4$\pm$0.1 &  0.4$\pm$0.1 & +111$\pm$1  &\\
1999.41&{\bf XG}&  0.020$\pm$0.001 & 7.0$\pm$0.2  & -97$\pm$2 & 2.9$\pm$0.2 &  0.6$\pm$0.1 & +115$\pm$5  &\\
1999.41&{\bf XH}&  0.017$\pm$0.001 & 9.4$\pm$0.2  & -97$\pm$1 & 3.6$\pm$0.3 &  0.6$\pm$0.1 & +84$\pm$5  &\\
\hline \noalign{\smallskip}
&&&&{\small 8.55 GHz} &&&&\\
\hline \noalign{\smallskip}
1994.52&{\bf 1}&  1.91 & 0.00  &  & 0.19 &  0.63 & -81  & Fey et al. 1996\\
1994.52&{\bf 2}&  0.38 & 1.0  & -92 & 1.53 &  0.28 & 83  &\\
1994.52&{\bf 3}&  0.05 & 3.5  & -86 & 1.26 &  1.00 &   &\\
1994.52&{\bf 4}&  0.05 & 6.6  & -95 & 2.57 &  1.00 &   &\\
\hline \noalign{\smallskip}
&&&&{\small 86 GHz} &&&&\\
\hline \noalign{\smallskip}
1993.29&{\bf C}&  0.448$\pm$0.063 & 0.00  &  & 0.08$\pm$0.03 &   &   & Lobanov et al. 2000\\
1993.29&{\bf J1}&  0.376$\pm$0.056 & 0.27$\pm$0.06  & -0.9$\pm$8.0 & 0.08$\pm$0.03 &   &   &\\
\noalign{\smallskip}
\hline
\end{supertabular}
\twocolumn
\begin{table*}[htb]
\setcounter{table}{6}
\begin{center}
\caption{Discrete cross-correlations (based on components from the 8 GHz model fits) and
Pearson's correlation coefficients (from 15 GHz model fits), calculated between the time series of various pairs of
jet parameters and/or the total flux densities. Column 1 gives the
frequency of the analyzed data set; colum 2 lists the
individual jet component from which the parameters to correlate are taken; columns 3 and 4 list the parameters correlated.  For 8 GHz data, the DCF peak and
$\tau$ are the peak of
discrete cross-correlation function and the associated time lag in years. 
For 15 GHz data, Corr.coeff.
is the value of the Pearson's correlation coefficient, and $p$ is
the probability of getting such a high correlation by chance.}
\label{dcf_total}
\medskip
\begin{tabular}{llrrcccc}
\hline
$\nu$ &Comp. & Par. 1 & Par. 2& DCF & $\tau$ & Corr. & $p$ \\
 $\rm [GHz]$& ID &  & & peak& & coeff. & \\
 \hline
 8&Ca & core sep. & total flux 8 GHz & -0.68 $\pm$ 0.06 & -1.12 $\pm$ 0.05 & & \\
 &   & p.a. & total flux 8 GHz  & -0.77 $\pm$ 0.04 & 0.72 $\pm$ 0.06 & & \\
 &   C2$_{1}$ & core sep. & total flux 8 GHz & 0.56 $\pm$ 0.06 & -0.07 $\pm$ 0.08 & & \\
 &   C2$_{3}$ & p.a. & total flux 8 GHz  & 0.60 $\pm$ 0.07 & -0.15 $\pm$ 0.08 & & \\
 &   C1 & core sep. & total flux 14.5 GHz & 0.41 $\pm$ 0.07 & -0.07 $\pm$ 0.04 & & \\
 &      & p.a. & total flux 14.5 GHz  & 0.61 $\pm$ 0.05 & 0.06 $\pm$ 0.06 & & \\
       \hline
       8& C1 & p.a. & core sep. & 0.54 $\pm$ 0.09 & 0.20 $\pm$ 0.05 & & \\
       &    & p.a. & flux & 0.53 $\pm$ 0.08 & -0.08 $\pm$ 0.04 & & \\
       &       & core sep. & flux & 0.41 $\pm$ 0.06 & -0.58 $\pm$ 0.06 & & \\
       &       Ca & p.a. & core sep. & -0.59 $\pm$ 0.06 & -0.59 $\pm$ 0.05 & & \\
       &          & p.a. & flux & 0.42 $\pm$ 0.06 & -0.56 $\pm$ 0.06 & & \\
       &        & core sep. & flux  & -0.64 $\pm$ 0.06 & -0.03 $\pm$ 0.04 & & \\
       &        C2$_{1}$ & p.a. & core sep. & -0.53 $\pm$ 0.09 & 0.31 $\pm$ 0.08 & & \\
       &        C2$_{2}$ & p.a. & core sep. & -0.61 $\pm$ 0.08 & -0.42 $\pm$ 0.04 & & \\
       &        C4 & p.a. & core sep. & 0.97 $\pm$ 0.10 & 0.09 $\pm$ 0.03 & & \\
                \hline
15&C0 & p.a. & core sep. &&& -0.44 $\pm$ 0.09 & 0.131 \\
&   & core sep. & flux &&& -0.52 $\pm$ 0.08 & 0.067 \\
&      C1 & p.a. & core sep. &&& 0.40 $\pm$ 0.10 & 0.198 \\
&            & core sep. & flux &&& 0.46 $\pm$ 0.09 & 0.164 \\
&             Ca & p.a. & core sep. &&& -0.73 $\pm$ 0.05 & 0.008 \\
&                  & core sep. & flux &&&  0.57 $\pm$ 0.07 & 0.054 \\
&                       C2 & core sep. & flux &&& -0.83 $\pm$ 0.03 & 0.002 \\
&                            C4 & p.a. & core sep. &&& 0.58 $\pm$ 0.07 & 0.064 \\
&                                    & p.a. & flux &&& -0.51 $\pm$ 0.08 & 0.107 \\
\hline
\end{tabular}
\end{center}
\end{table*}

\begin{table*}[htb]
\setcounter{table}{7}
\begin{center}
\caption{Discrete cross-correlations (based on components from the 8 GHz model fits) and Pearson's correlation
coefficients (from 15 GHz model fits), calculated for various pairs
jet parameters and the core flux density.
Column 1 gives the frequency of the analyzed data set; columns 2 and 3 lists the two jet
components participating in the correlation;
column 4 shows the circular Gaussian parameter correlated.  When "core" is
the first component, its parameter is always flux-density.
For 8 GHz data, DCF peak and $\tau$ are the peak of the discrete
correlation function and the associated time lag in years.  For 15 GHz
data, Corr.coeff.\ is
the value of the Pearson's correlation coefficient, and $p$ the
probability of getting such a high correlation by chance.}
\label{dcf_comp}
\medskip
\begin{tabular}{lllrcccc}
\hline
$\nu$ &Comp.  & Comp. & Par. & DCF & $\tau$ & Corr. & $p$ \\
$\rm [GHz]$& ID 1 &  ID 2 &  & peak & & coeff. & \\
\hline
8&C0 &  Ca  & core sep. & -0.48 $\pm$ 0.12 & 0.54 $\pm$ 0.07 & & \\
&C0 &  C2$_{1}$ & p.a. & -0.49 $\pm$ 0.14 & -0.32 $\pm$ 0.06 & & \\
&C0 &  C2$_{3}$ & p.a. & 0.69 $\pm$ 0.18 & 0.14 $\pm$ 0.08 & & \\
&C1 &  Ca  & core sep. & 0.47 $\pm$ 0.06 & 0.08 $\pm$ 0.06 & & \\
&   &      & p.a. & -0.48 $\pm$ 0.07 & 0.48 $\pm$ 0.07 & & \\
&C1 &  C2$_{1}$ & core sep. & -0.59 $\pm$ 0.09 & -0.80 $\pm$ 0.09 & & \\
&   &      & p.a. & -0.45 $\pm$ 0.09 & -0.29 $\pm$ 0.09 & & \\
&C1 &  C2$_{2}$  & core sep. & 0.53 $\pm$ 0.14 & -0.02 $\pm$ 0.06 & & \\
&   &      & p.a. & 0.96 $\pm$ 0.03 & -0.02 $\pm$ 0.02 & & \\
&C1 &  C2$_{3}$  & core sep. & 0.83 $\pm$ 0.06 & 0.81 $\pm$ 0.02 & & \\
&   &      & p.a. & 0.72 $\pm$ 0.09 & -0.06 $\pm$ 0.06 & & \\
&Ca &  C2$_{1}$  & core sep. & -0.48 $\pm$ 0.09 & -0.09 $\pm$ 0.06 & & \\
&Ca &  C2$_{3}$  & core sep. & 0.62 $\pm$ 0.09 & 0.55 $\pm$ 0.06 & & \\
&   &      & p.a. & -0.61 $\pm$ 0.07 & 0.49 $\pm$ 0.08 & & \\
&C2$_{1}$ &  C2$_{2}$  & p.a. & -0.53 $\pm$ 0.08 & 0.42 $\pm$ 0.05 & & \\
&C2$_{1}$ &  C2$_{3}$  & core sep. & 0.62 $\pm$ 0.09 & -0.42 $\pm$ 0.06 & & \\
&   &      & p.a. & -0.78 $\pm$ 0.09 & 0.17 $\pm$ 0.02 & & \\
&C2$_{2}$ &  C2$_{3}$  & p.a. & 0.79 $\pm$ 0.05 & -0.21 $\pm$ 0.05 & & \\
\hline
8&Core &  C0  & core sep. & 0.62 $\pm$ 0.09 & -0.23 $\pm$ 0.07 & & \\
&     &      & p.a.      & 0.77 $\pm$ 0.10 & -0.09 $\pm$ 0.05 & & \\
&          Core &  C1  & core sep. & -0.47 $\pm$ 0.07 & 0.35 $\pm$ 0.08 & & \\
&               &      & p.a.      & -0.45 $\pm$ 0.07 & -0.31 $\pm$ 0.04 & & \\
&                     Core &  Ca  & flux & 0.59 $\pm$ 0.06 & 0.49 $\pm$ 0.10 & & \\
&                                &      & core sep. & -0.91 $\pm$ 0.04 & 0.23 $\pm$ 0.03 & & \\
&                            &      & p.a.      & -0.83 $\pm$ 0.04 & -0.92 $\pm$ 0.04 & & \\
&                            Core &  C2$_{1}$  & core sep. & 0.64 $\pm$ 0.07 & 0.02 $\pm$ 0.05 & & \\
&                            Core &  C2$_{2}$  & p.a.      & -0.47 $\pm$ 0.08 & -0.29 $\pm$ 0.05 & & \\
&                    Core &  C2$_{3}$  & core sep. & 0.77 $\pm$ 0.05 & -0.36 $\pm$ 0.04 & & \\
&                              &      & p.a.      & 0.54 $\pm$ 0.08 & 0.37 $\pm$ 0.10 & & \\
\hline
15&Core &  C0  & core sep. & & & 0.65 $\pm$ 0.05 & 0.021 \\
&Core &  Ca  & core sep. & & & -0.72 $\pm$ 0.05 & 0.008 \\
&     &      & p.a.      & & & 0.49 $\pm$ 0.09 & 0.146 \\
&          Core &  C2  & core sep. &&& -0.49 $\pm$ 0.08 & 0.124 \\
&          Core &  C4  & p.a.      &&& -0.56 $\pm$ 0.08 & 0.074 \\
\hline
15&C0 &  C2  & p.a. &&&  0.77 $\pm$ 0.04 & 0.006 \\
&C1 &  Ca  & flux &&& -0.84 $\pm$ 0.03 & 0.001 \\
&   &      & p.a. &&& -0.67 $\pm$ 0.06 & 0.025 \\
&      Ca &  C2  & flux &&& -0.60 $\pm$ 0.07 & 0.051 \\
&            &      & core sep. &&& 0.69 $\pm$ 0.06 & 0.019 \\
&                &      & p.a. &&& 0.54 $\pm$ 0.08 & 0.084 \\
             \hline
                          \end{tabular}
                               \end{center}
                                    \end{table*}

\end{document}